\documentclass[11pt, a4paper]{article}
\usepackage[top=2.54cm,bottom=2.54cm,right=2.54cm,left=2.54cm]{geometry}
\pdfoutput=1
\usepackage[utf8]{inputenc}
\usepackage{amsmath}
\usepackage{dsfont}
\usepackage{amssymb}
\usepackage{booktabs}
\usepackage{epstopdf}
\usepackage{graphicx}
\usepackage[center]{caption}
\usepackage{subcaption}
\usepackage{verbatim}
\usepackage{bbm}
\usepackage{xcolor}
\usepackage{enumitem}
\usepackage[title]{appendix}
\usepackage{multirow}
\usepackage{hyperref}
\usepackage{apacite}
\usepackage[explicit]{titlesec}
\usepackage{xfrac}
\usepackage{nicefrac}

\usepackage{algpseudocode}
\usepackage{algorithm}
\algrenewcommand\algorithmicrequire{\textbf{Input:}}
\algrenewcommand\algorithmicensure{\textbf{Output:}}

\newcommand{\Q}{\mathbb{Q}}
\newcommand{\E}{\mathbb{E}}
\newcommand{\R}{\mathbb{R}}

\newcommand{\diff}{\mathrm{d}}
\newcommand{\dt}{\mathrm{dt}}

\DeclareMathOperator*{\plim}{plim}

\newtheorem{proposition}{Proposition}
\newtheorem{assumption}{Assumption}
\newtheorem{lemma}{Lemma}

\title{iCOS: Option-Implied COS Method\footnote{I am very grateful to Caio Almeida, Torben Andersen, Peter Boswijk, Zhenyu Cui, Fang Fang, Gustavo Freire, Roger Laeven, Olivier Scaillet, Viktor Todorov, and conference participants at the Financial Econometrics Conference at Lancaster University, the 2023 Annual SoFiE conference, and the 5th International Workshop in Financial Econometrics for helpful comments and suggestions. 
I am also thankful for the financial support received from the SoFiE travel grant and the Amsterdam University Fund.
}
}

\author{Evgenii Vladimirov\footnote{Correspondence to: Econometric Institute, Erasmus University Rotterdam, Burgemeester Oudlaan 50, 3062~PA, Rotterdam, The Netherlands. Email address: \href{mailto:Vladimirov@ese.eur.nl}{\tt vladimirov@ese.eur.nl}.} \\
\vspace{-0.2cm} \\
\small{Amsterdam School of Economics, University of Amsterdam}\\
\small{and Tinbergen Institute}
}

\date{ 
\vskip 0.6cm
\today 
}

\begin{document}

\maketitle

    \begin{abstract}
        This paper proposes the option-implied Fourier-cosine method, iCOS, for non-parametric estimation of risk-neutral densities, option prices, and option sensitivities. The iCOS method leverages the Fourier-based COS technique, proposed by \citeA{fang2008novel}, by utilizing the option-implied cosine series coefficients. 
        Notably, this procedure does not rely on any model assumptions about the underlying asset price dynamics, it is fully non-parametric, and it does not involve any numerical optimization.
        These features make it rather general and computationally appealing. Furthermore, we derive the asymptotic properties of the proposed non-parametric estimators and study their finite-sample behavior in Monte Carlo simulations. Our empirical analysis using S\&P 500 index options and Amazon equity options illustrates the effectiveness of the iCOS method in extracting valuable information from option prices under different market conditions. 
        Additionally, we apply our methodology to dissect and quantify observation and discretization errors in the VIX index.
    \end{abstract}



    \section{Introduction}

    Option prices play a crucial role in financial markets, providing investors and researchers with valuable insights into market expectations, risks, and dynamics associated with underlying instruments. This information is essential for effective risk management, portfolio optimization, and academic research aimed at understanding market dynamics. However, accurate extraction of this valuable information is a challenging task due to the complex nature of option prices and the presence of various sources of uncertainty.

    Traditional option pricing models, commonly used to extract embedded information from options, rely on parametric assumptions about market dynamics and the distribution of asset returns. While these models have been widely used and have provided valuable insights, they often fail to capture the complexity and nuances of real-world market behavior and are often subject to model misspecification.
    To address these limitations, non-parametric methods have gained increasing attention in option pricing research. These methods aim to extract information about underlying dynamics directly from market data without making explicit assumptions about the underlying asset price dynamics.

    In this paper, we propose a unified non-parametric estimation procedure for the risk-neutral density (RND), option prices, and option sensitivities, which are the key objects of interest in the option pricing literature. Our approach leverages the Fourier-based cosine technique, the COS method, proposed by \citeA{fang2008novel}, in a model-free way by \textit{implying} information from observed option contracts. Therefore, we refer to this method as the (option-)implied COS method, or, in short, iCOS. The proposed estimation method is fully non-parametric and does not require any optimization routines, offering a flexible and computationally appealing alternative to traditional techniques.

    Fourier-based methods are widely used numerical techniques for option evaluation that exploit the relation between the probability density function (PDF) and the characteristic function (CF). Examples of these methods include those proposed by \citeA{CM1999}, \citeA{lewis2001simple}, and \citeA{fang2008novel}. 
    The popularity of these methods is due to the CFs -- the Fourier transforms of the PDFs -- that can be obtained in a (semi-)closed form for a large class of parametric models, such as the affine jump-diffusion class as defined in \citeA{DPS2000}. As such, the main requirement for these methods is a fully parametric specification of the CF, often induced from a parametrized asset price dynamics.

    In contrast, the method proposed in this paper only requires the availability of observed option prices, which is a natural setup when working with option data. This allows us to bypass the unnecessarily restrictive requirement of a parametric CF and then utilize the flexibility of the COS machinery. In particular, we exploit the option spanning result of \citeA{carr2001optimal} to estimate the cosine series coefficients -- the projection coefficients of the orthonormal Fourier-cosine basis. These implied cosine coefficients are the building blocks for constructing non-parametric estimators of the RND, option prices with strikes that are not observed in the market, and option sensitivities.

    In particular, by utilizing the spanning result of \citeA{carr2001optimal} and the COS machinery, we provide portfolio representations for the RND, option prices and option deltas. These portfolios consist of option contracts with a continuum of strike prices and offer model-free representations of the objects of interest. Since, in practice, we observe only a finite number of option contracts subject to observation errors, we develop non-parametric estimators based on these spanning results. 
    We derive the asymptotic properties of these non-parametric estimators in an asymptotic setting in which the mesh of the strike grid shrinks to zero, while maturity and largest and smallest strike prices remain fixed. The resulting limiting distributions allow constructing confidence intervals for our estimates.

    The existing semi- and non-parametric methods for the estimation or interpolation of option prices include parametric curve fitting (\citeNP{shimko1993bounds}, \citeNP{gatheral2014arbitrage}), local polynomial estimators (\citeNP{ait2003nonparametric}), (penalized) cubic splines (\citeNP{bliss2002testing}, \citeNP{malz2014simple}), and kernel-based methods (\citeNP{ait1998nonparametric}, \citeNP{grith2012nonparametric}) among many others. These methods are also often used to obtain the RND via the famous \citeA{breeden1978prices} formula. In comparison, our iCOS approach does not rely on (local) parametric representations and can be considered as a `global' non-parametric method. In other words, the developed estimators utilize all available option prices via the portfolio spanning results, while kernel-based approaches or local polynomial regression use only limited information. 
    Furthermore, the proposed procedure provides a unified framework for estimating the RND, option prices and option sensitivities, without the need to take derivatives of estimated option functions.

    There are also various methods that estimate the RND directly without invoking the \citeA{breeden1978prices} formula. Examples of such methods include the mixture of distributions (\citeNP{melick1997recovering}, \citeNP{gemmill2000useful}), the Gram–Charlier approximation (\citeNP{jarrow1982approximate}, \citeNP{rompolis2007retrieving}), and the Hermite expansion (\citeNP{xiu2014hermite}, \citeNP{lu2021sieve}), among many others. For a comprehensive review of various methods, see \citeA{figlewski2018risk}. In this paper, we utilize the Fourier-cosine expansion of the RND and imply the cosine expansion coefficients directly from the observed option prices. This offers a flexible, computationally appealing, and model-free alternative to traditional techniques. A similar direction is taken in \citeA{cui2021model}, \citeA{cui2022new}, and \citeA{bossu2022static}, who also imply the expansion coefficients from the observed option prices. In fact, \citeA{cui2021model} use the Fourier cosine method in a model-free way to extract the RND. However, our paper goes beyond these studies by proposing a unified framework to estimate the RND together with the option prices and option sensitivities. Furthermore, we explicitly control the truncation of the RND to a finite interval, allow for observation errors in options, and derive asymptotic results for the proposed estimators.

    Our paper also contributes to the literature on model-free Greeks estimation (\citeNP{bates2005hedging}, \citeNP{alexander2007model}). Similar to these studies, our approach enables non-parametric estimation of option sensitivities, such as option deltas. However, unlike the methods proposed by \citeA{bates2005hedging} and \citeA{alexander2007model}, our approach does not require knowledge of the derivative of option prices with respect to strike and, hence, does not involve fitting the implied volatility curve. Instead, we estimate the option-implied deltas using an approach similar to the estimation of the RND and option prices, utilizing the portfolio spanning result based on the Fourier expansion. This eliminates the need for optimization and calibration to the market, thereby reducing the impact of model misspecification and calibration errors. 

    Our paper is also closely related to several studies that propose non-parametric approaches to estimate various risk measures from short-dated options, such as the Levy density in \citeA{qin2019nonparametric}, spot volatility in \citeA{todorov2019nonparametric}, and jump variation in \citeA{todorov2022nonparametric}. Like these papers, our approach is fully non-parametric and accounts for errors in the observed option prices. 
    Our analysis differs in terms of the specific information extracted from the option contracts, and our methodology is not restricted to short-dated options, making it more broadly applicable.
    
    We conduct extensive simulation experiments to assess the finite-sample properties of the proposed non-parametric estimators. We consider the classical and well-understood \citeA{black1973pricing} model and the more realistic `double-jump' stochastic volatility model of \citeA{DPS2000} as data generating processes. We find good finite-sample performance of all three non-parametric estimators for different maturities and show superiority of our procedure to kernel-based smoothing methods.



    Finally, in our empirical application, we demonstrate the effectiveness of the iCOS method across various market settings. For that, we analyze the performance of the method in the highly liquid and well-studied market of S\&P 500 index (SPX) options. We also consider Amazon (AMZN) options during an Earnings Announcement Day (EAD) in a unique, high-volatility conditions with short maturities and a distinct bimodal density pattern due to the EAD effect.
    Additionally, we apply our methodology to dissect and quantify errors in the VIX index, one of most popular measure of market volatility. We find that observation errors in the VIX, that arise from the imperfect observation of option prices, are centered around zero and have a small magnitude, reaching up to 0.04 percentage points of the index value. In contrast, discretization errors lead to positive biases, reaching up to 0.7 percentage points during periods of high volatility.

    The rest of the paper is organized as follows. In Section \ref{sec:iCOS}, we describe the option-implied COS method. The non-parametric estimators for the RND, option prices, and option deltas along with their asymptotic properties are discussed in Section \ref{sec:estimation}. Section \ref{sec:simulation} provides the Monte Carlo simulation results. The empirical applications are in Section \ref{sec:empirics}. Section \ref{sec:conlcusion} concludes the paper. The proofs of the propositions are collected in Appendix \ref{appendix:proofs} and some additional results are in Appendix \ref{appendix:additional}.
    
    \section{Option-implied COS method}
\label{sec:iCOS}

    In this section, we start with the discussion of the COS method, proposed by \citeA{fang2008novel}. Then, we show how the option-implied information can be incorporated into this machinery. The option-implied COS allows us to get the expansion expressions for the RND, option prices and option sensitivities that serve as the building blocks for the non-parametric estimators introduced in the next section.

\subsection{The COS method}
\label{subsec:COS}

    The COS method introduced by \citeA{fang2008novel} is based on the idea that the conditional density function $f(y)$ on an interval $[a,b] \subset \R$ can be represented via its Fourier cosine series expansion:
    \begin{align}
        f(y) &= \frac{1}{b-a}A_0 + \frac{2}{b-a} \sum_{m=1}^{\infty}A_m \cos\left(m \pi \frac{y-a}{b-a}\right) \notag \\ 
        \label{cos-pdf}
        &=\frac{2}{b-a}\ \sideset{}{'}\sum_{m=0}^{\infty } A_m \cos\left(u_m y-u_m a\right),
    \end{align}
    where $\sideset{}{'}\sum$ indicates the sum with the first term weighted by one-half, $u_m := \frac{m \pi}{b-a}$, and the cosine coefficients
    \begin{align}\label{Am}
        A_m = \int_a^b  \cos(u_m y - u_m a) f(y) \diff y, \quad m=0,1,\dots\ .
    \end{align}

    \citeA{fang2008novel} showed that the cosine coefficients \eqref{Am} can be calculated via the (`truncated') characteristic function (CF). In fact, let us denote the CF of the density function restricted to the interval $[a,b]$ by
    \begin{align*}
        \phi^{[a,b]}(u) := \int_a^b e^{\mathrm{i} u y} f(y) \diff y.
    \end{align*}
    Then, premultiplying $\phi^{[a,b]}(u_m) $ by $e^{-\mathrm{i} u_m a}$ and taking the real part we find that
    \begin{align*}
        A_m  = \Re \left\{\int_a^b e^{\mathrm{i} u_m (y - a)} f(y) \diff y \right\}
        = \Re\left\{\phi^{[a,b]}(u_m) e^{-\mathrm{i} u_m a} \right\}.
    \end{align*}

    Working with the regular CF of the density $\phi(u_m)$ turns out to be more convenient than with its `truncated' version $\phi^{[a,b]}(u_m)$ since the CF is often available in a semi-closed form for many parametric option pricing models. Fortunately, the truncated version is well approximated by its infinite counterpart $\phi(u_m)$ for sufficiently wide interval $[a,b]$, and so are the cosine coefficients:
    \begin{align}\label{Am_cf}
        A_m \approx \Re\left\{\phi(u_m) e^{-\mathrm{i} u_m a} \right\} =: \widetilde{A}_m.
    \end{align}
    
    Therefore, the COS method allows for efficient pricing of options under any parametric model when a closed-form CF is available.
    In particular, consider a European-style option with a general payoff $v(y,T)$ as a function of the state variable $y$ at a maturity time $T$.  
    Define the cosine series coefficients of the payoff function $v(y,T)$ as 
    \begin{align}
        H_m := \frac{2}{b-a} \int_a^b v(y,T) \cos(u_m y - u_m a) \diff y.
    \end{align}
    Then, the COS formula to evaluate the price of this contract at time $t=0$ is derived by plugging the Fourier cosine series expansion of the conditional density into the risk-neutral valuation. That is, under the risk-neutral measure $\Q$ with the deterministic interest rate $r$, we have:
    \begin{equation}\label{COS}
        \begin{aligned}
            v_0
            &= e^{-rT}  \E^{\Q}[v(y,T) ] 
            \stackrel{\scriptscriptstyle\mathrm{(1)}}{\approx} e^{-rT} \int_a^b v(y,T) f(y) \diff y\\
            &= e^{-rT} \int_a^b v(y,T) \left( \frac{2}{b-a}\ \sideset{}{'}\sum_{m=0}^{\infty} A_m \cos(u_m y - u_m a) \right) \diff y \\
            %
            &= e^{-rT}\  \sideset{}{'}\sum_{m=0}^{\infty} A_m  H_m 
            \stackrel{\scriptscriptstyle\mathrm{(2)}}{\approx} e^{-rT}\  \sideset{}{'}\sum_{m=0}^{\infty} \widetilde{A}_m  H_m \\
            &\stackrel{\scriptscriptstyle\mathrm{(3)}}{\approx} e^{-rT}\  \sideset{}{'}\sum_{m=0}^{N - 1} \widetilde{A}_m  H_m,
        \end{aligned}
    \end{equation}
    where by $\stackrel{\scriptscriptstyle\mathrm{(i)}}{\approx}$ we denote the subsequent numerical approximation. Note that the product of the two functions $v(y,T)$ and $f(y)$ is represented by the product of their Fourier-cosine series coefficients $A_m$ and $H_m$. The coefficients $H_m$ can be calculated analytically for many types of options. In Appendix \ref{appendix:additional}, we provide the analytic formulas for call options.

    The COS method allows for fast option evaluation using the Fourier cosine expansion. There are three numerical approximations involved: \textit{(1)} truncation of the integration range in the risk-neutral expectation, \textit{(2)} usage of the CF $\phi(u_m)$ (and hence, $\widetilde{A}_m$) instead of the truncated counterpart $\phi^{[a,b]}(u_m)$, and \textit{(3)} the cosine series truncation. Moreover, what is more important, this method requires a parametric model assumption, which is unknown a priori.

    In contrast to the traditional COS method, the approach proposed in this paper does not rely on a parametric specification of the CF. Instead, we use a finite number of plain vanilla option prices observed in the market.
    Given these observable prices, we can extract the risk-neutral density, price European options with strike prices that are not observed in the market (i.e.\ perform interpolation), and compute the option sensitivities. Furthermore, as we show in the next subsection, our approach does not require a proper choice of the interval $[a,b]$, i.e.,\ it entirely eliminates the first two numerical approximation errors.

\subsection{Option-implied information}

    Let us denote by $S_t$ the underlying price at time $t$ for a stock or an index under consideration and by $F_t$ the futures price at time $t$ for this underlying asset with some fixed maturity. Let us further denote by $C_0(K)$ and $P_0(K)$ the call and put option prices at time $t=0$ maturing at date $T>0$ with a strike price $K$. Assuming the existence of an arbitrage-free financial market and denoting with $\Q$ the risk-neutral measure, the prices of out-of-the-money (OTM) options at time $t=0$ are given as the discounted risk-neutral conditional expectations of the corresponding payoff functions:
    \begin{align*}
        O_0(K) = 
		\begin{cases}
			C_0(K) = e^{-rT} \E^\Q[\max(S_T - K,0) ],  \quad \mbox{if } K>F_{0},\\
			P_0(K) = e^{-rT} \E^\Q[\max(K - S_T,0) ],  \quad \mbox{if } K\leq F_{0},
        \end{cases}   
    \end{align*}
    where $r$ is a deterministic interest rate. Throughout the paper we consider option contracts with a fixed maturity, thus we eliminate the dependence on $T$ in our notations.

    The payoff spanning result of \citeA{carr2001optimal} allows expressing the value of any European-style contingent contract with a general payoff function $v(S_T)$ as a weighted portfolio of a risk-free bond and plain vanilla OTM options:
    \begin{align}\label{CM-spanning}
        v_0 &=  e^{-rT} \E^{\Q}[v(S_T)] = e^{-rT} \int_0^\infty v(S_T) f_S(S_T) \diff S_T \notag \\
        &= e^{-rT} v(F) + \int_0^{\infty} v''(K) O_0(K) \diff K,
    \end{align}
    where $f_S(\cdot)$ is the risk-neutral density of the future underlying price, and $F:=F_0$ is the futures price at time $t=0$. This result has been extensively used in the literature to, e.g.\ construct the VIX index (\citeNP{CBOE}), extract the risk-neutral expectations (\citeNP{bakshi2003stock}), and imply the characteristic function in a model-free way (\citeNP{todorov2019nonparametric}, \citeNP{blv}).

    For our purposes, it turns out to be convenient to modify this spanning formula to find values of contingent claims restricted to a finite interval $[\alpha, \beta] \subset \R$. For that, let us denote the risk-neutral valuation on a finite interval $[\alpha, \beta]$ as
    \[
        v_0^{[\alpha, \beta]} 
            = e^{-rT}\E^{\Q}[v(S_T) \mathbf{1}_{\{\alpha \leq S_T \leq \beta\}} ] = e^{-rT} \int_\alpha^\beta v(S_T) f_S(S_T) \diff S_T.
    \]
    Then we have the following spanning result for $v_0^{[\alpha, \beta]}$.

    \begin{proposition}\label{prop:spanning}
        The risk-neutral expectation of a European-style option contract with a twice continuously-differentiable payoff function $v(S_T) \in C^2([\alpha, \beta])$ restricted to a finite interval $[\alpha, \beta]$ with $\alpha \leq F \leq \beta$ can be replicated as follows:
        \begin{align*}
            v_0^{[\alpha, \beta]} 
            &= e^{-rT} v(F) + \int_{\alpha}^\beta v''(K) O_0(K) \diff K\\
            & \quad + v(\beta)C_K'(\beta) - v(\alpha)P_K'(\alpha) - v'(\beta)C_0(\beta) + v'(\alpha)P_0(\alpha),
        \end{align*}
        where $C_K'(x)$ and $P_K'(x)$ are derivatives of the call and put option prices with respect to strike evaluated at $x$. 
    \end{proposition}
    The proof follows from the application of the general payoff spanning formula \eqref{CM-spanning} to a function $v(S_T) \mathbf{1}_{\{\alpha \leq S_T \leq \beta\}}$ and using properties of the Dirac Delta function. Like the spanning result of \citeA{carr2001optimal}, Proposition \ref{prop:spanning} allows to replicate $v_0^{[\alpha, \beta]}$ by constructing a portfolio of risk-free asset and plain vanilla OTM options with strike prices from the interval $ [\alpha, \beta]$. Due to the restriction to the interval, the weights at the boundary option contracts, $C_0(\beta)$ and $P_0(\alpha)$, and risk-free asset position are adjusted.
    Note that with $\alpha \to 0$ and $\beta \to \infty$, the value $v_0^{[\alpha, \beta]}$ converges to the unrestricted contract value $ v_0$. 
    
    A natural choice for the truncated interval is the range of observable strike prices, i.e., we can set $\alpha$ to be the smallest observable strike price $\underline{K}$ and $\beta$ to be the largest observable strike $\overline{K}$. Alternatively, we can restrict the estimation to the interval with the most liquid options, i.e., $(\alpha, \beta) \subset (\underline{K}, \overline{K})$, which can be practically more appealing. This choice prevents the truncation errors, which are inevitable in the standard Carr-Madan spanning result \eqref{CM-spanning}.

    Now we can find the replicating portfolio for the (discounted) cosine coefficients $A_m$. For that, we consider the transformed\footnote{The motivation for this transformation comes from the COS method, where $x$ is the option's strike price.
    } variable $y= \log\frac{S_T}{x}$ (and, thus, $a= \log\frac{\alpha}{x}$ and $b= \log\frac{\beta}{x}$) with some $x>0$ and notice that 
    \begin{align}
        A_m &= \int_a^b  \cos(u_m y - u_m a) f(y) \diff y 
        = \int_\alpha^\beta  \cos\left(u_m \log \frac{S_T}{\alpha} \right) f_S(S_T) \diff S_T \notag \\
        &= \E^{\Q}\left[\cos\left(u_m \log \frac{S_T}{\alpha} \right) \mathbf{1}_{\{\alpha \leq S_T \leq \beta\}} \right]. 
    \end{align}
    Applying Proposition \ref{prop:spanning} to the function $v(S_T) = \cos\left(u_m \log \frac{S_T}{\alpha} \right)$ and denoting the second-order derivative of this function with respect to $S_T$ as 
    \begin{align}\label{psi_m}
        \psi_m(s) := v''(s) = \frac{u_m}{s^2} \left(\sin\left(u_m \log \frac{s}{\alpha} \right) - u_m \cos\left(u_m \log \frac{s}{\alpha}  \right)\right),
    \end{align}
    we get the (discounted) option-implied cosine coefficients as a portfolio of options:
    \begin{align}\label{iAm}
        e^{-rT} A_m = 
        \underbrace{e^{-r T} \cos\left(u_m \log \frac{F}{\alpha} \right)  +  \int_\alpha^\beta  \psi_m(K) O_0(K)  \diff K }_{\strut =:D_m}
        + \underbrace{ \vphantom{ \int_\alpha^\beta } 
        \cos(m\pi)C'_K(\beta) - P'_K(\alpha)}_{\strut =:b_m} .
    \end{align}
    Here, the term $b_m$ adjusts $D_m$ to account for the restriction to a finite interval, and with $\alpha \to 0$ and $\beta \to \infty$, the adjustment $b_m \to 0$. Therefore, we will also refer to $D_m$ as the cosine coefficient. It is important to emphasize that the cosine coefficient $A_m$ in equation \eqref{iAm} is completely model-free. This is in contrast to the standard COS method, where one needs to specify a parametric assumption on the dynamics of the underlying asset to get the cosine expansion coefficients via parametric CF. Furthermore, the option-implied coefficients $A_m$ are exact, thus, approximation~\textit{(2)} in equation \eqref{COS} is avoided.

    After having implied the cosine coefficients, we can obtain the risk-neutral density (RND) using equation \eqref{cos-pdf}. For instance, setting $x=1$ gives us the RND of the log future price $\log S_T$:
    \begin{align}\label{icos-rnd}
        f(y)
        &=\frac{2}{b-a}\ \sideset{}{'}\sum_{m=0}^{\infty } A_m \cos\left(u_m y-u_m a\right) \notag \\ 
        %
        &= \nu_f  \sideset{}{'}\sum_{m=0}^{\infty } (D_m + b_m) \cos\left(u_m y - u_m \log \alpha \right),
    \end{align}
    where $ \nu_f := \tfrac{2e^{rT}}{\log\left(\beta/\alpha \right) }$.
    Equation \eqref{icos-rnd} is an important representation of a portfolio of option prices with strike prices within the interval $[\alpha, \beta]$. 
    Like the cosine coefficients $A_m$, the spanning result for the RND in \eqref{icos-rnd} is exact and model-free. 
    This serves as a basis for our non-parametric estimator of the RND, which we discuss in Section \ref{sec:estimation}. 

    It is worth noting that equation \eqref{icos-rnd} provides the values of the RND for any $y \in [a,b]$, although the density itself may have support on $\mathbb{R}$. This restriction to the finite interval is  coherent with the availability of option data: if there are no options traded with strike prices $K < \alpha$, then it is difficult to infer information about the density accurately for $y < \log \alpha$ without making further (often parametric) assumptions.

\subsection{Risk-neutral valuation for plain vanilla options}

    After having extracted the option-implied cosine coefficients $A_m$, the same COS machinery can be used to price options but in a model-free way.
    While there is generally no need to price options that are already observed in the market, the developed iCOS method can be used to further evaluate option prices with the strikes that are not listed in the market. In other words, this approach allows interpolating plain vanilla options within the interval $[\alpha, \beta]$ in a completely model-free way. Additionally, as we discuss in Section \ref{sec:estimation}, option evaluation can be helpful in estimating unobserved quantities and constructing a feasible limiting distribution of the estimated RND and option sensitivities.

    For the accurate pricing/interpolation, it is important to take into account the truncation levels, i.e., we shall separate the information available in the interval $[\alpha, \beta]$ from the information outside of this range. For that, we can decompose the risk-neutral valuation of a contract with the general payoff function $v(S_T)$ as follows:
    \begin{align}\label{payoff-decomposition}
        v_0 &=  e^{-rT} \E^{\Q}[v(S_T)] = e^{-rT} \int_0^\infty v(S_T) f_S(S_T) \diff S_T \notag \\
        &= e^{-rT} \left[ \int_0^\alpha v(S_T) f_S(S_T) \diff S_T + \int_\alpha^\beta v(S_T) f_S(S_T) \diff S_T + \int_\beta^\infty v(S_T) f_S(S_T) \diff S_T \right] \notag\\
        &= v_0^{(0,\alpha)} + v_0^{[\alpha,\beta]} + v_0^{(\beta,\infty)}.
    \end{align}
    That is, we can express the price of the contract as a sum of values over the three non-overlapping intervals. The motivation for this is to approximate the infinite counterpart by a value on $[\alpha, \beta]$, while possibly taking into account the values outside this interval. In fact, the COS method of \citeA{fang2008novel} assumes that the value of a contract on $[\alpha, \beta]$, $v_0^{[\alpha, \beta]}$, represents the contract value $v_0$ well, i.e., it assumes that the values $v_0^{(0,\alpha)}$ and $v_0^{(\beta,\infty)}$ are negligible.

    It turns out that for the plain vanilla options, the values outside this finite interval $[\alpha, \beta]$ can be well controlled, completely eliminating the integration range truncation errors, represented by approximation \textit{(1)} in equation \eqref{COS}.
    In particular, for a call option with a strike price $x \in [\alpha, \beta]$, the value on the interval $(\beta,\infty)$ is given by 
    \begin{align*}
        C_0^{(\beta,\infty)} (x) &= e^{-rT}\E^{\Q}[\max(S_T - x, 0)\mathbf{1}_{\{S_T > \beta\}} ] = e^{-rT} \int_\beta^\infty \max(S_T - x, 0) f_S(S_T) \diff S_T\\
        &= e^{-rT} (\beta - x) \int_{\beta}^\infty f_S(S_T) \diff S_T + e^{-rT} \int_0^\infty \max(S_T - \beta, 0) f_S(S_T) \diff S_T \\
        &= -(\beta - x) C_K'(\beta) + C_0(\beta),
    \end{align*}
    where $C_0(\beta)$ is the call price with the strike $\beta$ and $C_K'(\beta)$ is its derivative with respect to the strike price evaluated at $\beta$. $C_0^{(\beta,\infty)} (x)$ is the price of a so-called \textit{gap call option} with a strike price $x$ and a trigger price $\beta$. A similar relation can be found with the \textit{gap put option} for the value of a put option evaluated on the interval $(0,\alpha)$. See Appendix \ref{appendix:additional} for the details about the put options.

    Therefore, we have the following relations for the call and put option prices with strike prices $x$ such that $\alpha \leq x \leq \beta$:
    \begin{align}
        C_0(x) = C_0^{[\alpha, \beta]} (x) +(x -\beta ) C_K'(\beta) + C_0(\beta),\\ 
        P_0(x) = P_0^{[\alpha, \beta]} (x) +(x -\alpha)P_K'(\alpha) + P_0(\alpha).
    \end{align}

    These decompositions allow us to take into account the truncation of the integral in the risk-neutral valuation.
    In particular, by setting $\alpha$ and $\beta$ to the smallest and largest observable strike prices $\underline{K}$ and $\overline{K}$ respectively, we can account for the price of the gap call option using information from the observable range of strike prices. 
    This means that our method is not affected by the choice of $[\alpha, \beta]$, and, hence, $[a,b]$, unlike the COS method.

    Therefore, to interpolate call options, we compute $C_0^{[\alpha, \beta]} (x)$ using the COS machinery with the option-implied information from the corresponding interval and add the price of the gap call option.
    The value of the call contract truncated to the interval $[\alpha, \beta]$ is obtained using the COS formula as
    \begin{align}
        C_0^{[\alpha, \beta]} (x) &= e^{-rT} \sideset{}{'}\sum_{m=0}^{\infty} A_m  H_m(x) = \sideset{}{'}\sum_{m=0}^{\infty} D_m  H_m(x) + \sideset{}{'}\sum_{m=0}^{\infty} b_m  H_m(x) \notag \\ 
        &= \sideset{}{'}\sum_{m=0}^{\infty} D_m  H_m(x) + \left(\sideset{}{'}\sum_{m=0}^{\infty} (-1)^m  H_m(x) \right) C_K'(\beta) - \left(\sideset{}{'}\sum_{m=0}^{\infty}  H_m(x) \right) P_K'(\alpha),
    \end{align}
    where $H_m(x)$ are the cosine series coefficients specific to the call payoff function with the strike price $x$. The closed-form expression for $H_m(x)$ is provided in Appendix \ref{appendix:additional}.

    Therefore, the price of a call option with strike price $x \in [\alpha, \beta]$ can be represented as 
    \begin{align}
        C_0(x) &= C_0^{[\alpha, \beta]} (x) +(x -\beta ) C_K'(\beta) + C_0(\beta) \notag \\
        &= \underbrace{\sideset{}{'}\sum_{m=0}^{\infty} D_m  H_m(x)}_{\strut =:\overline{C}_0(x)} + C_0(\beta) 
        + \underbrace{\left( x - \beta + \sideset{}{'}\sum_{m=0}^{\infty} (-1)^m  H_m(x) \right)}_{\strut =:Z_c(x) }  C_K'(\beta) 
        \underbrace{ - \left(\sideset{}{'}\sum_{m=0}^{\infty}  H_m(x) \right)}_{\strut =:Z_p(x)} P_K'(\alpha) \notag \\
        \label{iCall}
        &= \overline{C}_0(x) + C_0(\beta) + Z_c(x) \theta_c + Z_p(x) \theta_p,
    \end{align}
    where we additionally denote $\theta_c := C_K'(\beta) $ and $ \theta_p := P_K'(\alpha) $. Equation \eqref{iCall} represents the price of a call option as a portfolio of a continuum of OTM contracts with strike prices from the interval $[\alpha, \beta]$, with additional hedging terms due to the truncation on the finite interval. 
    It is important to note that the decomposition \eqref{iCall} is exact, i.e., it does not involve any numerical approximations and integration range truncation errors thanks to the gap options. In practice, however, we only observe a finite number of OTM options and truncate the cosine series expansion with a finite number of terms $N$. We address these issues in the next section.

    Although this decomposition is circular (to find the price of a single option we need to know the prices of a continuum of options), it is essential in practice, where we observe only a finite number of option prices but might be interested in pricing options with strikes that are not observed in the market, i.e., we use \eqref{iCall} to perform the interpolation. 

    Finally, the call and put price first-order derivatives with respect to the strike price, $\theta_c$ and $\theta_p$, are not directly observable in the market. However, we can approximate them using, e.g., the finite-difference approach. Alternatively, as we show in the next section, we can estimate them from the observed cross-section of option contracts as they are linearly loaded on the call prices. After having estimated these derivatives, we can use them to non-parametrically estimate the RND and option sensitivities.

    \subsection{Implied delta}

    In a similar model-free way, we can replicate some option sensitivities such as the option delta, $\delta$, the first derivative of the option price with respect to the underlying asset price~$S_0$. For that, we first note that 
    \begin{align*}
        \frac{\partial A_m}{\partial S_0} 
        &= \frac{\partial }{\partial S_0} \E^{\Q}\left[\cos\left(u_m \log \frac{S_T}{\alpha} \right) \mathbf{1}_{\{\alpha \leq S_T \leq \beta\}} \right]\\
        &= \E^{\Q}\left[\frac{\partial }{\partial S_T}  \left(\cos\left(u_m \log \frac{S_T}{\alpha} \right) \mathbf{1}_{\{\alpha \leq S_T \leq \beta\}} \right) \frac{\partial S_T}{\partial S_0}\right] \\
        %
        %
        &= -\frac{u_m}{S_0} \E^{\Q}\left[ \sin\left(u_m \log \frac{S_T}{\alpha} \right) \mathbf{1}_{\{\alpha \leq S_T \leq \beta\}} \right],
    \end{align*}
    where we additionally assume that $\frac{\partial S_T}{\partial S_0} = \frac{S_T}{S_0}$, i.e.\ the solution to the stochastic process $S_T$ is homogenous of degree one as a function of initial stock price $S_0$.

    Let us further denote $B_m := e^{-rT} \E^{\Q}\left[ \sin\left(u_m \log \frac{S_T}{\alpha} \right) \mathbf{1}_{\{\alpha \leq S_T \leq \beta\}} \right]$. Then, $B_m$ can be replicated similar to the cosine coefficients terms $A_m$ using Proposition \ref{prop:spanning}:
    \begin{align}\label{iBm}
        B_m = 
        e^{-r T} \sin\left(u_m \log \frac{F}{\alpha} \right)  +  \int_\alpha^\beta  \widetilde{\psi}_m(K) O_0(K)  \diff K 
        - \frac{u_m}{\beta} (-1)^m C_0(\beta) + \frac{u_m}{\alpha} P_0(\alpha),
    \end{align}
    with    
    \begin{align}
        \widetilde{\psi}_m(s) := -\frac{u_m}{s^2} \left(\cos\left(u_m \log \frac{s}{\alpha} \right) + u_m \sin\left(u_m \log \frac{s}{\alpha}  \right)\right).
    \end{align}

    Hence, the delta of the call option restricted to the interval $[\alpha, \beta]$ is given by
    \begin{align*}
        \delta^{[\alpha, \beta]}(x) 
        &= \frac{\partial C_0^{[\alpha, \beta]}(x)}{\partial S_0} 
        = e^{-rT} \sideset{}{'}\sum_{m=0}^{\infty} \frac{\partial A_m}{\partial S_0}  H_m(x)\\
        &= -\frac{1}{S_0} \sum_{m=1}^{\infty} u_m B_m  H_m(x).
    \end{align*}
    
    Furthermore, the delta of the gap call option for $x < \beta$ can be expressed as 
    \begin{align*}
        \delta^{(\beta, \infty)}
        &= \frac{\partial C_0^{(\beta, \infty)}(x)}{\partial S_0} = 
        e^{-rT} \E^\Q\left[\frac{\partial}{\partial S_0} \max(S_T - x, 0) \mathbf{1}_{\{S_T > \beta\}} \right]\\
        &=\frac{1}{S_0} e^{-rT}\int_\beta^\infty \mathbf{1}_{\{S_T > x\}} S_T f_S(S_T) \diff S_T \\
        &= \frac{1}{S_0} e^{-rT} \int_\beta^\infty (S_T - \beta) f_S(S_T) \diff S_T + \frac{1}{S_0} e^{-rT} \int_\beta^\infty \beta f_S(S_T) \diff S_T \\ 
        &= \frac{1}{S_0} \left(C_0(\beta) - \beta C_K'(\beta) \right),
    \end{align*}
    which does not depend on $x$.
    Therefore, combining two deltas, the implied delta for a European call is given as following: 
    \begin{align}\label{iDelta}
        \delta(x) = \delta^{[\alpha, \beta]}(x) + \delta^{(\beta, \infty)} 
        = -\frac{1}{S_0} \sum_{m=1}^{\infty} u_m B_m  H_m(x) + \frac{1}{S_0}\left(C_0(\beta) - \beta \theta_c \right).
    \end{align}

    As the RND expansion \eqref{icos-rnd} and option evaluation \eqref{iCall}, the replication result for the delta~\eqref{iDelta} is exact and model-free for all strike values $x \in [\alpha, \beta]$. A similar spanning result can be derived for option gamma, the second order sensitivity. However, one can also easily obtain gamma by properly scaling the option-implied RND \eqref{icos-rnd} (see, e.g., \citeNP{bates2005hedging} and \citeNP{alexander2007model}).
    
    It is important to emphasize that all three spanning results (given by equations \eqref{icos-rnd}, \eqref{iCall}, and \eqref{iDelta}) are exact, and the restriction to the finite interval does not lead to the associated truncation errors due to the usage of the gap call option. Furthermore, the choice of the interval is data-driven. This is in contrast to the traditional COS method, where one has to select wide interval to minimize the integration range truncation error.
    
    All three option-implied quantities depend on a continuum number of option contracts. In practice, however, we observe only a finite number of option contracts. Nevertheless, all three expressions are easy to approximate using a limited number of observable option prices. Additionally taking into account the observation errors, in the next section, we develop feasible non-parametric estimators for the RND, option prices, and deltas.

    \section{Estimation}\label{sec:estimation}

    In this section, we introduce the computationally feasible estimators based on the option-implied COS valuation method and derive their asymptotic properties. 

    \subsection{The observation scheme}\label{sec:estimation-observation scheme}

    Unlike the COS method, our approach does not require a parametric specification of model dynamics under the risk-neutral measure. Instead, we use a cross-section of option prices observed in the market. However, these options are observed across a finite set of strike prices and prone to observation errors due to, e.g., bid-ask spread, tick sizes of quotes, and liquidity issues. Therefore, we first describe the observation option scheme.

    Our data consists of $n$ OTM option prices observed at time $t=0$ and expiring at a fixed time $T>0$ with a deterministic sequence of strike prices: 
    \[ 
        0< \underline{K}:= K_1 < K_2 <\dots < K_n =: \overline{K} <\infty. 
    \]
    For the asymptotic analysis developed below, we assume that the smallest and the largest strike prices, $\underline{K}$ and $\overline{K}$, are fixed, and the number of options $n$ with strike prices between them goes to infinity by shrinking the strike mesh. We could also consider the joint asymptotic scheme as in \citeA{todorov2019nonparametric} and \citeA{blv}, where $\underline{K} \to 0$ and $ \overline{K} \to \infty$ at certain rates as $n\to \infty$. However, we fix $\alpha=\underline{K}$ to be the smallest strike price and $\beta = \overline{K}$ to be the largest strike price since the choice of the interval $[\alpha, \beta]$ in the spanning results does not introduce an integration range truncation error, as discussed in the previous section.

    \begin{assumption}\label{assumption:fixed_grid}
        The smallest and largest strike prices are fixed at $\alpha = \underline{K}$ and $\beta = \overline{K}$, and the strike prices between them are equidistant, i.e.\ 
        $$\Delta_n := \Delta K_i = K_i - K_{i-1} = \frac{\beta - \alpha}{n-1},\quad i=2,\dots,n.$$
    \end{assumption}

    Assumption \ref{assumption:fixed_grid} can be relaxed to allow for a non-equidistant grid. However, using an equidistant grid simplifies our analysis and enables us to use various numerical approximation methods under one umbrella. Table \ref{tab:integrations} summarizes several popular numerical integration methods for an equidistant grid. Later, we comment on our results with a non-equidistant grid of strike prices.

    We emphasize again that we do not require an increasing range of strike prices since we account for the restriction to a finite interval. This is different from other methods that estimate RNDs using expansion series, such as in \citeA{lu2021sieve} and \citeA{cui2021model}. Fixing the interval $[\alpha, \beta]$ comes as a great advantage in practice since it allows us to choose different intervals of strike prices without introducing additional approximation errors. For instance, we can consider an interval with only actively traded options.

    \begin{table}[ht]
        \centering
        \caption{Numerical integration methods and their errors}
        \begin{tabular}{lcc}
            \toprule
            Methods  &  Coefficients $w_i$  & Error order\\
            \midrule
            Left Riemann sum & $w_1=\dots =w_{n-1}=1,\ w_n = 0$ & $\psi'(s) \mathcal{O}(n^{-1})$ \\ 
            Right Riemann sum & $w_1=0$,\ $w_2=\dots =w_{n}=1$ & $\psi'(s) \mathcal{O}(n^{-1})$ \\
            Trapezoidal rule & $w_1=w_n=\frac{1}{2},\ w_2=\dots=w_{n-1}=1$ & $\psi''(s)  \mathcal{O}(n^{-2})$ \\
            Simpson's 1/3 rule & $w_1=w_n=\frac{1}{3},\ w_i = \frac{1}{3}(3 +(-1)^i),\ i=2,\dots,n{-}1$ & $\psi^{(4)}(s) \mathcal{O}(n^{-4})$ \\
            \bottomrule
        \end{tabular}
        \label{tab:integrations}

        \medskip
        \begin{minipage}{0.9\textwidth}\scriptsize
            Note: This table provides the list of popular numerical approximations for definite interval in the form of $\int_\alpha^\beta \psi(x) \diff x \approx \sum_{i=1}^n w_i \psi(x_i) \Delta_n $. For each method, we report the coefficients $w_i$ assuming equidistant grid $\Delta_n$ and corresponding orders of the error terms with $s$ being some number between $\alpha$ and $\beta$.
        \end{minipage}

    \end{table}

    \begin{assumption}\label{assumption:observation_errors}
        Option prices are observed with an additive error term: 
        \[
        O(K_i) = O_0(K_i) + \varepsilon_i, \quad i=1,\dots,n,  
        \]
        where the observation errors $\varepsilon_i$ are such that: (1) $\E[\varepsilon_i ] = 0$, (2) $\E[\varepsilon_i^2 ] = \sigma_i^2$ are positive and finite-valued, (3) $\E[\varepsilon_i^4] < \infty $, and (4) $\varepsilon_i$ and $\varepsilon_j$ are conditionally independent whenever $i\neq j$.
    \end{assumption}

    As common in the option pricing literature, Assumption \ref{assumption:observation_errors} imposes an additive error structure form with independent but possibly heteroskedastic error terms (see, e.g.,\ \citeNP{andersen2015parametric}, \citeNP{todorov2019nonparametric}, \citeNP{blv}). The independence assumption can be further relaxed by considering a spatial dependence as in \citeA{andersen2021spatial} at the cost of more complex expressions for the limiting distributions. This would, however, play a secondary role in the developed estimation procedure.

    Note that we drop the null index to denote the observed OTM prices. 
    Furthermore, due to the put-call parity, the same observation errors translate into the counterpart in-the-money contracts, i.e., both $ C(K_i) = C_0(K_i) + \varepsilon_i$ and $P(K_i) = P_0(K_i) + \varepsilon_i$ for the call and put contracts with the same strike price $K_i$ and error term $\varepsilon_i$.

\subsection{Option prices estimator}

    Using $n$ observable option prices, we can estimate (part of the) cosine coefficients $D_m$ defined in equation \eqref{iAm}, by using a numerical approximation of the integral: 
    \begin{align}\label{iDm_hat}
        \widehat{D}_m := e^{-r T} \cos\left(u_m \log \frac{F}{\alpha} \right) 
        + \sum_{i=1}^n w_i \psi_m(K_i) O(K_i) \Delta_n, 
    \end{align}
    where $w_i$ are the coefficients of a chosen numerical integration method, as listed in Table \ref{tab:integrations}.

    The deviation of the estimated cosine expansion coefficient $\widehat{D}_m$ from its true value $D_m$ stems from the observation and discretization errors. These errors also arise in the VIX calculation (see, e.g., \citeNP{jiang2005model} and \citeNP{jiang2007extracting}). \citeA{todorov2019nonparametric} and \citeA{blv} also analyze these errors in their estimation procedures along with the truncation errors that arise due to integration over a finite interval. However, in our setting, there are no truncation errors for the cosine coefficients $D_m$ since we take this truncation further into account. See the discussion in Section \ref{sec:iCOS}.

    Given the fixed smallest and largest strike prices $\alpha$ and $\beta$, we have the following asymptotic result for the cosine coefficients.
    \begin{proposition}\label{prop:Dm}
        Under Assumptions \ref{assumption:fixed_grid}--\ref{assumption:observation_errors}, the computationally feasible estimator $\widehat{D}_m$ with fixed $m>0$ is such that
        \[
        \E\left[\widehat{D}_m - D_m \right] = \zeta^D_{m,n},
        \]
        where $\zeta^D_{m,n} = \mathcal{O}\left( \frac{m^{2+\iota}}{n^{\iota}} \right)$ is the discretization error with the order controlled by the chosen numerical integration scheme $\iota \geq 1$, and as $n \to \infty$
        \[
        \frac{\widehat{D}_m - D_m }{\sigma_D(m)} \xrightarrow{d} \mathcal{N}(0, 1),
        \]
        with $\sigma_D^2(m) = \sum_{i=1}^n w_i^2 \psi_m^2(K_i) \sigma_i^2 \Delta_n^2 $.
    \end{proposition}

    The proof of Proposition \ref{prop:Dm} is provided in Appendix \ref{appendix:proofs}. The proposition states that although the estimator $\widehat{D}_m$ with fixed $m$ based on a finite number of option prices is biased, it is asymptotically unbiased as the number of option prices $n$ increases. This serves as a building block for the non-parametric estimators introduced below.


    Next, we introduce the computationally feasible option-implied call price estimator $\widehat{\overline{C}}(x)$ of the error-free counterpart $\overline{C}_0(x)$, defined in equation \eqref{iCall}, with a strike $x$ and the payoff restricted to the interval $[\alpha, \beta]$. It can be expressed as a linear combination of asymptotically unbiased estimators $\widehat{D}_m$ with $m= 1, \dots, N-1$ as follows:
    \begin{align}
        \widehat{\overline{C}}(x) := \sideset{}{'}\sum_{m=0}^{N-1} \widehat{D}_m  H_m(x), 
    \end{align}
    where $N$ is the number of expansion terms in the Fourier-cosine expansion and $\widehat{D}_0 = e^{-rT}$. 

    Unlike its error-free counterpart $\overline{C}_0(x)$, the option-implied call price estimator $\widehat{\overline{C}}(x)$ is based on a finite number of noisy option prices and is prone to three types of errors. These errors can be decomposed as follows: 
    \begin{align}\label{icall_bar}
        \widehat{\overline{C}}(x) - \overline{C}_0(x) = \xi(x) + \zeta(x) + \overline{\eta}(x),
    \end{align}
    where $\xi(x),\ \zeta(x),$ and $\overline{\eta}(x)$ are observation, discretization and series truncation errors, respectively, all formally defined in Appendix \ref{appendix:proofs}. The series truncation error $\overline{\eta}(x)$ refers to the truncation of the cosine expansion to the finite number of terms $N$.
    To get an order of this truncation error, we additionally impose the assumption on the smoothness of the RND.
    \begin{assumption}\label{assumption:smoothness}
        The RND of the future prices $f_S(s) \in C^p\left([\alpha, \beta]\right)$ with $p>1$ and $[ \alpha, \beta]~{\subset}~\mathcal{D}$, where $\mathcal{D} \subseteq \mathbb{R}^+$ is the support of the RND.
    \end{assumption}

    \begin{proposition}\label{prop:iCall_bar}
        Under Assumptions \ref{assumption:fixed_grid}--\ref{assumption:smoothness}, the computationally feasible option-implied call price estimator $\widehat{\overline{C}}(x)$ with a strike price $x \in [\alpha, \beta]$ is such that
        \[
            \E\left[ \widehat{\overline{C}}(x) - \overline{C}_0(x) \right] = \zeta(x) + \overline{\eta}(x), 
        \]
        where the accumulated discretization error $\zeta(x) = \sum_{m=1}^{N-1} \zeta^D_{m,n} H_m(x) = \mathcal{O}\left( \frac{N^{1+\iota}}{n^{\iota}} \right) $ with $\iota \geq 1$ and the series truncation error $\overline{\eta}(x) = \mathcal{O}\left(N^{1-p}\right)$. Furthermore, as $n \to \infty$ and $N \to \infty$ with $Nn^{-1/2} \to 0$, we have
        \[
            \frac{\widehat{\overline{C}}(x) - \overline{C}_0(x)}{\overline{\sigma}_c(x)}  \xrightarrow[]{d} \mathcal{N}\left(0, 1 \right),
        \]
        where 
        \[
            \overline{\sigma}_c^2(x) 
            = \sum_{i=1}^n w_i^2 \psi^2(x, K_i) \sigma_i^2 \Delta_n^2
        \]
        with $\psi(x, K_i):= \sum_{m=1}^{N-1}  \psi_m(K_i) H_m(x)$.
    \end{proposition}

    The proof of Proposition \ref{prop:iCall_bar} is provided in Appendix \ref{appendix:proofs}. 
    

    The evaluation of options introduces two types of biases: the discretization error bias $\zeta(x)$ and the series truncation bias $\overline{\eta}(x)$, which arises from the truncation of the cosine expansion as in the original COS method. The former vanishes with an increase in the number of option prices $n$, while the latter decreases with an increase in the number of expansion terms $N$ as in the COS method. The joint asymptotic for $n$ and $N$, with $N$ increasing slower than $\sqrt{n}$, guarantees that the option-implied call price estimator $\widehat{\overline{C}}(x)$ is asymptotically unbiased. However, in a finite setting with noisy option prices, an increase in expansion terms can lead to a higher variance of the estimators. 
    We address this bias-variance trade-off in the next subsection by choosing an optimal number of expansion terms $N^*$.

    We also note that, like in the COS method, for a sufficiently large interval $[\alpha, \beta]$, the estimator $\widehat{\overline{C}}(x)$ gives a good approximation for the call price with a payoff unrestricted to this interval, $C_0(x)$. 
    However, to accurately evaluate call options, we also need the first-order derivatives of call and put options $\theta_c$ and $\theta_p$ evaluated at the boundaries of this interval (see equation \eqref{iCall}). Although we could use finite differences to estimate the first-order derivatives non-parametrically, here we use a simple linear relation of observed option prices on these derivatives instead.
    In particular, for the observed call price with the strike price $K_i$, we can get the following decomposition:
    \begin{align*}
        C(K_i) &= C_0(K_i) + \varepsilon_i \\
        &= \overline{C}_0(K_i) + C_0(\beta) + Z_{c}(K_i) \theta_c + Z_{p}(K_i) \theta_p + \varepsilon_i\\
        &= \widehat{\overline{C}}(K_i) - \xi(K_i) - \zeta(K_i) - \overline{\eta}(K_i) + C(\beta) - \varepsilon_n + Z_{c}(K_i) \theta_c + Z_{p}(K_i) \theta_p + \varepsilon_i \\
        &= \widehat{\overline{C}}(K_i) + C(\beta) - \zeta(K_i) - \overline{\eta}(K_i) - \varepsilon_n  + Z_{c}(K_i) \theta_c + Z_{p}(K_i) \theta_p - \xi(K_i) +  \varepsilon_i,
    \end{align*}
    where the second equality follows from equation \eqref{iCall} and the third one from the decomposition \eqref{icall_bar}.
    Therefore, given $n$ observed option prices, the first derivatives of call and put options $\theta_c$ and $\theta_p$ can be estimated using a simple linear regression of $ C(K_i) - \widehat{\overline{C}}(K_i) - C(\beta)$ on $Z_{c}^N(K_i)$ and $Z_{p}^N(K_i)$ with an intercept $\bar{\theta}$, where $Z_{c}^N(K_i)$ and $Z_{p}^N(K_i)$ are the partial sum counterparts of $Z_{c}(K_i)$ and $Z_{p}(K_i)$, respectively, defined in \eqref{iCall}. Adding an intercept into this regression reduces finite-sample biases due to the discretization and truncation errors.

    Finally, the option-implied call price estimator for any strike price $x \in [\alpha, \beta]$ is given by
    \begin{align}\label{iCall-estimator}
        \widehat{C}(x) := \widehat{\overline{C}}(x) + C(\beta) + Z_{c}^{N}(x) \widehat\theta_c + Z_{p}^{N}(x) \widehat\theta_p + \widehat{\bar{\theta}}, 
    \end{align}
    where $\widehat{\bar{\theta}},\ \widehat\theta_c$ and $\widehat\theta_p$ are the OLS estimates of the aforementioned regression. We emphasize here that while the OLS estimates are obtained using a finite number of observed option prices, the estimator \eqref{iCall-estimator} can be obtained for any strike price $x \in [\alpha, \beta]$. Therefore, the call price estimator~$\widehat{C}(x)$ can be seen as an interpolation-approximation method, for which we have the following asymptotic result.

    \begin{proposition}\label{prop:iCall}
        Under Assumptions \ref{assumption:fixed_grid}--\ref{assumption:smoothness}, the computationally feasible option-implied call price estimator $\widehat{C}(x) $ with a strike price $x \in [\alpha, \beta]$ is such that as $n \to \infty$ and $N \to \infty$ with $Nn^{-1/2} \to~0$,
        \[
            \frac{ \widehat{C}(x) - C_0(x) }{\sigma_c(x)}  \xrightarrow[]{d} \mathcal{N}\left(0, 1 \right),
        \]
        with 
        the variance $\sigma_c^2(x)$ given by equation \eqref{var call} in Appendix \ref{appendix:proofs}.

    \end{proposition}

    The proof of Proposition \ref{prop:iCall} can be found in Appendix \ref{appendix:proofs}. 
    Like the estimator $\widehat{\overline{C}}(x)$, the call price estimator $\widehat{C}(x)$ is asymptotically unbiased when the number of expansion terms grows slower than $\sqrt{n}$.
    

   The developed call price estimator is related to non-parametric kernel smoothing methods that are widely used in the literature (see, e.g., \citeNP{ait1998nonparametric}, \citeNP{grith2012nonparametric}, \citeNP{dalderop2020nonparametric}), but it can be considered as a `global' smoother. While kernel methods are typically local smoothers (bandwidth parameters control the locality of these estimators), our call price estimator uses all available option prices via the portfolio spanning result \eqref{iCall} discussed in Section \ref{sec:iCOS}. This difference allows our method to provide a more flexible approximation of option prices. In the simulation section, we compare these two approaches and demonstrate the superiority of our method.

   \begin{figure}[htbp]
    \centering
    \caption{iCOS weights for the at-the-money call price estimator}
    \label{fig: weights}
        \hspace*{-0.5cm}
        \includegraphics[scale=0.13]{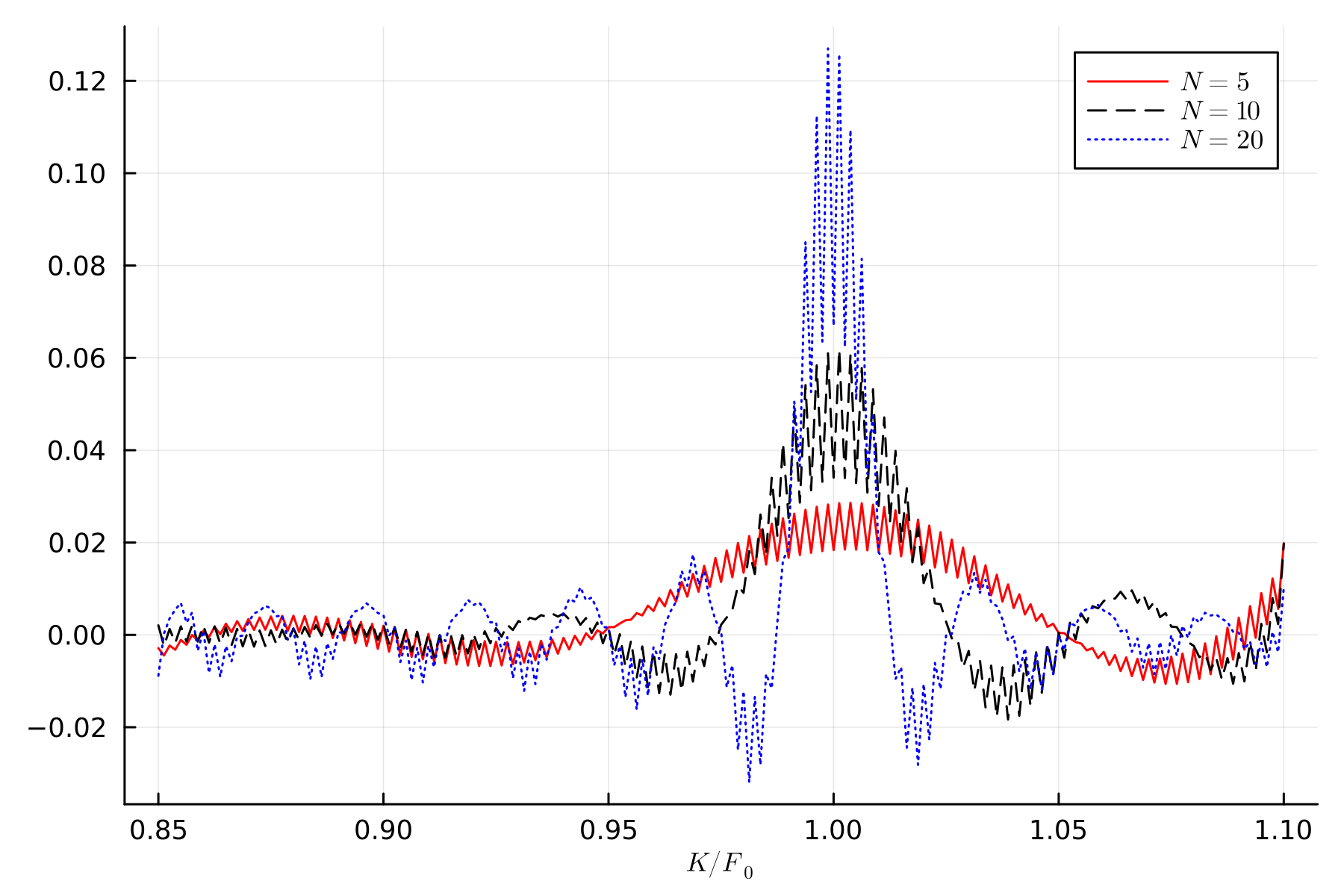}
        \medskip
        \begin{minipage}{0.8\textwidth}\scriptsize
            Note: This figure plots the weights of the option portfolios for the at-the-money call price estimator for different number of expansion terms $N$. The illustration is based on the simulated Black-Scholes model with strike prices between 85\% and 110\% of the spot price with equidistant increments and 201 option contracts. The further simulation details are discussed in Section~\ref{sec:simulation}
        \end{minipage}
    \end{figure}

    To demonstrate the `global' nature of our approach, in Figure \ref{fig: weights} we display the weights of option portfolios for the at-the-money call option for different number of expansion terms $N$. The illustration is based on the Black-Scholes model and the simulation set-up is outlined in Section~\ref{sec:simulation}. As shown in the figure, our call price estimator for the strike price $K=F_0$ utilizes all available option contracts instead of restricting information to a local neighborhood. When the number of expansion terms increases, the weights concentrate more around the target strike price while still incorporating information from all available contracts.

    \subsection{RND estimator}

    After having estimated the cosine coefficients $\widehat{D}_m$ and the first order derivatives $\widehat{\theta}_c$ and  $\widehat{\theta}_p$, we can get the non-parametric estimator for the RND of the log price:
    \begin{align}\label{icos-rnd-estimator}
        \widehat{f}(y)
        =\nu_f \sideset{}{'}\sum_{m=0}^{N-1} \left(\widehat{D}_m + (-1)^m \widehat{\theta}_c - \widehat{\theta}_p \right) \cos\left(u_m y - u_m \log \alpha \right),
    \end{align}
    where $\nu_f = \frac{2e^{rT}}{\log\left(\beta/\alpha \right) }$. The RND of the future price $S_T$ is obtained by the appropriate transformation of the log price density \eqref{icos-rnd-estimator}.
    A similar asymptotic result carries over to the non-parametric RND estimator.
    \begin{proposition}\label{prop:iRND}
        Under Assumptions \ref{assumption:fixed_grid}--\ref{assumption:smoothness}, the computationally feasible option-implied RND estimator $\widehat{f}(y) $ is such that for any fixed $y \in [\log\alpha, \log\beta]$ as $n \to \infty$  and $N \to \infty$ with $Nn^{-1/6} \to~0$
        \[
            \frac{\widehat{f}(y) - f(y) }{ \nu_f \sigma_f(y)}  \xrightarrow[]{d} \mathcal{N} \left( 0 , 1 \right),
        \]
        where  
        $\sigma_f^2(y)$ is the variance term formally defined in equation~\eqref{var rnd} in Appendix \ref{appendix:proofs}.
    \end{proposition}

    The proof of Proposition \ref{prop:iCall} can be found in Appendix \ref{appendix:proofs} as well. Unlike the option-implied call price estimator, here we require the number of expansion terms $N$ to grow slower than $n^{-1/6}$ due to the different cosine function. Nevertheless, when this condition is met, the option-implied RND remains asymptotically unbiased.

    We emphasize again that this estimator is not for the truncated density, but for the full RND evaluated at any $y$ within the interval $[\log\alpha, \log\beta]$. If one wishes to estimate the RND outside of this interval, additional, often parametric assumptions have to be made about the behavior of the density of options in areas where no option prices are observed. 
    For instance, one possible approach to estimating the RND outside of this interval is to extrapolate option prices beyond the observable range of strike prices using a parametric form based on no-arbitrage conditions. This extrapolated data can then be used to estimate the RND based on the same estimator~\eqref{icos-rnd-estimator}. In practice, for sufficiently liquid options, the observed range of strike prices covers almost an entire distribution.

    \subsection{Delta estimator}

    The non-parametric estimator for the option delta can also be derived in a similar way using the spanning result \eqref{iDelta} given a finite number of option prices:
    \begin{align}\label{iDelta-estimator}
        \widehat{\delta}(x) 
        = -\frac{1}{S_0}  \sum_{m=1}^{N-1} u_m \widehat{B}_m  H_m(x) + \frac{1}{S_0}\left(C(\beta) - \beta \widehat{\theta}_c \right),
    \end{align}
    where
    \begin{align*}
        \widehat{B}_m := e^{-r T} \sin\left(u_m \log \frac{F}{\alpha} \right) 
        + \sum_{i=1}^n w_i \widetilde{\psi}_m(K_i) O(K_i) \Delta_n - \frac{u_m}{\beta} (-1)^m C(\beta) + \frac{u_m}{\alpha} P(\alpha) . 
    \end{align*}

    An analogous asymptotic result holds for the delta estimator \eqref{iDelta-estimator}, but with an additional assumption.

    \begin{assumption}\label{assumption:delta}
        The solution to the stochastic process $S_T$ is homogenous of degree one as a function of initial stock price $S_0$, i.e., $\frac{\partial S_T}{\partial S_0} = \frac{S_T}{S_0}$.
    \end{assumption}

    Assumption \ref{assumption:delta} is required to derive the expansion for the delta given in equation \eqref{iDelta}. The same assumption is imposed for the other non-parametric delta estimators (see, e.g., \citeA{bates2005hedging} and \citeA{alexander2007model}).

    \begin{proposition}\label{prop:iDelta}
        Under Assumptions \ref{assumption:fixed_grid}--\ref{assumption:delta}, the computationally feasible option-implied delta estimator $\widehat{\delta}(x) $ is such that for any fixed $x \in [\alpha, \beta]$ as $n \to \infty$ and $N \to \infty$ with $Nn^{-1/2} \to 0$
        \[
            \frac{ \widehat{\delta}(x) - \delta(x)}{ \tfrac{1}{S_0} \sigma_\delta(y)}  \xrightarrow[]{d} \mathcal{N} \left( 0 , 1 \right),
        \]
        where $\sigma_\delta^2(y)$ is the variance term formally defined in Appendix~\ref{appendix:proofs}.
    \end{proposition}

    It is worth noting, that in the current formulations, all limiting results are self-scaling. This implies that equidistant strike price Assumption \ref{assumption:fixed_grid} can be easily relaxed without affecting the limiting distributions.

    \subsection{Optimal number of expansion terms}
    \label{sec:optimal N}

    As discussed earlier, the number of expansion terms $N$ controls the bias-variance tradeoff in the developed non-parametric estimators. 
    An increase in the number of terms reduces the bias resulting from the series truncation error, but increases the variance of the estimators.
    To find the optimal number of expansion terms $N$ for the Fourier-cosine expansion in the iCOS method, we consider the expansion for the RND\footnote{Depending on the purposes, one could also determine the optimal $N$ that minimizes the difference between the observed and estimated option prices. However, since option prices are observed with noise, this approach can potentially lead to severe arbitrage violations.}.

    To assess the impact of truncation on the Fourier-cosine expansion, it is convenient to consider the fit of the density based on the Mean Integrated Squared Error (MISE), defined as 
    \begin{align}\label{eq mise}
        \mbox{MISE}_N := \E\left[\int_a^b (\widehat{f}(y) - f(y))^2 \diff y \right],
    \end{align} 
    where $\widehat{f}(y)$ is the density estimate based on $N$ expansion terms.
    Following \citeA{leitao2018data} and \citeA{kronmal1968estimation}, the MISE can be decomposed as follows:
    \begin{align*}
        \mbox{MISE}_N
        = \sum_{m=N}^\infty A_m^2  + \sum_{m=1}^{N-1}  \E\left[ \left( \widehat{A}_m - A_m \right)^2\right] ,
    \end{align*}
    where $A_m$ is the $m$-th Fourier-cosine coefficient, and $\widehat{A}_m$ is its estimate. 
    Since the discretization errors contribute to the asymptotically vanishing bias, we consider the Asymptotic MISE (AMISE), where the second moment equals the variance of $A_m$.
    Hence, the optimal number of expansion terms $N$ trades off the bias, given by the first part, and the variance of the estimator.

    We can derive a recursive relationship in $N$ as follows:
    \begin{align*}
        \mbox{AMISE}_{N+1} &= \sum_{m=N+1}^\infty A_m^2 + \sum_{m=1}^{N} \mbox{Var} \left( \widehat{A}_m \right) \\
        &= \sum_{m=N}^\infty A_m^2 - A_N^2 + \sum_{m=1}^{N-1} \mbox{Var} \left( \widehat{A}_m \right)  
        + \mbox{Var} \left( \widehat{A}_N \right)  \\
        &= \mbox{AMISE}_{N} - \left( A_N^2 - \mbox{Var}\left( \widehat{A}_N \right) \right),
    \end{align*}
    from which we can see that if $A_N^2 - \mbox{Var}\left( \widehat{A}_N \right) >0$, then $\mbox{AMISE}_{N+1} < \mbox{AMISE}_N$. We can use this inequality as a rule to determine the optimal number of expansion terms.

    The variance of the cosine coefficient estimators depends on the estimates $\widehat{D}_N$ and $\widehat{\theta}$, and is provided in a closed-form in Appendix \ref{appendix:additional}. A true value of $A_N$ is, however, unknown a priori.  Furthermore, due to the presence of discretization error bias in finite settings, this inequality may not accurately represent the MISE.
    Nevertheless, we can operationalize this inequality to obtain a rule-of-thumb for the optimal number of expansion terms $N$ given the feasible estimates $\widehat{A}_N$. We provide such a rule-of-thumb algorithm in Appendix \ref{appendix:additional}.

    Finally, we note that the MISE given by equation \eqref{eq mise} is for the Fourier-cosine expansion. Hence, the delta estimator, which essentially utilizes the Fourier-sine expansion, may require a different\footnote{The sine expansion is known to have a slower rate of convergence than the cosine series. In fact, this is the main reason for popularity of the Fourier cosine expansions rather than the Fourier or sine series.} optimal choice of expansion terms $\widetilde{N}$. In this case, a similar rule can be applied but with the sine coefficients $\widehat{B}_N$ instead.

	\section{Monte Carlo study}\label{sec:simulation}

In this section, we investigate the finite-sample performance of the developed non-parametric estimators.
We consider two models to generate data: the \citeA{black1973pricing} model and the `double-jump' stochastic volatility model of \citeA{DPS2000}. The former has closed-form solutions for the true quantities of interests, while the latter offers a more realistic depiction of options data that features two stylized facts -- stochastic volatility and jump components in returns and volatility. 

For each model, we set the initial spot price $S_0=4000$, the interest rate $r=0$, and the strike prices between 85\% and 110\% of the spot price with equidistant increments of 5, similar to the available S\&P 500 index option data. This results in $n=201$ option contracts for each maturity.

We distort the true option prices of each model by adding homoskedastic observation errors, i.e.,
\begin{align*}
    O(K_i) = O_0(K_i) + 0.025\cdot \epsilon, \quad i=1,\dots,n,
\end{align*}
where $\epsilon$ is an i.i.d.\ standard normal random variable.
This error structure roughly matches the dispersion of errors in empirical applications, where option errors are typically within the tick size of \$0.05. Note that the smallest and the largest strike prices are fixed and the corresponding OTM options are strictly positive.

\subsection{Black-Scholes model}

First, we consider the Black-Scholes model with short (30 days) and long (1 year) maturities, and set the volatility parameter $\sigma=0.3$. The true option prices are generated via the Black-Scholes formula and then distorted with the additive error terms as described above. Table~\ref{tab:options-BS-mc} provides the simulation results of the estimated option call prices for a selection of strike prices, along with the estimates of $\boldsymbol{\widehat{\theta}}$. The latter includes the intercept $\bar{\theta}$ and the first-order derivatives $\theta_c$ and $\theta_p$, which have closed-form solutions in the Black-Scholes model. The number of expansion terms is set to $N=14$ for short maturity options and to $N=7$ for long maturity options. This choice is motivated by the rule-of-thumb discussed in Section \ref{sec:optimal N} and Appendix~\ref{appendix:additional}. The numerical integration scheme is set to Simpson's 1/3 rule throughout the simulations.

\begin{table}[!ht]
    \centering
    \caption{Monte Carlo results for the call prices based on the Black-Scholes model}
    \footnotesize
    \begin{tabular}{lcccccc|ccc}
        \toprule 
        $K/F_0$ & 0.86 & 0.9 & 0.95 & 1.0 & 1.05 & 1.09 & $\bar{\theta}$ & $\theta_c$ & $\theta_p$ \\
        \multicolumn{10}{c}{$T = 30$ days} \\
        \midrule 
        $C_0(K)$ & 565.11 & 417.38 & 256.86 & 137.21 & 62.66 & 29.79 & 0.0 & -0.125 & 0.032 \\
        MC bias & 0.0002 & 0.0002 & -0.00032 & 0.00031 & 0.0007 & -0.00117 & 0.00347 & 0.00064 & -0.00045 \\
        MC std & 0.0087 & 0.0071 & 0.0066 & 0.0066 & 0.0072 & 0.0083 & 0.0205 & 0.0009 & 0.0011 \\
        As. std & 0.0084 & 0.007 & 0.0068 & 0.0068 & 0.0068 & 0.008 & 0.0141 & 0.0008 & 0.001 \\
        \multicolumn{10}{c}{$T = 1$ year} \\
        \midrule 
        $C_0(K)$ & 777.92 & 680.52 & 571.75 & 476.94 & 395.27 & 338.66 & 0.0 & -0.32 & 0.348 \\
        MC bias & -0.00065 & -0.00033 & -0.00014 & 0.00052 & 0.00014 & -0.00064 & 0.00157 & 0.00013 & 0.69513 \\
        MC std & 0.0063 & 0.0055 & 0.0048 & 0.0049 & 0.0049 & 0.0058 & 0.0231 & 0.0003 & 0.0004 \\
        As. std & 0.006 & 0.0054 & 0.0047 & 0.0049 & 0.005 & 0.0057 & 0.0113 & 0.0003 & 0.0004 \\
        \bottomrule 
    \end{tabular}              
        
    \label{tab:options-BS-mc}%
		  
    \medskip
    \begin{minipage}{0.95\textwidth}\scriptsize
        Note: This table provides Monte Carlo simulation results for the call option price estimates, based on 1000 replications from the Black-Scholes model. 
        Two settings with short (30 days) and long (1 year) maturities are considered. 
        For different strike prices, each panel lists the true call value ($C_0$), the Monte Carlo bias (MC bias), the Monte Carlo standard deviation (MC std), and the asymptotic standard deviation (As.\ std), defined as the square root of the average estimated asymptotic variance. 
        The number of expansion terms is $N=14$ and $N=7$ for $T = 30$ days and $T = 1$ year option contracts, respectively. The numerical integration scheme is Simpson's 1/3 rule.
    \end{minipage}

\end{table}

Although we estimate option prices from the set of OTM prices themselves, the simulation results indicate the convergence of the considered approach. Furthermore, the Monte Carlo results show no significant bias for all levels of strikes and maturities and a reduction in the variance of option prices. In fact, the Monte Carlo standard deviations of the estimated prices are smaller than the standard deviations used to simulate option errors, indicating the smoothing effect of the estimation procedure. The asymptotic standard deviations, defined as the square root of the average estimated asymptotic variance, roughly correspond to the Monte Carlo standard errors, which indicates the validity of the constructed standard errors.

The estimated parameters $\boldsymbol{\widehat{\theta}}$ also exhibit good finite-sample performance. The estimated intercept $\bar{\theta}$, which collects the average of finite-sample biases, indicates that these errors are of rather small order. The good finite-sample performance of the first-order derivatives $\theta_c$ and $\theta_p$ is crucial for the RND and delta estimators considered below.

\begin{table}[!h]
    \centering
    \caption{Monte Carlo results for the RND based on the Black-Scholes model}
    \footnotesize
    \begin{tabular}{lcccccc}
        \toprule 
        $K/F_0$ & 0.86 & 0.9 & 0.95 & 1.0 & 1.05 & 1.09 \\
        \multicolumn{7}{c}{$T = 30$ days} \\
        \midrule 
        $f(\log K)$ & 1.07 & 2.31 & 3.98 & 4.63 & 3.85 & 2.69 \\
        MC bias & -0.002 & -0.0029 & 0.0037 & -0.0018 & -0.0052 & 0.0083 \\
        MC std & 0.0574 & 0.0239 & 0.02 & 0.0212 & 0.0241 & 0.0615 \\
        As. std & 0.055 & 0.0235 & 0.0203 & 0.0213 & 0.0229 & 0.0571 \\
        \multicolumn{7}{c}{$T = 1$ year} \\
        \midrule 
        $f(\log K)$ & 1.25 & 1.3 & 1.33 & 1.31 & 1.27 & 1.21 \\
        MC bias & 0.0046 & 0.0009 & 0.0002 & -0.001 & 0.0001 & 0.0074 \\
        MC std & 0.0185 & 0.006 & 0.0033 & 0.0042 & 0.005 & 0.015 \\
        As. std & 0.0176 & 0.0058 & 0.0033 & 0.0043 & 0.005 & 0.0149 \\
        \bottomrule 
    \end{tabular}                     
        
    \label{tab:rnd-BS-mc}%
		  
    \medskip
    \begin{minipage}{0.95\textwidth}\scriptsize
        Note: This table provides Monte Carlo simulation results for the RND estimates, based on 1000 replications from the Black-Scholes model. 
        Two settings with short (30 days) and long (1 year) maturities are considered. 
        For different strike levels, each panel lists
        the true RND value for log future price ($f(\log K)$), the Monte Carlo bias (MC bias), the Monte Carlo standard deviation (MC std), and the asymptotic standard deviation (As.\ std), defined as the square root of the average estimated asymptotic variance. 
        %
        The number of expansion terms is $N=14$ and $N=7$ for $T = 30$ days and $T = 1$ year option contracts, respectively. The numerical integration scheme is Simpson's 1/3 rule.
    \end{minipage}

\end{table}

The Monte Carlo simulation results for the RND estimates are reported in Table \ref{tab:rnd-BS-mc}. Note that the RND estimator is for the log price $\log S_T$ and evaluated at a few strike levels after the corresponding log transformation. The true RND for the Black-Scholes model is the normal distribution with mean $\log S_0 - \tfrac{1}{2}\sigma^2 T$ and variance $\sigma^2 T$. Similar to the option price estimates, the estimated RNDs exhibit good finite-sample performance. 
The asymptotic standard errors roughly match the Monte Carlo standard errors, and both tend to increase towards the bounds of the considered interval.

\begin{table}[!ht]
    \centering  
    \caption{Monte Carlo results for the call deltas based on the Black-Scholes model}
    \footnotesize
    \begin{tabular}{lcccccc}
        \toprule 
        $K/F_0$ & 0.86 & 0.9 & 0.95 & 1.0 & 1.05 & 1.09 \\
        \multicolumn{7}{c}{$T = 30$ days} \\
        \midrule 
        $\delta(K)$ & 0.964 & 0.898 & 0.739 & 0.517 & 0.3 & 0.169 \\
        MC bias & -0.00622 & -0.0064 & -0.00622 & -0.00634 & -0.00669 & -0.00762 \\
        MC std & 0.00122 & 0.00118 & 0.0012 & 0.00116 & 0.00106 & 0.00125 \\
        As. std & 0.0014 & 0.00138 & 0.00142 & 0.00137 & 0.00129 & 0.00141 \\
        \multicolumn{7}{c}{$T = 1$ year} \\
        \midrule 
        $\delta(K)$ & 0.74 & 0.69 & 0.63 & 0.56 & 0.49 & 0.45 \\
        MC bias & -0.0027 & -0.00313 & -0.00308 & -0.00312 & -0.00324 & -0.00387 \\
        MC std & 0.00138 & 0.00134 & 0.00134 & 0.00135 & 0.00119 & 0.00131 \\
        As. std & 0.00139 & 0.00136 & 0.00136 & 0.00136 & 0.00122 & 0.00133 \\
        \bottomrule 
    \end{tabular}                  
        
    \label{tab:delta-BS-mc}%
		  
    \medskip
    \begin{minipage}{0.95\textwidth}\scriptsize
        Note: This table provides Monte Carlo simulation results for the delta estimates, based on 1000 replications from the Black-Scholes model. 
        Two settings with short (30 days) and long (1 year) maturities are considered. 
        Each panel lists, for different strike levels,
        the true delta value, the Monte Carlo bias and standard deviation.
        For different strike levels, each panel lists
        the true delta value ($\delta(K)$), the Monte Carlo bias (MC bias), the Monte Carlo standard deviation (MC std), and the asymptotic standard deviation (As.\ std), defined as the square root of the average estimated asymptotic variance.
        The number of expansion terms is $\widetilde{N}=25$ and the numerical integration scheme is Simpson's 1/3 rule.
    \end{minipage}

\end{table}

Finally, the simulation results for the delta estimates are reported in Table \ref{tab:delta-BS-mc}. The true values are obtained in the closed-form under the Black-Scholes assumptions. The number of expansion terms for the delta estimator is set to $\widetilde{N}=25$ since the sine series exhibits a slower convergence rate (see discussion in Section \ref{sec:optimal N}).
Unlike the RND and call price estimates, the estimation of the delta exhibits small bias terms, which are economically likely to be negligible.

\subsection{SVCJ model}

The `double-jump' stochastic volatility model of \citeA{DPS2000}, labeled as SVCJ, allows for stochastic volatility and jumps in returns and volatility, and under the risk-neutral measure $\Q$ is given by the following system of stochastic differential equations:
\begin{align*}
    \diff \log S_t &= (r -\tfrac{1}{2}v_t - \mu\lambda) \dt + \sqrt{v_t} \diff W_{1,t} + J_t \diff N_t,\\
    \diff v_t &= \kappa(\bar{v} - v_t) \dt + \sigma \sqrt{v_t} \diff W_{2,t} + J_t^v \diff N_t,
\end{align*}
where two Brownian motions $W_1$ and $W_2$ are correlated with coefficient $\rho$, and $N_t$ is a Poisson jump process with intensity $\lambda$. The jump sizes in returns are Gaussian, $J \sim \mathcal{N}(\mu_j, \sigma_j^2)$ with the expected relative jump size in returns $\mu = \exp(\mu_j + \frac{1}{2}\sigma_j^2) - 1$, while the co-jump sizes in volatility are exponentially distributed, $J^v \sim \exp(1/\mu_v)$, and independent of jump sizes in returns. We choose the following parameter values for the simulation:
\begin{align*}
    v_0 = 0.1^2,\ \kappa = 2.6,\ \bar{v}=0.02,\ \rho = -0.95,\ \sigma=0.3,\ \lambda = 1.0,\ \mu_j = -0.05,\ \sigma_j = 0.03,\ \mu_v=0.05.
\end{align*}

Since there is no closed-form solution for option prices under the SVCJ model, we simulate them using the COS method. We use the analytic solution for the CCF and a large number of expansion terms $N=1024$ with $[a,b] = [-4\sqrt{T}, 4\sqrt{T}]$. We then add observation errors to these option prices as described previously.

\begin{figure}[htb]
    \centering
    
    \caption{Simulated option prices from the SVCJ model on BSIV and log-OTM spaces}
    \begin{subfigure}{0.45\textwidth}
        \includegraphics[scale=0.35]{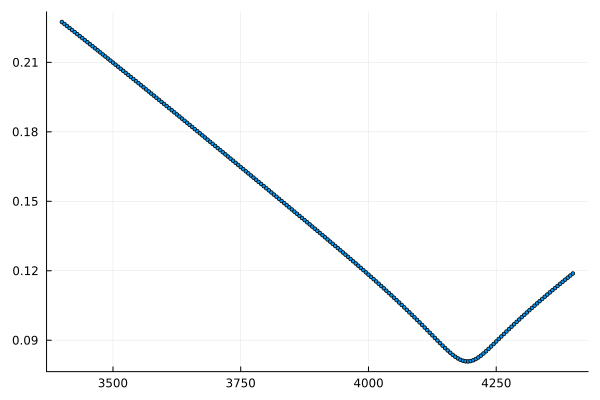}
        \caption{BSIV}
    \end{subfigure}
    \begin{subfigure}{0.45\textwidth}
        \includegraphics[scale=0.35]{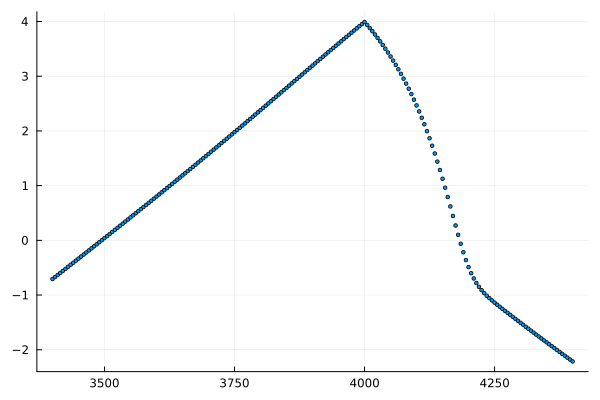}
        \caption{log OTM}
    \end{subfigure}
    
    \label{fig: sim-dps}
		  
    \medskip
    \begin{minipage}{0.95\textwidth}\scriptsize
        Note: This figure illustrates the simulated option prices from the SVCJ model displayed on BSIV (Panel (a)) and log-OTM (Panel (b)) spaces. The simulation details are provided in the main text.
    \end{minipage}
\end{figure}

Figure \ref{fig: sim-dps} displays the simulated option prices from the SVCJ model on BSIV and log OTM spaces for $T=30$ days. The chosen model parameters generate the so-called implied volatility `smile', which is commonly observed in the market, particularly for short-dated options. Capturing such pronounced `smiles' can be challenging for many parametric and non-parametric methods since they require the methods to be rather flexible. As a consequence, these methods often fail to accurately capture option prices, RND, and deltas.

For the SVCJ model, we compare the simulation results of the developed approach with the closest non-parametric and widely-used alternative, the kernel smoother. In fact, kernel smoothing methods are also model-free and do not require any optimization routines.
In particular, we consider the Nadaraya–Watson kernel estimator with the Gaussian kernel applied to the BSIV space, as in, e.g., \citeA{ait1998nonparametric} and \citeA{grith2012nonparametric}. After smoothing BSIV observations, we convert them into price dimension to obtain call price estimates. We then calculate the second-order derivatives to obtain the RND estimates due to \citeA{breeden1978prices}. Fitting option prices on implied volatility space is commonly used in practice (see, e.g., \citeA{ait1998nonparametric}, \citeA{andersen2015parametric} among many others).

Finding the bandwidth parameter $h$ is crucial for the kernel smoothing methods as it controls the bias-variance tradeoff. Since we are interested in estimating both option prices and the RND, we consider the kernel bandwidths, as in \citeA{ait1998nonparametric}, of the following form:
\begin{align*}
    h = \tfrac{c}{\log n} n^{-\frac{1}{2p+1}}
\end{align*}
with $p=2$ and some constant $c>0$. In practice, one can use the cross-validation to find the optimal bandwidth, but in simulations we vary the constant $c$ to illustrate the bias-variance tradeoff.

\begin{table}[htbp]
    \centering
    \caption{Monte Carlo results for the call prices under the SVCJ model}
    \footnotesize
    \begin{tabular}{llcccccc}
        \toprule 
         & $K/F_0$ & 0.86 & 0.9 & 0.95 & 1.0 & 1.05 & 1.09 \\
        \midrule 
        $C_0(K)$ &  & 560.66 & 402.23 & 210.81 & 54.13 & 0.61 & 0.14 \\[1ex]
        iCOS & MC bias & -0.00075 & 0.00039 & -0.00036 & 3.0e-5 & -0.00207 & -0.00055 \\
         & MC std & 0.0102 & 0.00969 & 0.00941 & 0.00926 & 0.009 & 0.00933 \\
         & As. std & 0.00981 & 0.00944 & 0.009 & 0.00909 & 0.00906 & 0.00893 \\
        \midrule 
        KS, $c=0.2$ & MC bias & -0.07909 & -0.00276 & -0.00383 & -0.10338 & 0.35918 & -0.04309 \\
         & MC std & 0.00442 & 0.00427 & 0.0044 & 0.00433 & 0.00637 & 0.00369 \\
        \midrule 
        KS, $c=0.1$ & MC bias & -0.01325 & 2.0e-5 & -0.00027 & -0.02496 & 0.11357 & -0.00912 \\
         & MC std & 0.00621 & 0.00582 & 0.00603 & 0.00589 & 0.00661 & 0.00594 \\
        \midrule 
        KS, $c=0.05$ & MC bias & -0.00078 & 0.0005 & 6.0e-5 & -0.00627 & 0.03224 & -0.00178 \\
         & MC std & 0.00828 & 0.00811 & 0.00835 & 0.00831 & 0.00839 & 0.00834 \\
        \midrule 
        KS, $c=0.03$ & MC bias & -0.00066 & 0.00068 & 0.00024 & -0.00249 & 0.01229 & -0.00133 \\
         & MC std & 0.01043 & 0.01051 & 0.01063 & 0.01061 & 0.01068 & 0.01082 \\
        \bottomrule 
    \end{tabular}                   
        
    \label{tab:options-DPS-mc}%
    \medskip
    \begin{minipage}{0.95\textwidth}\scriptsize
        Note: This table provides Monte Carlo simulation results for the call option price estimates under the SVCJ model based on the iCOS approach and the kernel smoothing (KS) with different bandwidths values.
        The number of expansion terms is $N=25$ and the numerical integration scheme is Simpson's 1/3 rule.
    \end{minipage}

\end{table}

Table \ref{tab:options-DPS-mc} provides simulation results for the call price estimates under the simulated SVCJ model. We compare the results obtained using our proposed iCOS method with the widely-used kernel smoothing approach with different bandwidth parameters. First, we observe that the simulation results for the call price estimates under the SVCJ model using our approach are similar to the results based on the Black-Scholes model discussed in the previous subsection. 
Second, as expected, the biases for the kernel smoother decline as the bandwidth parameter decreases, but this comes at the cost of increased variance. Notably, the biases for the kernel smoother are especially pronounced at the moneyness level of 1.05, which roughly corresponds to the `turning' point of the smile depicted in Figure~\ref{fig: sim-dps}. Finally, when comparing two methods, we notice that only the results with the parameter $c=0.03$ for the kernel smoother are comparable to the iCOS approach in terms of biases. However, such a small bandwidth value results in a non-smooth RND as we discuss below.

\begin{table}[htbp]
    \centering
    \caption{Monte Carlo results for the RND under the SVCJ model}
    \footnotesize
    \begin{tabular}{llcccccc}
        \toprule 
         & $K/F_0$ & 0.86 & 0.9 & 0.95 & 1.0 & 1.05 & 1.09 \\
        \midrule 
        $f(\log K)$ &  & 0.13 & 0.49 & 2.76 & 11.37 & 3.11 & 0.03 \\[1ex]
        iCOS & MC bias & 0.0211 & 0.0032 & 0.0058 & -0.0066 & 0.0711 & 0.0216 \\
         & MC std & 0.1503 & 0.1138 & 0.0994 & 0.0972 & 0.0954 & 0.0937 \\
         & As. std & 0.1462 & 0.1099 & 0.095 & 0.0933 & 0.091 & 0.0854 \\
        \midrule 
        KS, $c=0.2$ & MC bias & 0.002 & 0.0511 & 0.0169 & 0.319 & -0.2341 & 0.0161 \\
         & MC std & 0.0033 & 0.005 & 0.006 & 0.0702 & 0.0075 & 0.0043 \\
        \midrule 
        KS, $c=0.1$ & MC bias & -0.0999 & -0.0373 & -0.1284 & -0.2761 & -0.1874 & -0.0584 \\
         & MC std & 0.0211 & 0.0263 & 0.0286 & 0.0316 & 0.0327 & 0.0285 \\
        \midrule 
        KS, $c=0.05$ & MC bias & -0.0209 & -0.0203 & -0.0609 & -0.1186 & -0.1087 & -0.0186 \\
         & MC std & 0.1364 & 0.1469 & 0.1541 & 0.1627 & 0.1743 & 0.1844 \\
        \midrule 
        KS, $c=0.03$ & MC bias & 0.0063 & -0.0174 & -0.0296 & 0.0006 & -0.0703 & -0.0091 \\
         & MC std & 0.469 & 0.4929 & 0.5117 & 0.5418 & 0.5904 & 0.6125 \\
        \bottomrule 
    \end{tabular}                                          
        
    \label{tab:rnd-dps-mc}%
    \medskip
    \begin{minipage}{0.95\textwidth}\scriptsize
        Note: This table provides Monte Carlo simulation results for the RND estimates under the SVCJ model based on the iCOS approach and the kernel smoothing (KS) with different bandwidths values.
        The number of expansion terms is $N=25$ and the numerical integration scheme is Simpson's 1/3 rule.
    \end{minipage}

\end{table}

Table \ref{tab:rnd-dps-mc} provides the Monte Carlo simulation results for the RND of log future prices. We again compare our approach with the kernel smoothing method using the same set of bandwidth parameters. The developed iCOS approach demonstrates a good finite-sample performance, similar to the Black-Scholes case. In contrast, the results for the kernel smoothing approach are often biased or exhibit large variances, which is an indication of non-smooth and, hence, arbitrage-violating RND estimates. 

Analogously, Table \ref{tab: delta-dps-mc} provides the Monte Carlo results for the call delta estimates under the SVCJ model. The non-parametric iCOS method yields insignificant biases but larger variances than the kernel smoothing method. The latter, however, again exhibits biases at the moneyness level of 1.05 except for the parameter $c=0.03$, which corresponds to a non-smooth RND.

Overall, when comparing two approaches, we observe that the kernel smoothing method fails to fully capture the shape of the observed option data, which results in the biased estimates of the call prices, RND, and deltas. Decreasing the bandwidth parameter reduces the biases but at the cost of a non-smooth RND with large variance. In contrast, the iCOS method is able to simultaneously capture the shape of the observed option data, RND and deltas with insignificant biases and small variances.

\begin{table}[htbp]
    \centering
    \caption{Monte Carlo results for the delta under the SVCJ model}
    \footnotesize
    \begin{tabular}{llcccccc}
        \toprule 
         & $K/F_0$ & 0.86 & 0.9 & 0.95 & 1.0 & 1.05 & 1.09 \\
        \midrule 
        $\delta(K)$ &  & 0.9959 & 0.9852 & 0.9195 & 0.5959 & 0.0156 & 0.0011 \\[1ex]
        iCOS & MC bias & 5.6e-5 & 9.8e-5 & 7.6e-5 & 0.000129 & 0.000177 & 3.8e-5 \\
         & MC std & 0.0014 & 0.0012 & 0.0013 & 0.0012 & 0.0014 & 0.0011 \\
         & As. std & 0.0019 & 0.0019 & 0.002 & 0.0019 & 0.002 & 0.0019 \\
        \midrule 
        KS, $c=0.2$ & MC bias & -0.000997 & -2.1e-5 & 8.5e-5 & 0.001071 & 0.009063 & 0.000448 \\
         & MC std & 3.7e-5 & 5.2e-5 & 5.7e-5 & 7.9e-5 & 8.6e-5 & 4.5e-5 \\
        \midrule 
        KS, $c=0.1$ & MC bias & -0.000641 & -1.0e-5 & -4.1e-5 & 0.000165 & 0.003354 & 0.000397 \\
         & MC std & 0.000135 & 0.000143 & 0.000152 & 0.00016 & 0.000182 & 0.000168 \\
        \midrule 
        KS, $c=0.05$ & MC bias & -3.1e-5 & -0.0 & -3.5e-5 & 1.9e-5 & 0.001163 & 1.7e-5 \\
         & MC std & 0.000383 & 0.000382 & 0.00041 & 0.000432 & 0.000474 & 0.00047 \\
        \midrule 
        KS, $c=0.03$ & MC bias & 1.0e-6 & 7.0e-6 & -4.2e-5 & 7.0e-6 & 0.000594 & -1.1e-5 \\
         & MC std & 0.000772 & 0.000774 & 0.000833 & 0.000891 & 0.000948 & 0.00096 \\
        \bottomrule 
    \end{tabular}                                                         
        
    \label{tab: delta-dps-mc}%
    \medskip
    \begin{minipage}{0.9\textwidth}\scriptsize
        Note: This table provides Monte Carlo simulation results for the call delta estimates under the SVCJ model based on the iCOS approach and the kernel smoothing (KS) with different bandwidths values.
        The number of expansion terms is $\widetilde{N}=30$ and the numerical integration scheme is Simpson's 1/3 rule.
    \end{minipage}

\end{table}

	\section{Empirical applications}
\label{sec:empirics}

	In this section, we first demonstrate the proposed estimation procedures in two empirical illustrations. Then, we apply the developed method to analyse errors in the VIX index.

\subsection{SPX options}
    In the first empirical illustration, we consider options on the S\&P 500 stock market index (SPX) obtained from the Chicago Board Options Exchange (CBOE), which are commonly used in the literature. We consider a snapshot of options at 3:45 pm ET with a time-to-maturity $T = 29$ days traded on April~1, 2021. The forward price implied from the put-call parity is $F = \$ 4008.5$.
	For the developed estimation procedure, we utilize the mid-quote prices of OTM contracts. Additionally, we compare the pricing accuracy of our approach with the bid-ask spread of the corresponding contracts, which allows us to assess the performance of our method in the real-world options market context.

    We focus our analysis on the interval $[\alpha, \beta] = [2950, 4400]$, i.e., we use options with strikes from this interval. Although there are a few option contracts with strike prices beyond this range, they tend to be less liquid and spaced further apart\footnote{ The distance between strike prices outside the considered interval is \$50 or \$100, while the distance for close to ATM options is only \$5.}. We do not filter out any options, except for those with zero bid prices. This results in a total number of 239 OTM options with non-equidistant strike prices.

	\begin{figure}[htbp]
		\centering
		\caption{Estimated cosine coefficients}
		\label{fig: Dm}

		\begin{subfigure}{0.45\textwidth}
			\hspace*{-0.5cm}
			\includegraphics[scale=0.12]{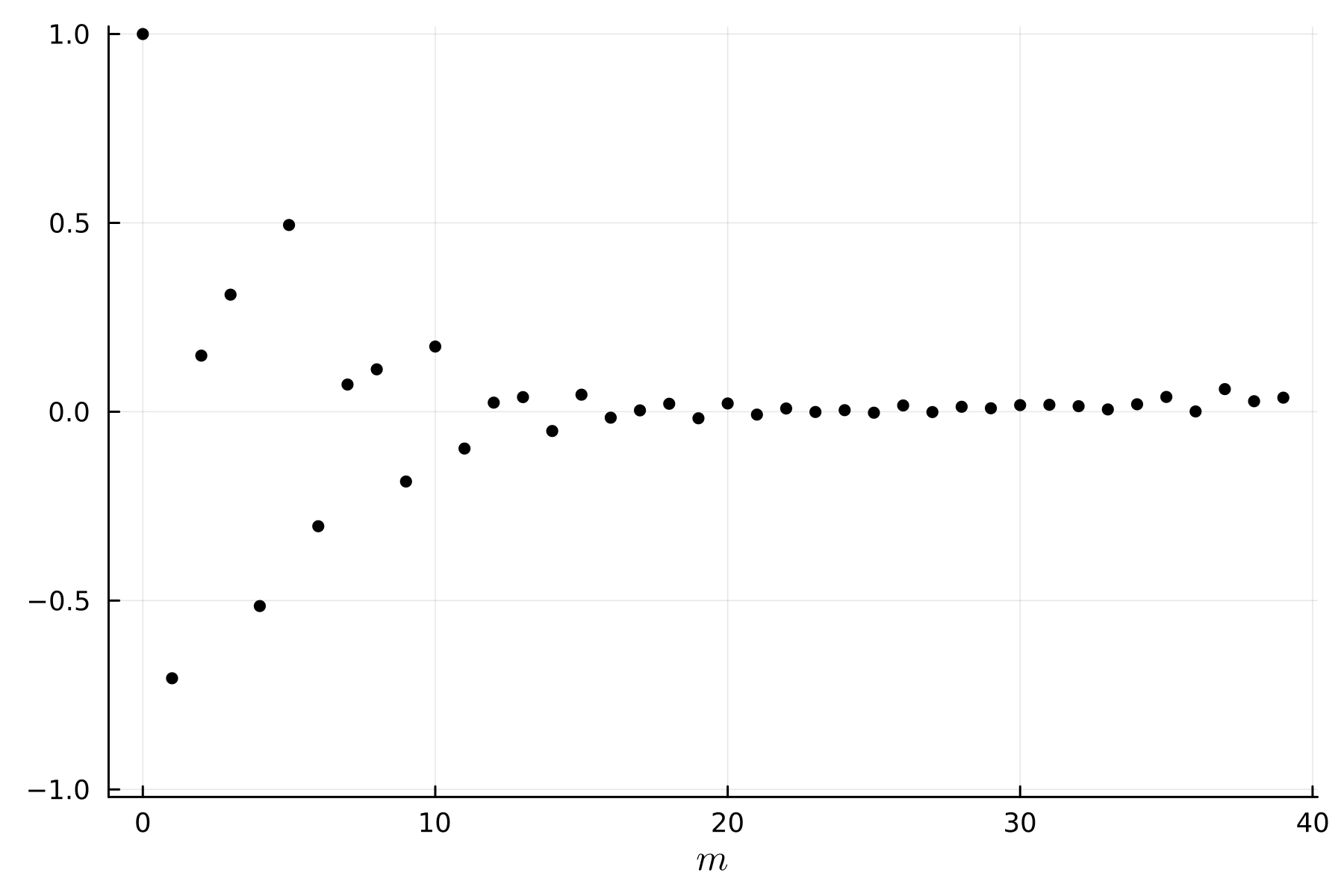}
            \caption{$\widehat{D}_m$}
		\end{subfigure}
		\begin{subfigure}{0.45\textwidth}
			\includegraphics[scale=0.12]{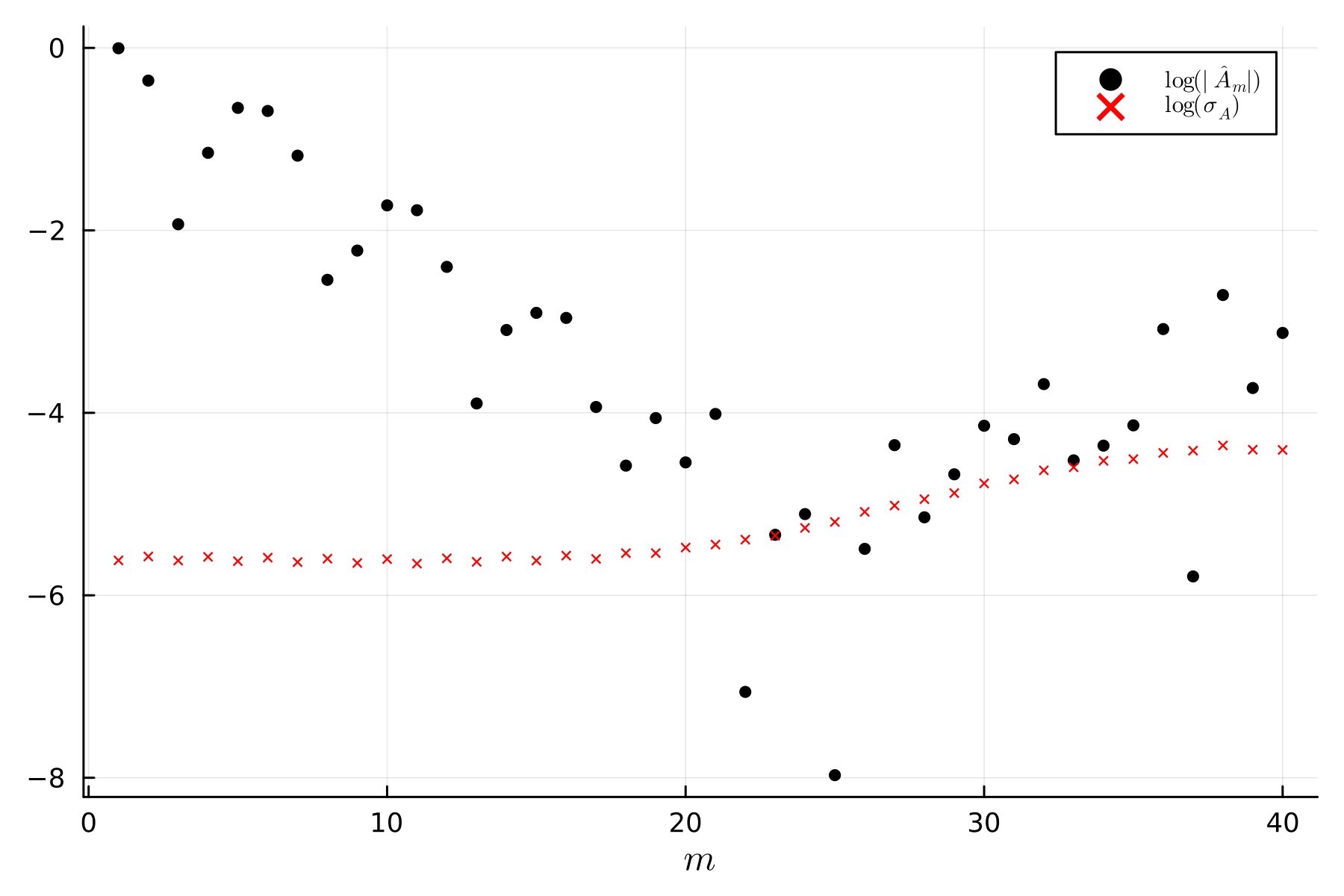}
            \caption{$|\widehat{A}_m|$ and $\sigma_A$ in log-scale}
		\end{subfigure}

		\medskip
		\begin{minipage}{0.9\textwidth}\scriptsize
			Note: This figure plots the option-implied cosine coefficient estimates $\widehat{D}_m$ and $\widehat{A}_m$ based on Simpson's 1/3 rule. The SPX options are with 29 days-to-maturity traded on April 1, 2021.
		\end{minipage}
	\end{figure}

	To approximate the integrals in the option-implied cosine coefficients \eqref{iDm_hat}, we use Simpson's~1/3 rule. Figure~\ref{fig: Dm} displays the estimated coefficients $\widehat{D}_m$ and $\widehat{A}_m$ plotted against the term number $m$. The coefficients $\widehat{A}_m$ are displayed after applying a logarithmic transformation, along with their corresponding standard deviations $\sigma_A$. As expected, these coefficients converge towards zero up to a certain point but then exhibit a divergent pattern due to the discretization errors and increased variance. The rule-of-thumb algorithm, detailed in Appendix \ref{appendix:additional}, selects $N^{*}=23$ as the optimal number of terms. This number roughly corresponds to the point where the cosine coefficients stabilize and their standard deviations surpass the magnitude of the coefficients themselves. We use this number of terms for the subsequent analysis.

\begin{figure}[h!]
		\centering
		\caption{SPX option price estimates based on the iCOS method}
		\begin{subfigure}{0.45\textwidth}
			\hspace*{-0.5cm}
			\includegraphics[scale=0.12]{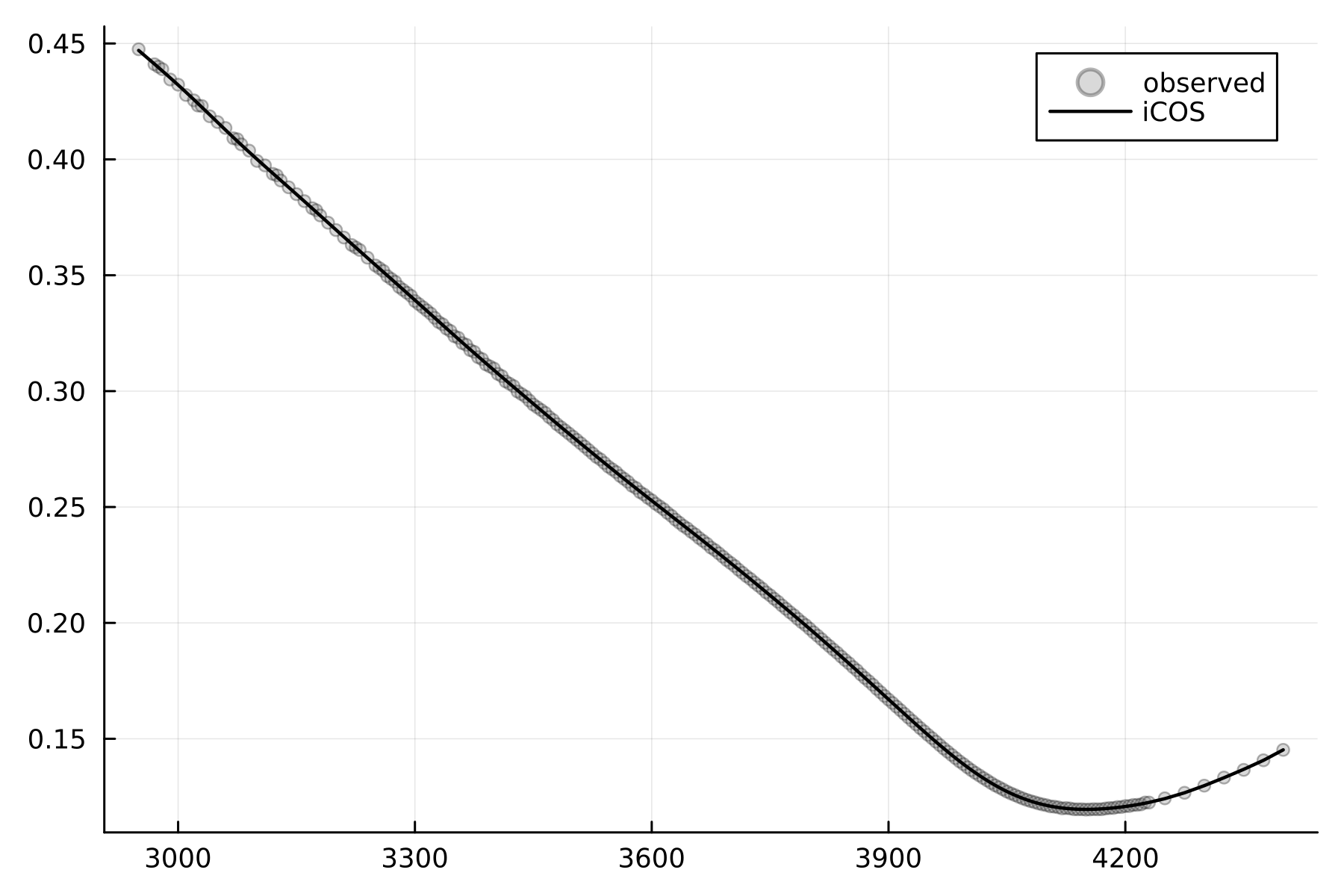}
            \caption{BSIV}
		\end{subfigure}
		\begin{subfigure}{0.45\textwidth}
			\includegraphics[scale=0.12]{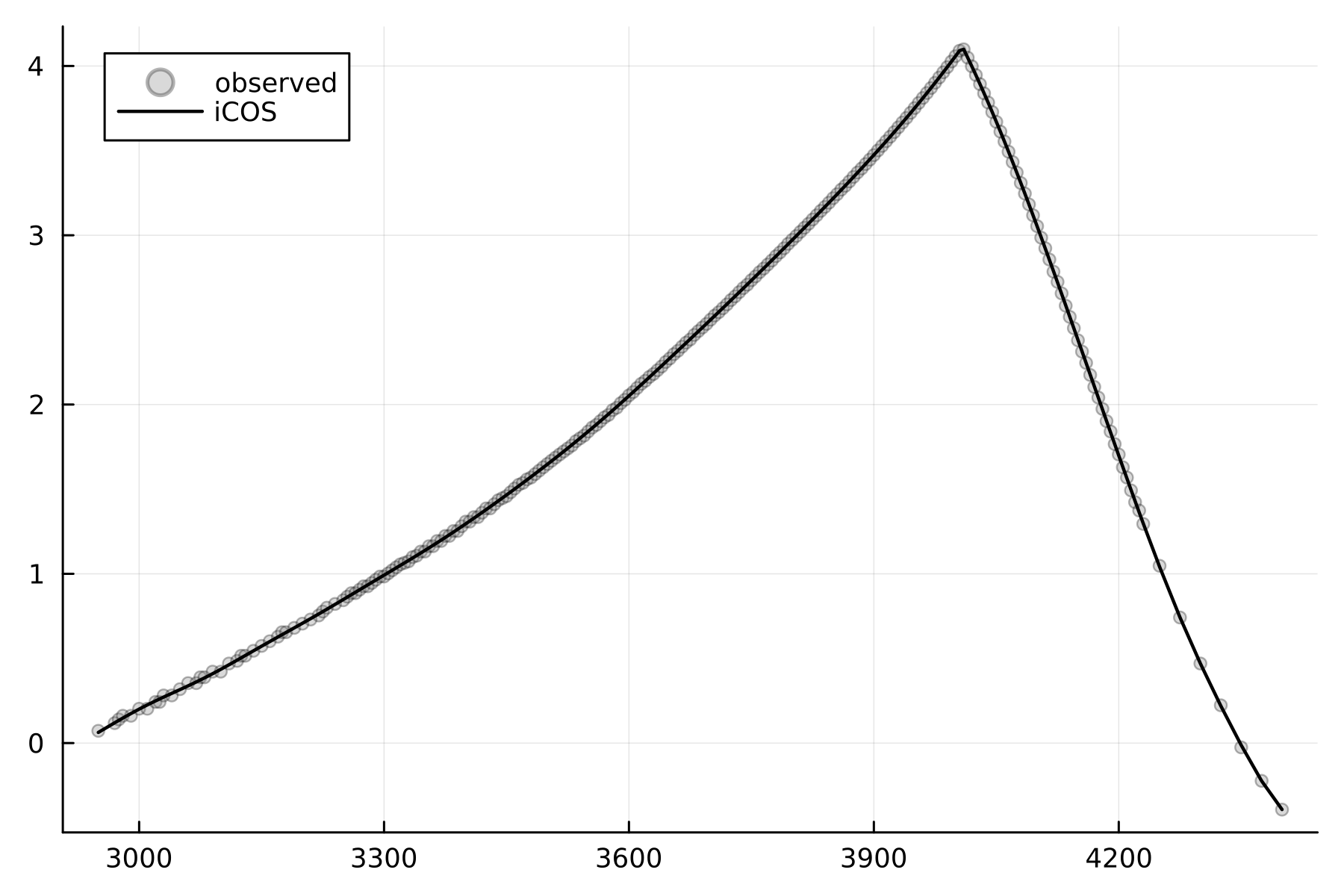}
            \caption{log OTM}
		\end{subfigure}
		\label{fig:interp}

		\medskip
		\begin{minipage}{\textwidth}\scriptsize
            Note: This figure illustrates the call price estimation of the proposed method based on the SPX options traded on April~1, 2021 with 29 days-to-maturity. Panel (a) displays the result of the estimation along with the observed option prices on BSIV space, while Panel (b) plots displays it on log OTM space. The estimation is conducted in terms of the dollar-amount OTM prices.
		\end{minipage}
	\end{figure}

	Figure \ref{fig:interp} depicts the option-implied call price estimates $\widehat{C}(x)$ given by equation \eqref{iCall-estimator} for the considered data. The figure displays the market prices in terms of the BSIV and the logarithm of OTM prices, although the estimation is performed in terms of dollar-amount OTM prices. As observed, the estimates closely capture the shape of the implied volatility smile and the log OTM prices. Note that the solid line does not pass exactly through all market prices, but provides an approximation of them.

\begin{figure}[h]
		\centering
		\caption{Pricing errors for call estimates}
		\begin{subfigure}{0.45\textwidth}
			\hspace*{-0.5cm}
			\includegraphics[scale=0.12]{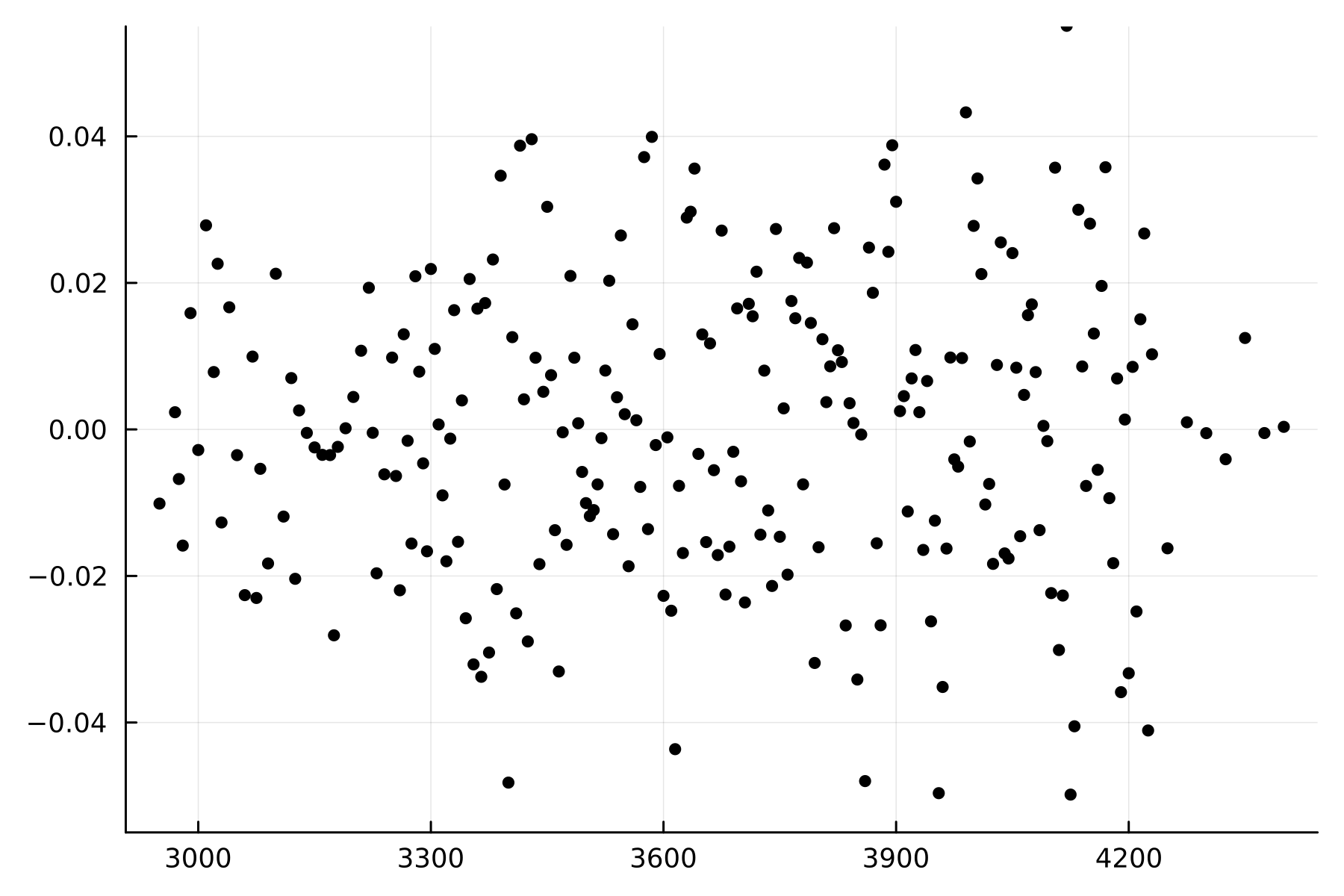}
            \caption{$\widehat{C}(x) - C(x)$}
		\end{subfigure}
		\begin{subfigure}{0.45\textwidth}
			\includegraphics[scale=0.12]{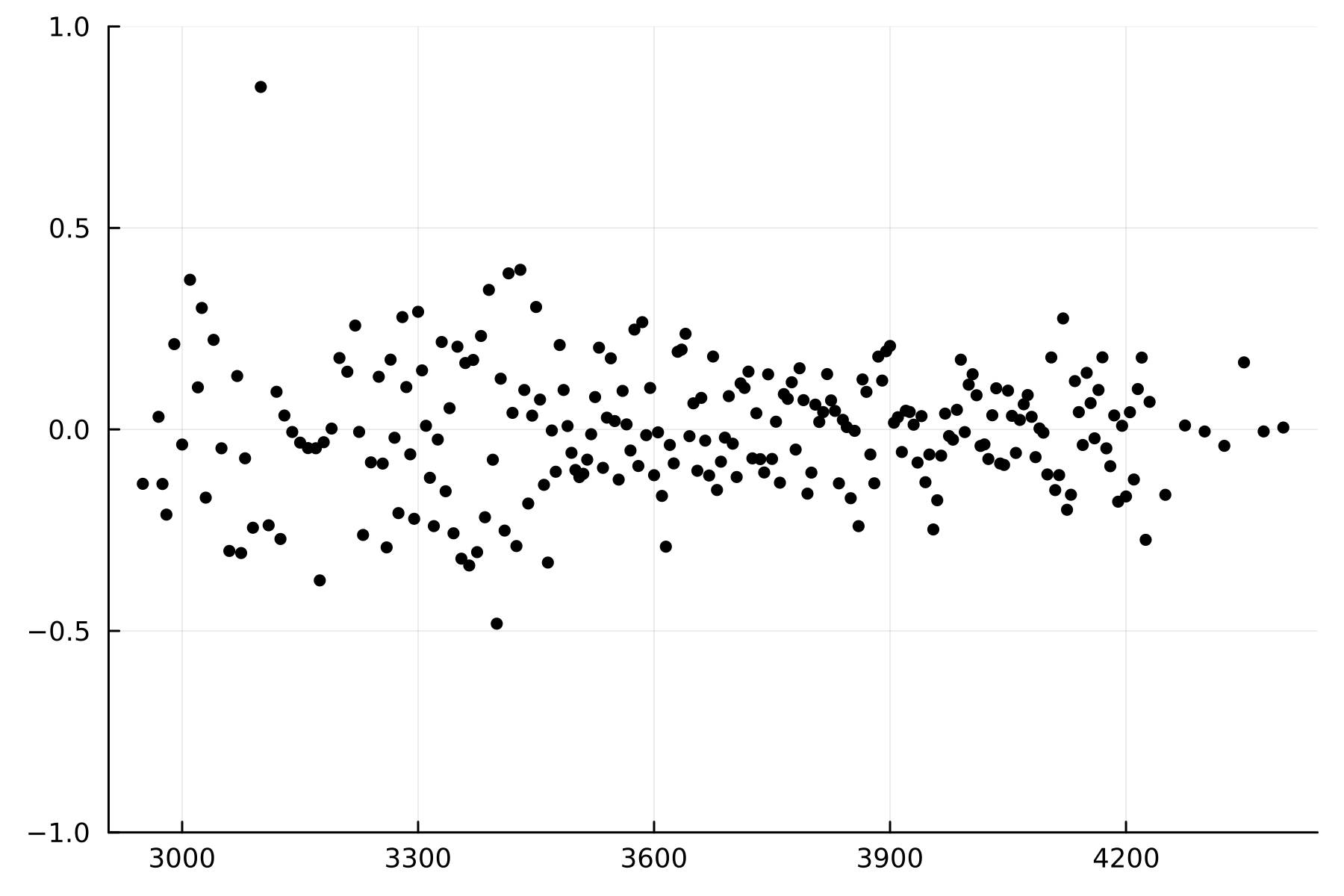}
            \caption{$\frac{\widehat{C}(x) - C(x)}{HS(x)}$}
		\end{subfigure}
		\label{fig:interp_erors}

		\medskip
		\begin{minipage}{\textwidth}\scriptsize
            Note: This figure illustrates the pricing errors of the proposed method based on the SPX options traded on April 1, 2021 with 29 days-to-maturity. Panel (a) plots the pricing errors $\widehat{C} - C$. Panel (b) plots  pricing errors relative to the half-spread~$\frac{\widehat{C} - C}{HS}$.
		\end{minipage}
	\end{figure}

	To investigate the accuracy of our estimation procedure, we plot in Figure \ref{fig:interp_erors}(a) the pricing errors as the difference between the call option price estimates $\widehat{C}(x)$ and the market observed call prices $C(x)$ for $x\in\{K_1, \dots, K_n\}$. Most of these differences range between \$-0.05 and \$0.05, corresponding to two ticks of size \$0.05 in this dataset. 
	Notably, these differences exhibit a nearly homoskedastic pattern with respect to strike prices. In other words, the error terms do not vary much with the price level of the options and strike prices, contrary to what is often assumed in the literature. 
	It suggests that the primary source of observation errors for these highly liquid SPX options can be attributed to the minimum tick size. 

	Panel (b) in Figure \ref{fig:interp_erors} displays the same pricing differences divided by the half-spread, which is defined for each option contract with strike price $x$ as 
	\[ HS(x) = \frac{AskO(x) - Bid O(x)}{2}, \]
	where $AskO(x)$ and $Bid O(x)$ are the ask and bid prices for the OTM option with strike price~$x$, respectively. The half-spread is calculated based on put option quotes if $x < F$ and on call options otherwise. Since we use mid-quote prices $O(x) = \frac{AskO(x) + BidO(x)}{2}$ for our estimation, these pricing errors indicate the percentage distance between the mid-quote price and the bid (if negative) or ask (if positive) prices. As shown in Figure \ref{fig:interp_erors}(b), most pricing errors fall within half of the bid-ask spread, indicating a very good pricing performance of the method.

\begin{figure}[h!]
        \centering
		\caption{Option-implied RND and call delta for SPX options}

		\begin{subfigure}{0.45\textwidth}
			\hspace*{-0.5cm}
			\includegraphics[scale=0.12]{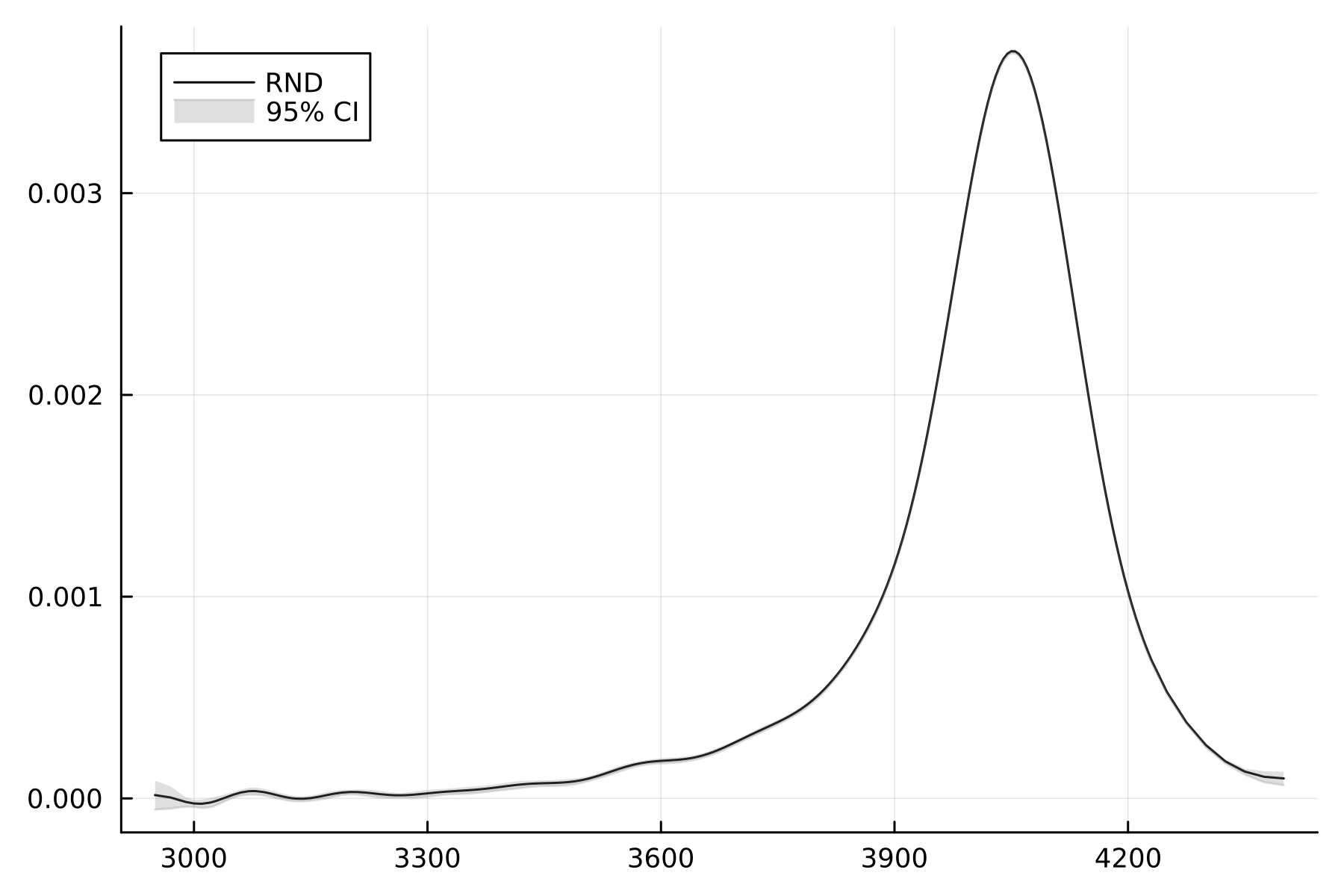}
            \caption{RND}
		\end{subfigure}
		\begin{subfigure}{0.45\textwidth}
			\includegraphics[scale=0.12]{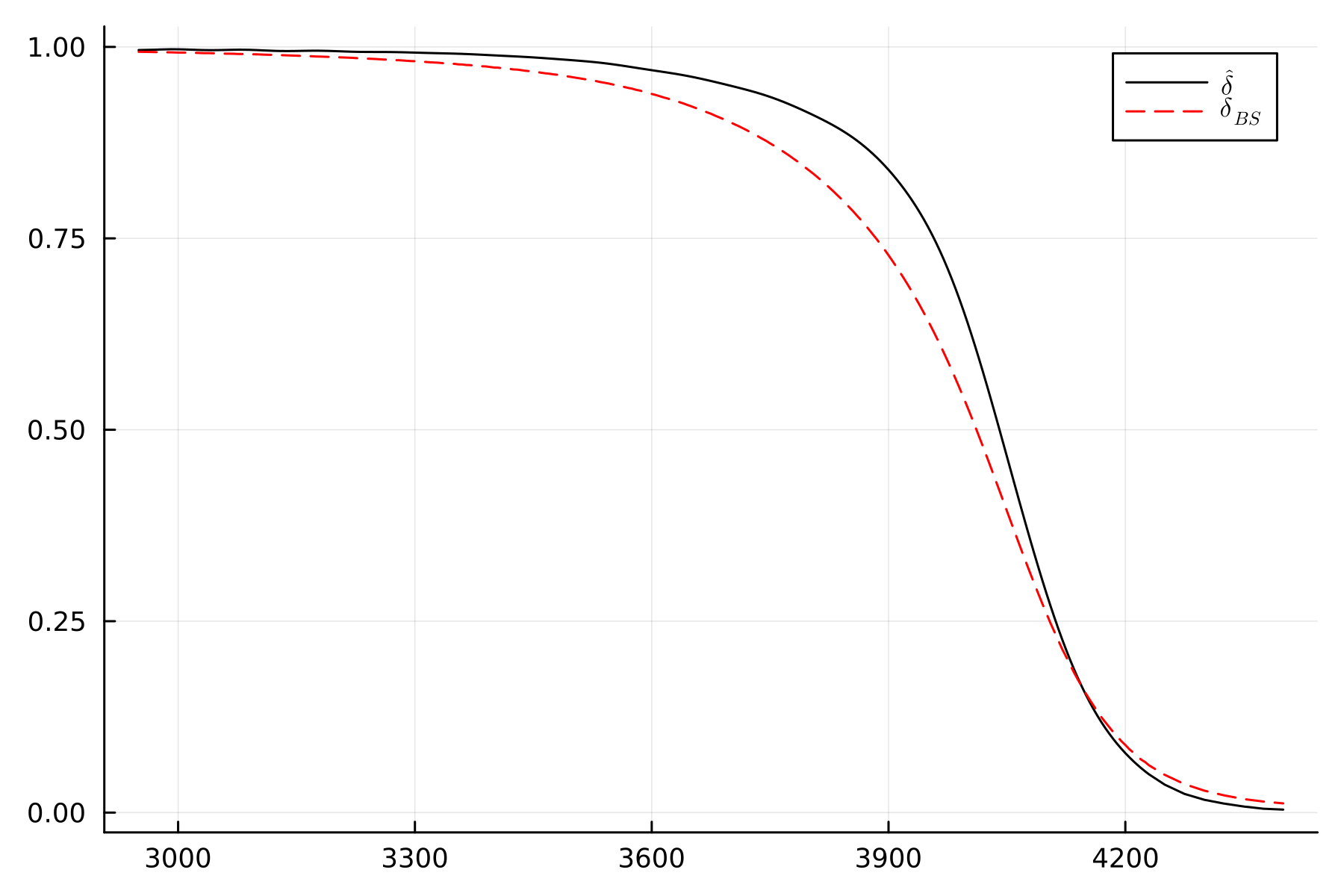}
            \caption{call delta}
		\end{subfigure}

        \label{fig: spx iRND and iDelta }

		\medskip
		\begin{minipage}{\textwidth}\scriptsize
            Note: This figure plots the estimated RND (Panel (a)) with the 95\% confidence interval and call deltas (Panel (b)) for the SPX options with 29 days-to-maturity traded on April 1, 2021. Panel (b) also depicts the Black-Scholes call deltas,~$\delta_{BS}$.
		\end{minipage}
    \end{figure}

	Finally, Figure \ref{fig: spx iRND and iDelta } plots the non-parametric estimates for the RND of future asset price and call deltas for the SPX options data. The RND estimate of a future price is obtained from the RND estimate of a log future price given by equation \eqref{icos-rnd-estimator}. Given the derived asymptotic theory (and the appropriate delta-rule), the confidence interval for the RND is displayed in a gray area. It is very narrow in the main part of the distribution and slightly widens towards the ends of the considered interval. We also note that we do not impose any arbitrage-free conditions on our RND and option price estimators. Thus, the RND estimates have negative values for some values of strikes. However, such minor arbitrage violations are unlikely to have any practical implications since all corresponding call price estimates fall within the minimum tick size and bid-ask spread. 
	
	The estimated call deltas $\widehat{\delta}$ are displayed alongside deltas based on the Black-Scholes model, $\delta_{BS}$. As observed, the Black-Scholes deltas can substantially underestimate the in-the-money call deltas. This might potentially result in hedging errors as discussed in \citeA{bates2005hedging} and \citeA{alexander2007model}.

	\subsection{AMZN options}

	In our second application, we examine equity options on Amazon with a very short time-to-maturity of $T = 1$ day. These options are traded on the Earning Announcement Day (EAD) of April 26, 2018, prior to the announcement itself. Compared to the SPX options, Amazon options are less liquid and are prone to larger observation errors due to their very short maturity. Moreover, the EAD introduces extra uncertainty about the stock price at the expiry.

	Similar to the SPX options, we use mid-quote prices of OTM contracts and filter out only zero-bid contracts. We concentrate our analysis on the interval $[\alpha, \beta] = [1250, 1760]$, which corresponds to approximately 18\% below and 16\% above the underlying spot price of \$1518.96 on this EAD. The availability of such a wide interval for short-dated options is attributed to the information uncertainty surrounding the EAD. Based on the rule-of-thumb for the optimal number of expansion terms, we set $N = 13$.

	\begin{figure}[ht]
		\centering
		\caption{AMZN option price estimates based on the iCOS method}
		\begin{subfigure}{0.45\textwidth}
			\hspace*{-0.5cm}
			\includegraphics[scale=0.12]{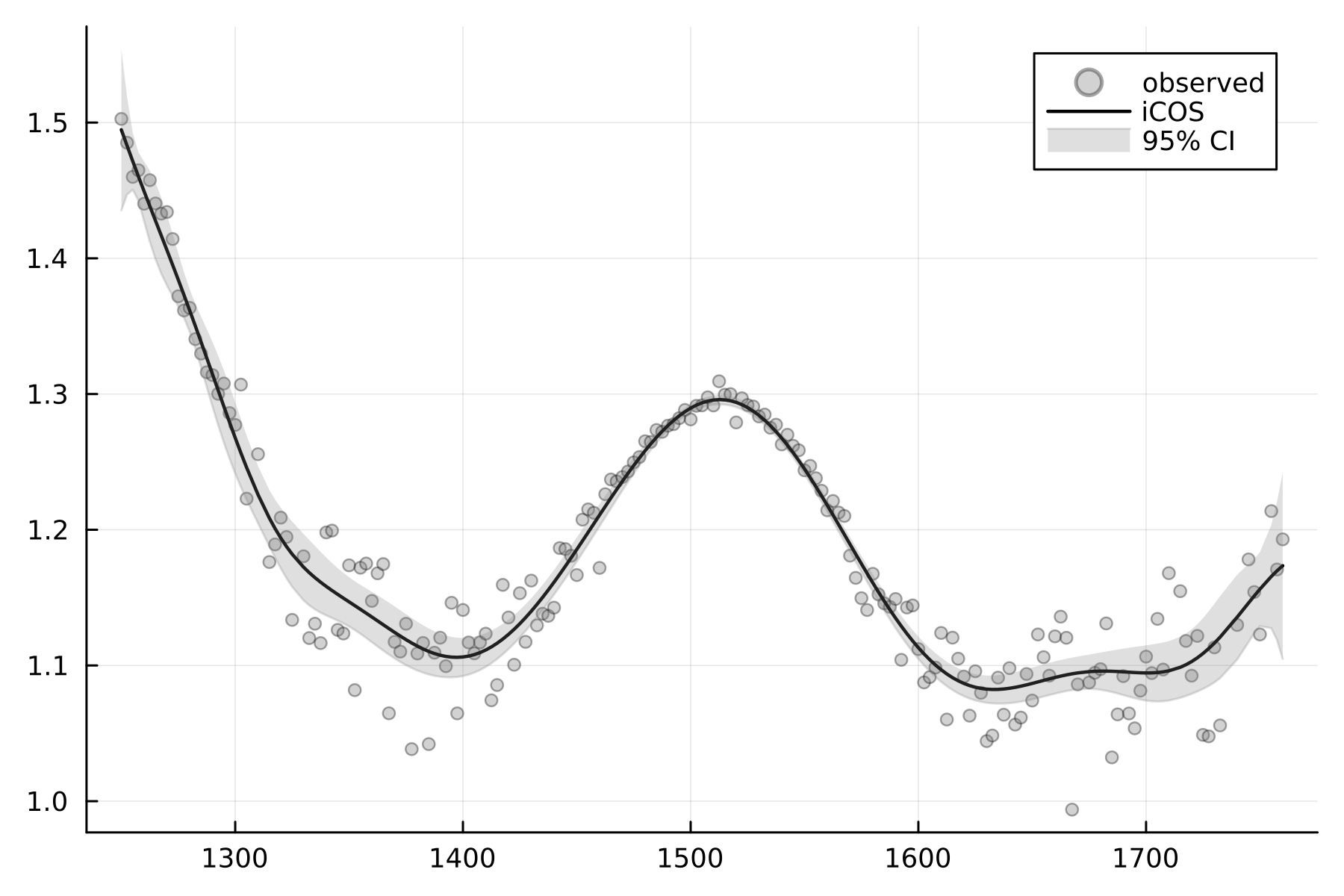}
            \caption{BSIV}
		\end{subfigure}
		\begin{subfigure}{0.45\textwidth}
			\includegraphics[scale=0.12]{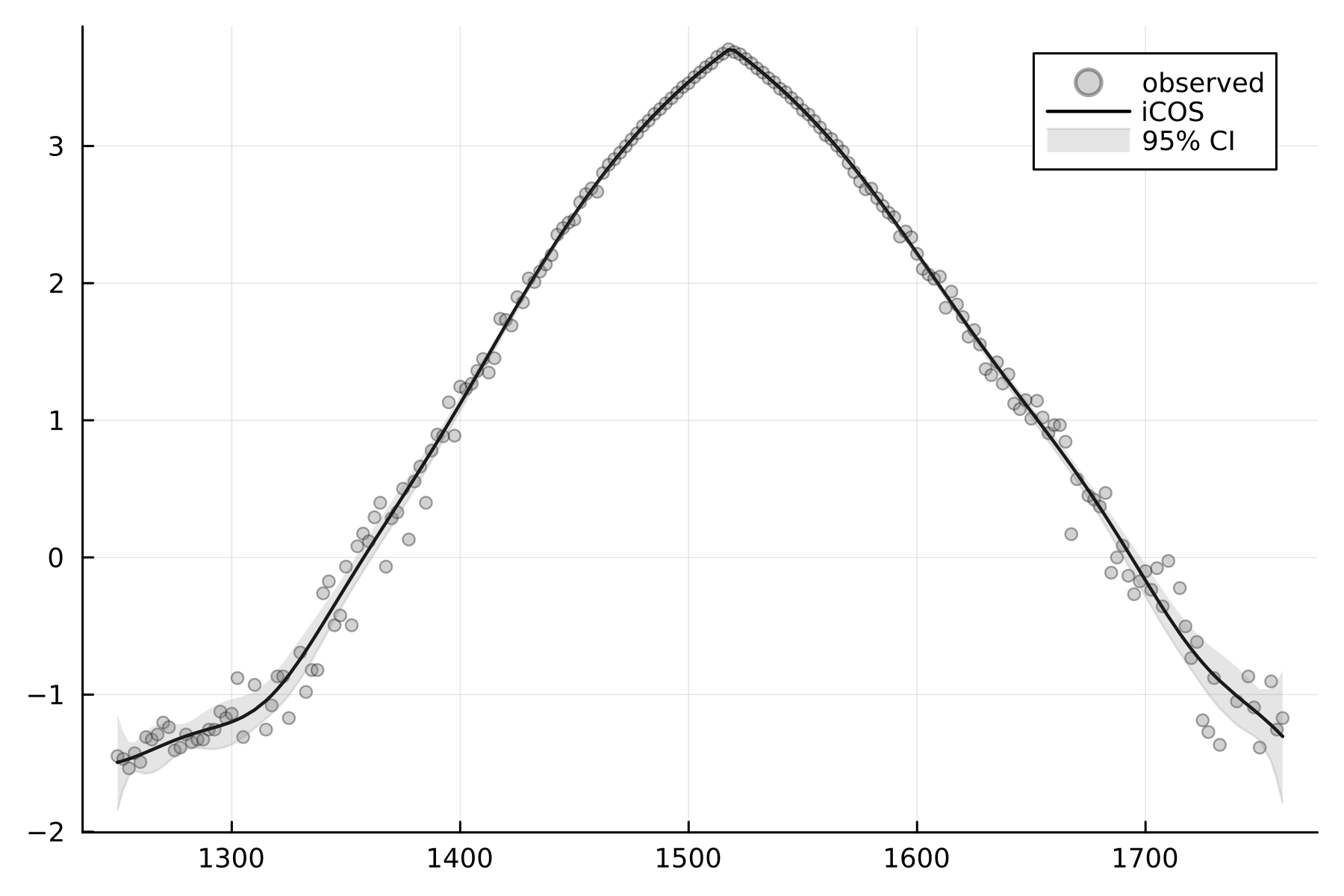}
            \caption{log OTM}
		\end{subfigure}
		\label{fig:amzn_interp}

		\medskip
		\begin{minipage}{\textwidth}\scriptsize
            Note: This figure illustrates the call price estimation of the proposed method based on the AMZN options traded on April 26, 2018 with 1 day-to-maturity. Panel (a) displays the result of the estimation along with the observed option prices on BSIV space, while Panel (b) displays it on log OTM space.
		\end{minipage}
	\end{figure}

	Figure \ref{fig:amzn_interp} presents the estimation result for the option-implied call prices displayed on BSIV and log OTM spaces. Notably, the BSIVs are exceptionally high, reaching approximately 130\% for ATM options expiring in just one day. Furthermore, these BSIVs exhibit a distinctive W-shaped pattern, which is atypical for implied volatility curves\footnote{Note that most parametric curves amd models commonly used in the literature would fail to capture this pattern, leading to large estimated errors.}. 
	The W-shape arises from the anticipation of a significant stock price jump following the earnings announcement release. \citeA{alexiou2021pricing} document frequent concave patterns in implied volatilities prior to the EAD for equity options.

	Additionally, we observe a large dispersion of option prices. However, our estimation procedure effectively smoothes out the noisy data, resulting in accurate price estimates. For this example, in Figure \ref{fig:amzn_interp} we also display the 95\% confidence interval around the estimated call prices, obtained by applying the appropriate delta rules for the derived asymptotic results. We emphasize that this confidence interval reflects the uncertainty around the estimates and not the observation errors in option prices. Therefore, it does not and need not cover the observed prices.

	\begin{figure}[ht]
		\centering
		\caption{Pricing errors for AMZN call estimates}
		\begin{subfigure}{0.45\textwidth}
			\hspace*{-0.5cm}
			\includegraphics[scale=0.12]{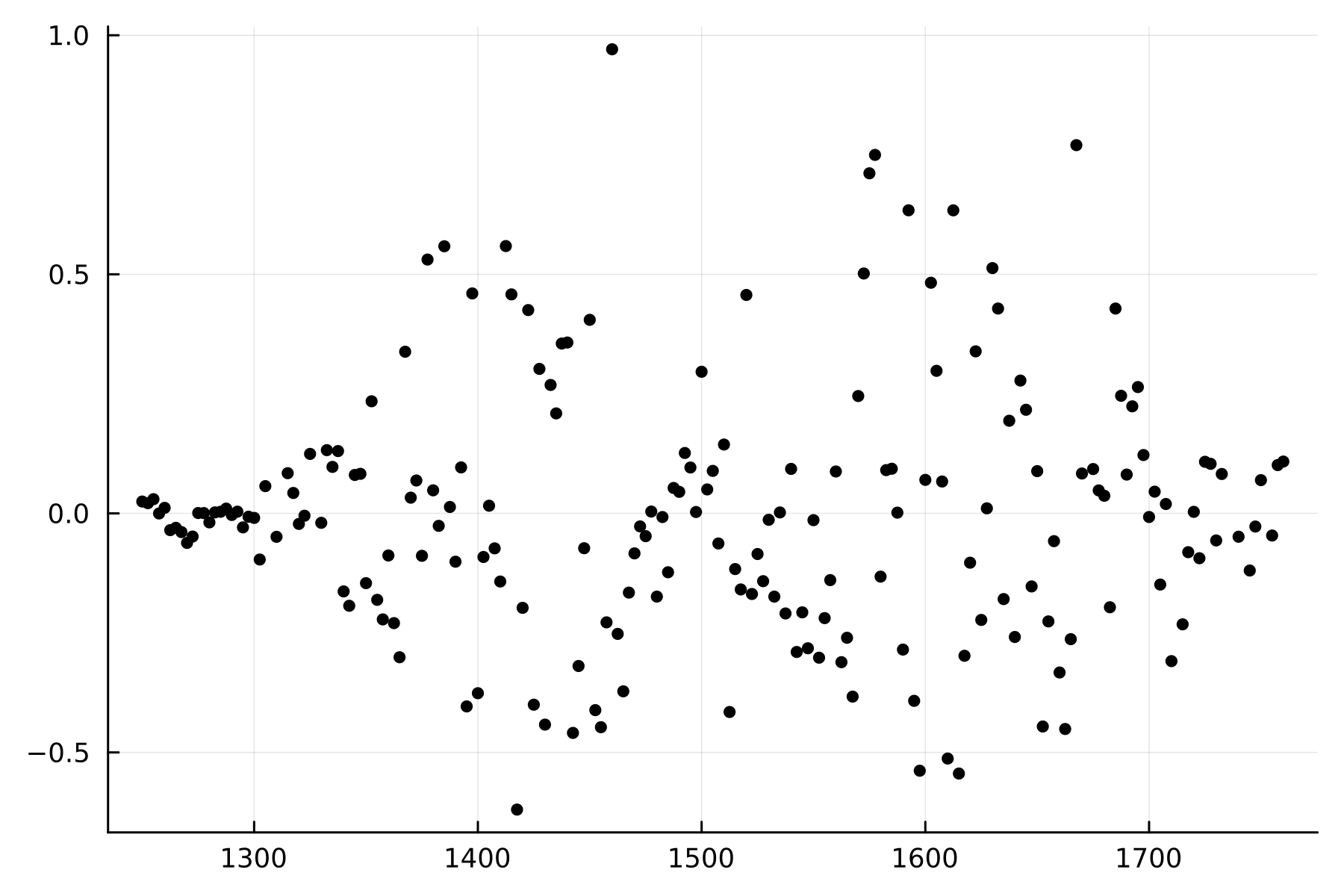}
            \caption{$\widehat{C}(x) - C(x)$}
		\end{subfigure}
		\begin{subfigure}{0.45\textwidth}
			\includegraphics[scale=0.12]{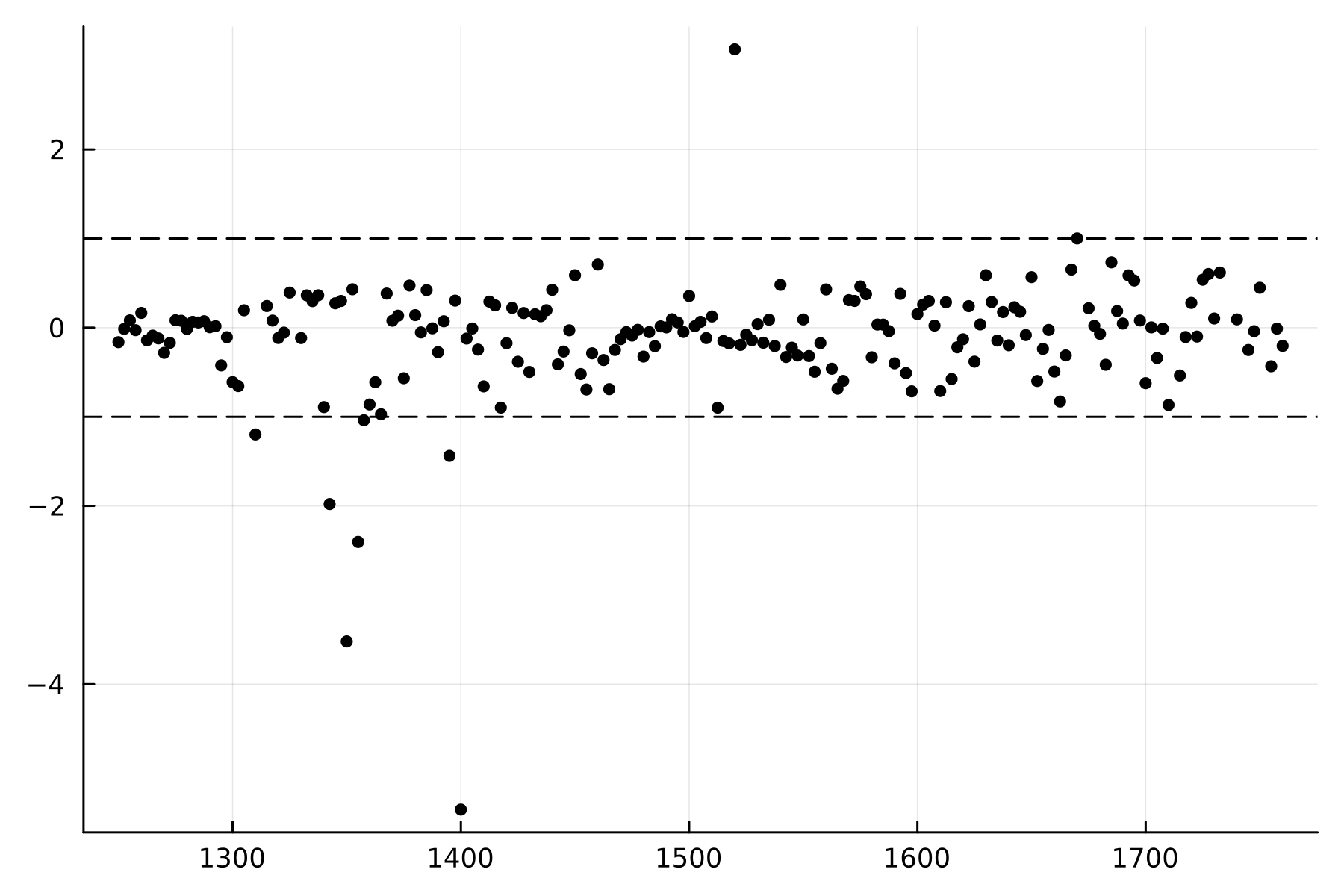}
            \caption{$\frac{\widehat{C}(x) - C(x)}{HS(x)}$}
		\end{subfigure}
		\label{fig:amzn_interp_erors}

		\medskip
		\begin{minipage}{\textwidth}\scriptsize
            Note: This figure illustrates the pricing errors of the proposed method based on the AMZN options traded on April 26, 2018 with 1 day-to-maturity. Panel (a) plots the pricing errors $\widehat{C} - C$. Panel (b) plots  pricing errors relative to the half-spread~$\frac{\widehat{C} - C}{HS}$.
		\end{minipage}
	\end{figure}

	Figure \ref{fig:amzn_interp_erors} displays the pricing errors for call price estimates. As with the SPX options, we plot the pricing errors and the errors relative to the half-spread. Consistent with Figure \ref{fig:amzn_interp}, the pricing errors are larger than those for the SPX options but are still centered around zero.

	\begin{figure}[ht]
		\centering
		\caption{Option-implied RND and call delta for AMZN options }
		\begin{subfigure}{0.45\textwidth}
			\hspace*{-0.5cm}
			\includegraphics[scale=0.12]{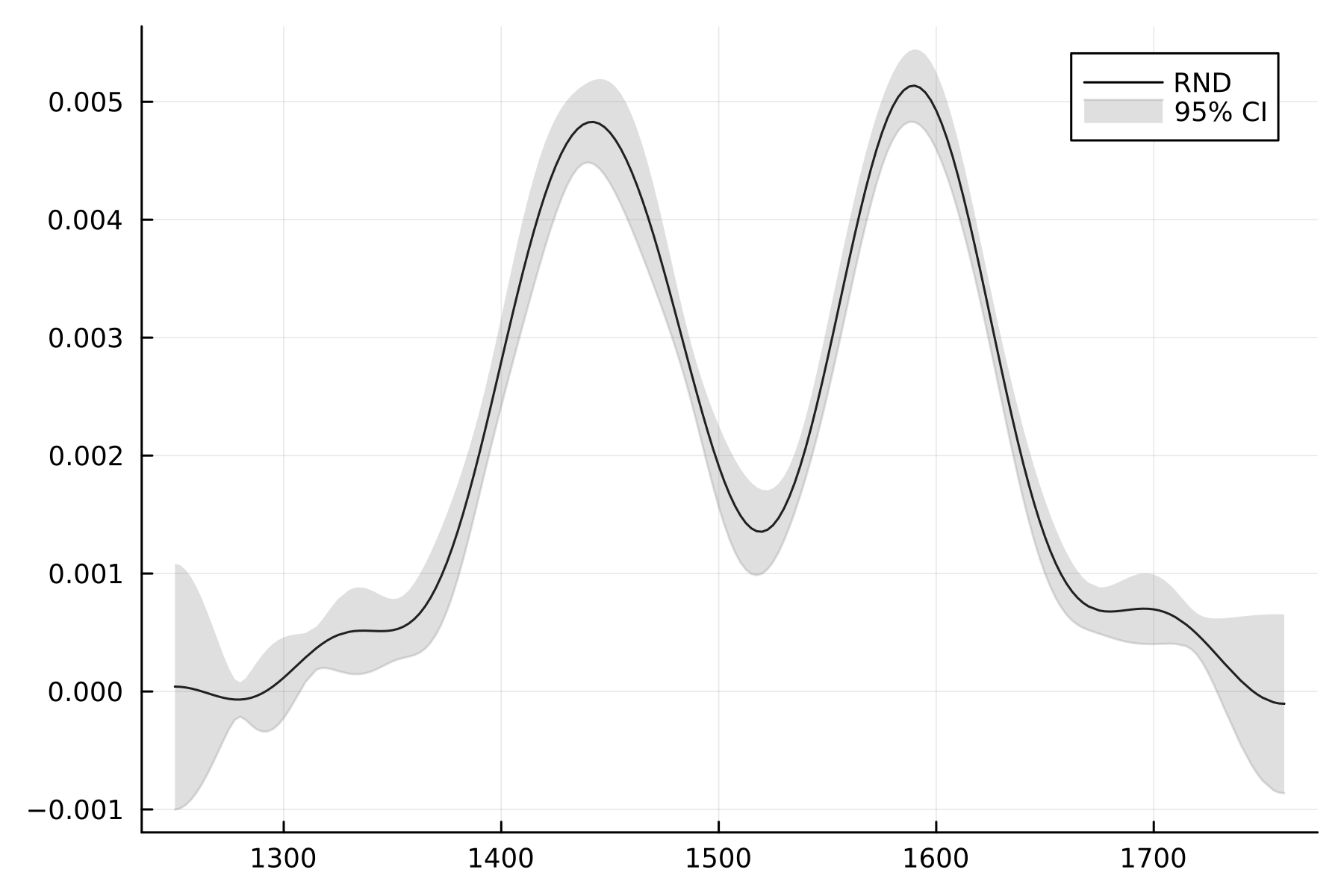}
            \caption{RND}
		\end{subfigure}
		\begin{subfigure}{0.45\textwidth}
			\includegraphics[scale=0.12]{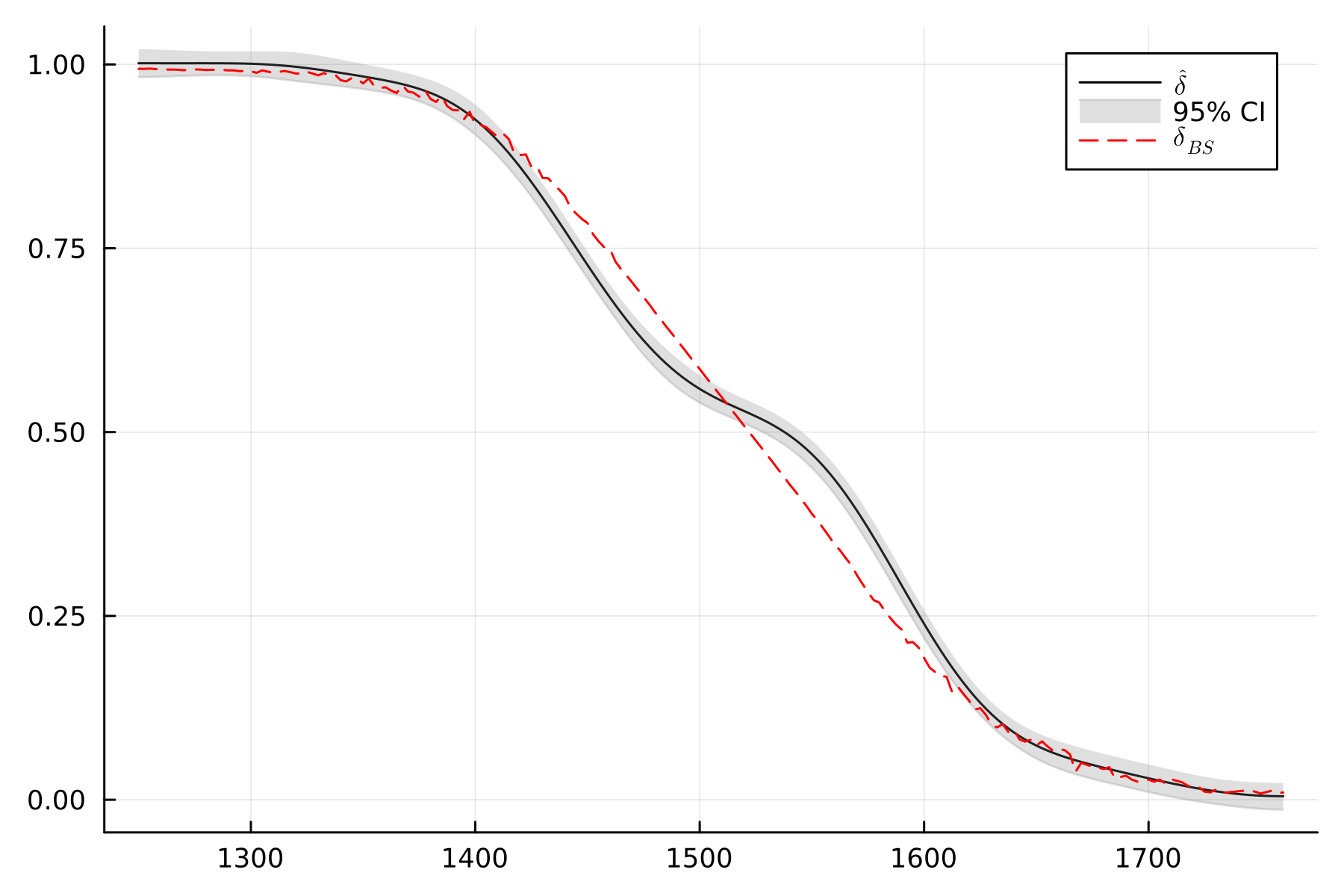}
            \caption{call delta}
		\end{subfigure}
		\label{fig:amzn_rnd_delta}

		\medskip
		\begin{minipage}{\textwidth}\scriptsize
            Note: This figure illustrates the estimated RND (Panel (a)) and call deltas (Panel (b)) with 95\% confidence intervals for the 1 day-to-maturity AMZN options traded on April~26, 2018.
		\end{minipage}
	\end{figure}

	Finally, Figure \ref{fig:amzn_rnd_delta} displays the estimated RND and deltas for Amazon options. The consequence of the W-shaped implied volatility curve is a bimodal RND, reflecting the market's anticipation of two possible outcomes. The two modes of the estimated RND are at \$1442 and \$1590, which corresponds to around 5\% down and 4.7\% up from the spot underlying stock price, respectively. After the announcement, the next day's opening price for Amazon was \$1634 and it closed at \$1574 (3.62\% up from the spot price). 
	The bimodality of the RND is also reflected in the estimated deltas. As shown in Figure~\ref{fig:amzn_rnd_delta}(b), the Black-Scholes deltas overestimate the deltas for the strikes around the first mode and underestimate them for strike prices close to the second mode.

	\subsection{Errors in VIX}

	The developed methodology allows us to analyse errors embedded in the VIX index. As noted by \citeA{jiang2007extracting}, the construction of the VIX is prone to several types of approximation errors, including truncation and discretization errors. The former arises from truncating the real line to the range of observed strike prices, and the latter is due to the discreteness of strike prices. On top of that, option prices used in the VIX construction are subject to observation errors since the true prices are not observed perfectly, as argued in Section \ref{sec:estimation-observation scheme}. This results in observation errors in the VIX index. 
	Our methodology enables us to estimate and disentangle observation and discretization errors in the VIX. 
	
	In particular, the CBOE calculates the VIX index as\footnote{For simplicity of notation, the exposition is based on a single maturity of 30 days. The CBOE averages (in total variances) the two VIX measures constructed using the near-term and the next-term options. In our empirical application, we follow the same procedure.}
	\begin{align}\label{eq: vix}
		\mbox{VIX} = 100 \cdot \sqrt{\frac{2}{T} e^{rT} \sum_{i=1}^n \frac{\Delta K_i}{K_i^2}  O(K_i) - \frac{1}{T} \left(\frac{F}{K_0} -1\right)^2 },
	\end{align}
	where $K_0$ is the largest strike price below the forward level $F$, $\Delta K_i = \frac{1}{2}(K_{i+1} - K_{i-1})$ for $i=2,\dots,n{-}1$, and $T=30$ days. For more details, see the \citeA{CBOE} white paper. Since the OTM option prices are observed with noise, the VIX itself contains measurement error. Given the consistent estimator of option prices $\widehat{O}(K_i)$, we can (re-)construct the VIX using these estimates and obtain estimates of the observation errors in the VIX. That is, we calculate
	\begin{align}\label{eq: ivix}
		\widehat{\mbox{VIX}} = 100 \cdot \sqrt{\frac{2}{T} e^{rT} \sum_{i=1}^n \frac{\Delta K_i}{K_i^2}  \widehat{O}(K_i) - \frac{1}{T} \left(\frac{F}{K_0} -1\right)^2 }
	\end{align}
	and define $\widehat{\xi}_{\mbox{vix}} := \mbox{VIX} - \widehat{\mbox{VIX}}$ as the estimator of the observation error in the VIX index.

	On the other side, the VIX is developed to approximate the model-free implied volatility. Since our method allows us to further consistently interpolate between observed strike prices, we can construct the measure of the model-free corridor implied volatility (CIV) as 
	\begin{align}\label{eq: cvs}
		\widehat{\mbox{CIV}} = 100 \cdot \sqrt{ \frac{2}{T} e^{rT} \int_{\alpha}^{\beta} \frac{1}{x^2} \widehat{O}(x) \diff x }.
	\end{align}
	The VIX can be seen as a measure of CIV with barriers fixed at the lowest and highest strike prices that the CBOE uses for calculating the index (\citeNP{andersen2015exploring}).
	Therefore, we define $\widehat{\zeta}_{\mbox{vix}} := \widehat{\mbox{VIX}} - \widehat{\mbox{CIV}}$ as the estimator of the discretization error in the VIX index.

	\begin{figure}[ht]
		\centering
		\caption{Errors in the VIX }
		\begin{subfigure}{0.45\textwidth}
			\hspace*{-0.5cm}
			\includegraphics[scale=0.12]{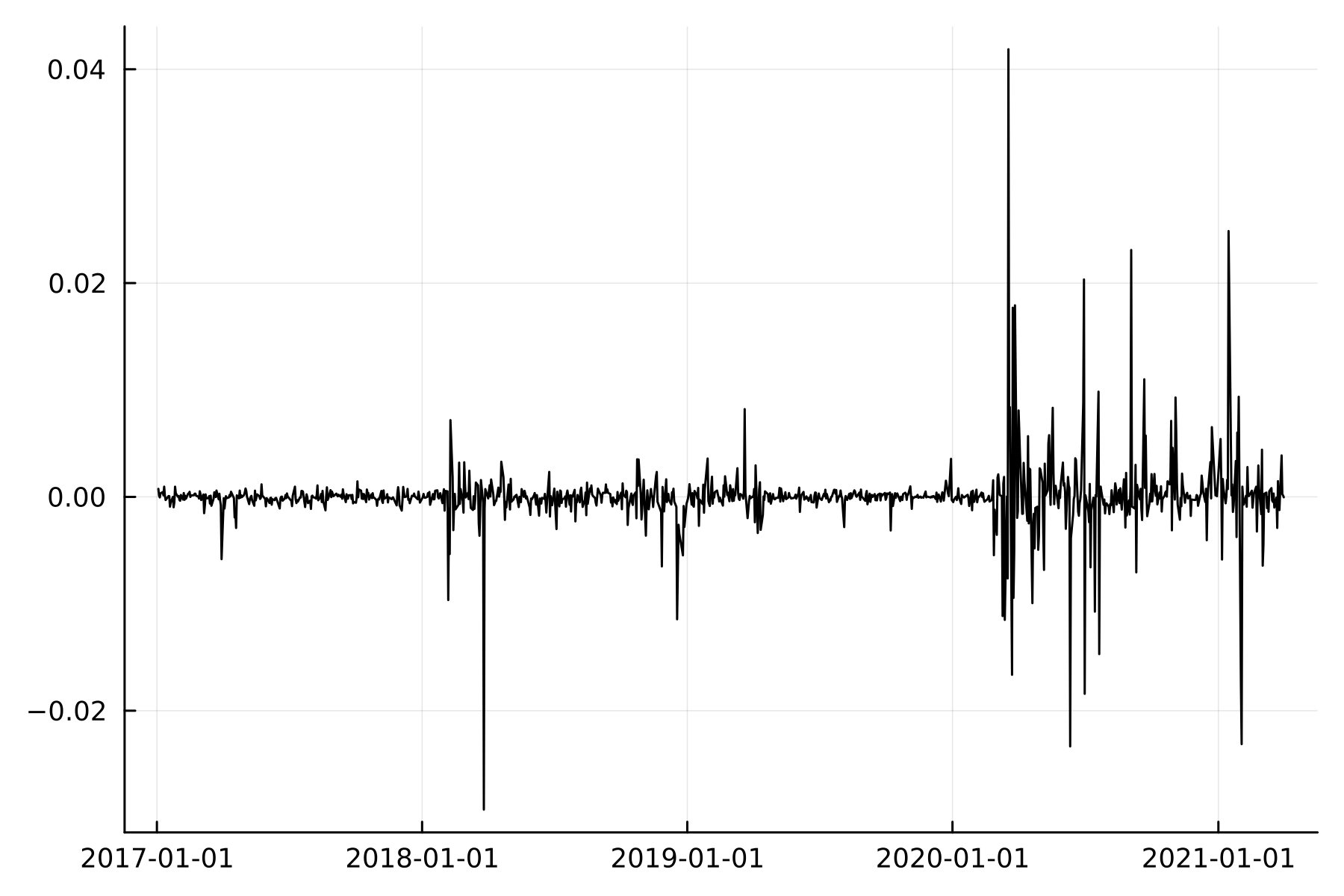}
            \caption{observation errors $\widehat{\xi}_{\mbox{vix}}$}
		\end{subfigure}
		\begin{subfigure}{0.45\textwidth}
			\includegraphics[scale=0.12]{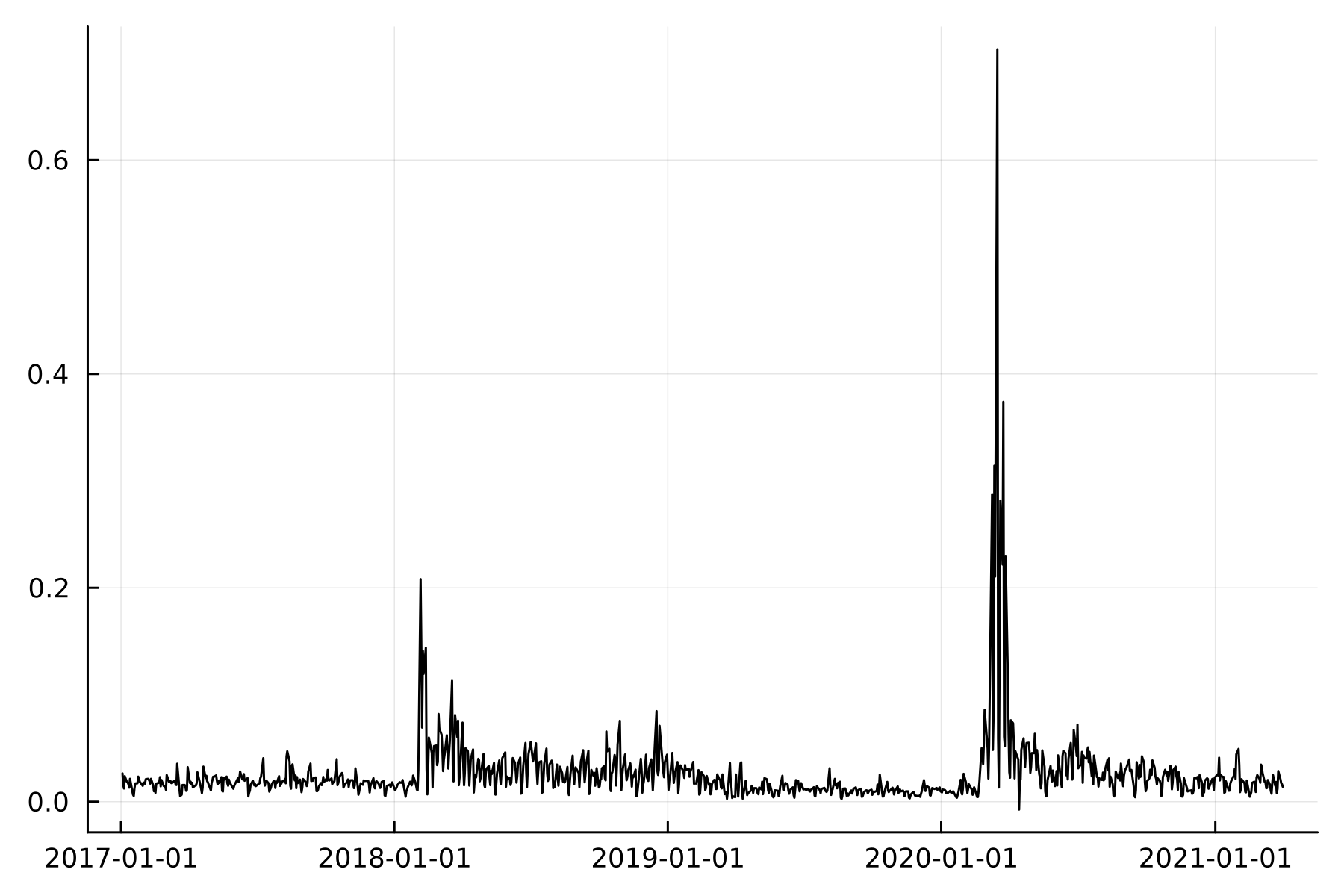}
            \caption{discretization errors $\widehat{\zeta}_{\mbox{vix}}$}
		\end{subfigure}
		\label{fig: errors in vix}

		\medskip
		\begin{minipage}{\textwidth}\scriptsize
            Note: This figure plots the time series of the estimated observation (Panel a) and discretization (Panel b) errors in the VIX index.
		\end{minipage}
	\end{figure}

	To estimate the observation and discretization errors in the VIX, we consider the SPX options obtained from the CBOE from January 3, 2017 until April 1, 2021. We follow the exact same procedure for the construction of the index as outlined in their white paper (\citeNP{CBOE}). To reduce the finite-sample bias in the iCOS procedure due to the discreteness of the observed option strikes, for each tenor, we interpolate option prices using cubic splines applied to implied volatilities. This can be seen as a bias-reduction procedure as motivated in \citeA{blv} in the context of option-implied CCFs.

	\begin{figure}[ht]
		\centering
		\caption{Histograms of percentage errors in the VIX }
		\begin{subfigure}{0.45\textwidth}
			\hspace*{-0.5cm}
			\includegraphics[scale=0.12]{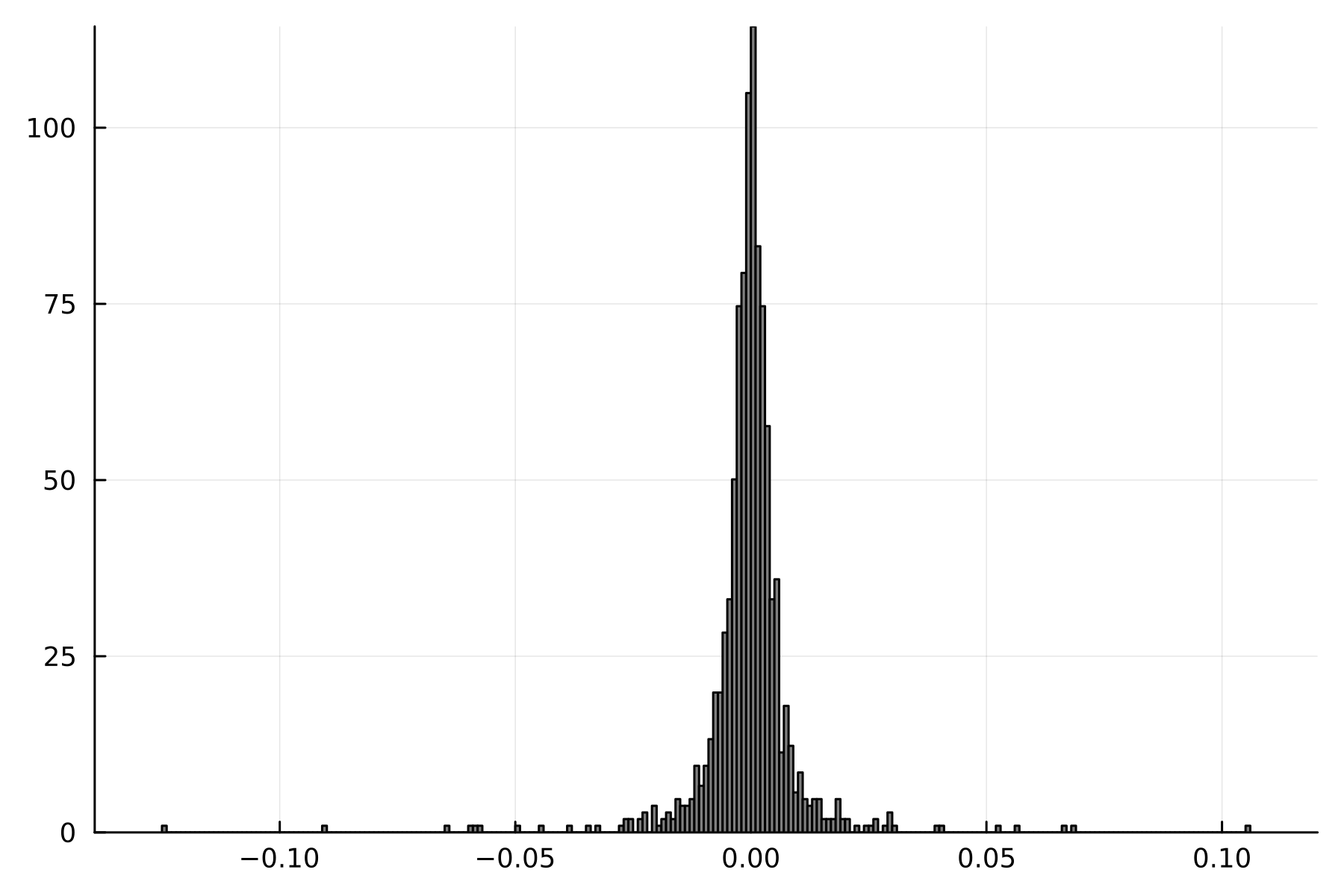}
            \caption{percentage observation errors}
		\end{subfigure}
		\begin{subfigure}{0.45\textwidth}
			\includegraphics[scale=0.12]{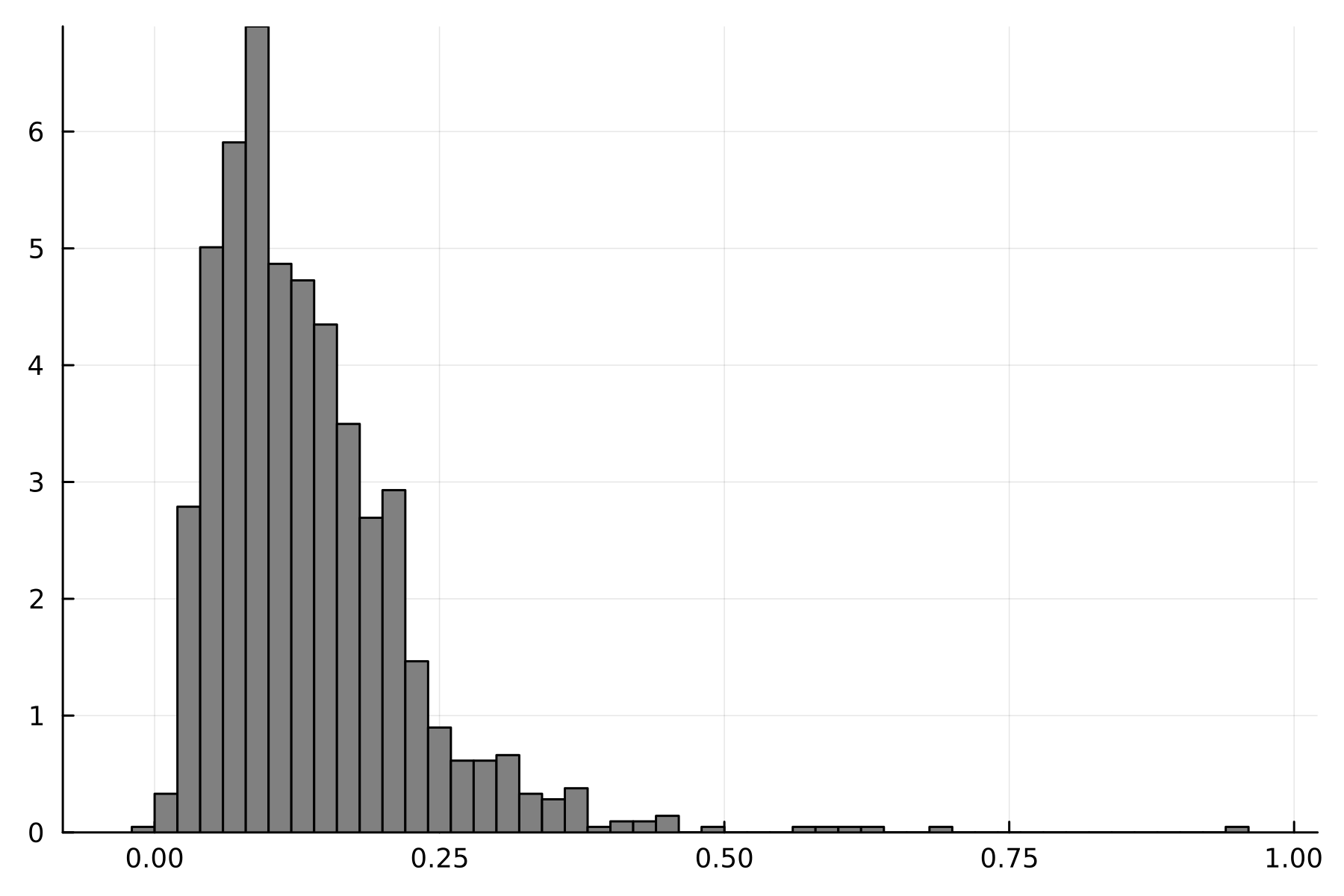}
            \caption{percentage discretization errors}
		\end{subfigure}
		\label{fig: hist errors in vix}

		\medskip
		\begin{minipage}{\textwidth}\scriptsize
            Note: This figure plots the normalized histograms of the percentage estimated errors in the VIX index. The percentage observation errors (Panel a) are defined as $100 \cdot \widehat{\xi}_{\mbox{vix}} / \mbox{VIX}$ and the percentage discretization errors (Panel b) are given by $100 \cdot \widehat{\zeta}_{\mbox{vix}} / \widehat{\mbox{VIX}}$.
		\end{minipage}
	\end{figure}

	Figure \ref{fig: errors in vix} displays the time series plots of the estimated observation and discretization errors in the VIX over the period of more than four years. Figure \ref{fig: hist errors in vix} complements it with histograms of the percentage errors over the same time period. We notice that the observation errors are centered around zero, while the discretization errors are mainly positive. This is expected since observation errors in option prices do not introduce biases in the VIX, while the discreteness of the strikes leads to a finite-sample bias in the constructed index. In fact, the sample average of the percentage observation errors is nearly zero, and the average of the percentage discretization errors is estimated at around 0.135\%. Furthermore, the magnitude of observation errors is rather low, reaching in absolute terms up to 0.04 percentage points. The discretization errors, on the other hand, can result in a substantial overestimation of the index, with the differences up to 0.7 percentage points during high volatility periods. 
	\section{Conclusion}
\label{sec:conlcusion}

In this paper, we proposed a non-parametric estimation procedure for option prices, RND, and option sensitivities. This method is based on the combination of Fourier-based cosine technique and the option spanning result of \citeA{carr2001optimal}. This combination allows for a flexible and accurate estimation of the density, option prices and option sensitivities without imposing parametric assumptions on the dynamics of underlying asset and on the shape of implied volatility surface. We have also established the asymptotic properties of the proposed estimators and demonstrated the finite sample properties through the Monte Carlo simulations. 

The usage of the proposed method is illustrated in empirical applications using options data on the S\&P 500 stock market index and Amazon equity options on the Earning Announcement Day. 
The empirical analysis demonstrates the effectiveness of the iCOS method in accurately estimating option prices and capturing important market features in different market conditions.
Additionally, we demonstrated the usefulness of our methodology to dissect and quantify errors in the VIX index, one of most popular measure of market volatility.
We found that observation errors in the VIX are centered around zero and have a small magnitude, while discretization errors can lead to positive and substantial biases in the VIX index.



    \clearpage
    \begin{appendices}
        \numberwithin{equation}{section}
		\numberwithin{assumption}{section}
		\numberwithin{figure}{section}
		\numberwithin{table}{section}

        \section{Proofs}
\label{appendix:proofs}

    \textit{Proof of Proposition \ref{prop:Dm}:}

    Given the option observation scheme outlined in Assumptions \ref{assumption:fixed_grid} and \ref{assumption:observation_errors}, the total measurement error in the option-implied cosine coefficients $\widehat{D}_m$ can be decomposed as follows:
    \begin{align*}
        \widehat{D}_m - D_m &= \sum_{i=1}^n w_i \psi_m(K_i)O(K_i) \Delta_n - \int_\alpha^\beta \psi_m(K) O_0(K) \diff K\\
        &= \underbrace{\sum_{i=1}^n w_i \psi_m(K_i) \varepsilon_i \Delta_n }_{=:\xi^D_{m,n}} 
        + \underbrace{\sum_{i=1}^n w_i \psi_m(K_i)O_0(K_i) \Delta_n - \int_\alpha^\beta \psi_m(K) O_0(K) \diff K}_{=:\zeta^D_{m,n} },
    \end{align*}
    where the error term $\xi^D_{m,n}$ represents the observation error due to the noisy observation of option prices, and $\zeta^D_{m,n}$ is the discretization error resulting from the numerical approximation of the integral.

    Under Assumptions \ref{assumption:fixed_grid} and \ref{assumption:observation_errors}, the observation error has zero mean $\E[\xi^D_{m,n}] = 0$ and $\xi^D_{m,n} = \mathcal{O}_p(n^{-1/2}) $ since
    \begin{align*}
        \E \left[|\xi^D_{m,n}|^2  \right] 
        & \leq \sum_{i=1}^n w_i^2 \psi_m^2(K_i) \E \left[\varepsilon_i^2 \right] \Delta_n^2 \\
        & = \sum_{i=1}^n w_i^2 \psi_m^2(K_i) \sigma_i^2 \Delta_n^2 \\ 
        & \leq \mathcal{C} n^{-1},
    \end{align*}
    where $\mathcal{C}$ is some constant. Furthermore, this implies convergence to zero in probability, i.e.,\ $\xi^D_{m,n} = o_p(1)$.

    To invoke the Lyapunov CLT for non-identical but independent random variables, we first verify the Lyapounov's condition. For that, we note that the ratio 
    \begin{align*}
        \frac{\sum_{i=1}^n \E\left[ \rvert w_i \psi_m(K_i) \varepsilon_i  \rvert^{2+\omega}\right]}{\left(\sum_{i=1}^n w_i^2 \psi_m^2(K_i) \sigma_i^2\right)^{1 + \frac{\omega}{2}}} 
        &= 
        \frac{\sum_{i=1}^n \E\left[ \rvert w_i \psi_m(K_i) \varepsilon_i  \rvert^{2+\omega}\right] \Delta_n}{\left(\sum_{i=1}^n w_i^2 \psi_m^2(K_i) \sigma_i^2 \Delta_n \right)^{1 + \frac{\omega}{2}} \left( \Delta_n\right)^{-\frac{\omega}{2}}} \\
        &\leq 
        \left( \Delta_n \right)^{\frac{\omega}{2}}
        \frac{\sum_{i=1}^n \rvert w_i \psi_m(K_i) \rvert^{2+\omega} \E\left[ \rvert  \varepsilon_i  \rvert^{2+\omega}\right] \Delta_n}{\left(\sum_{i=1}^n w_i^2 \psi_m^2(K_i) \sigma_i^2 \Delta_n \right)^{1 + \frac{\omega}{2}} }\\
        &\leq \mathcal{C}n^{-\frac{\omega}{2}},
    \end{align*}
    for some $\omega>0$ and another constant $\mathcal{C}$. The last inequality follows since the summations in the numerator and denominator converge to definite integrals as $n\to \infty$, and the smallest and largest strike prices (which define the integration range) are fixed by Assumption~\ref{assumption:fixed_grid}. Hence, the Lyapunov's condition is satisfied and we can use the Lyapunov CLT, which yields as $n \to \infty$
    \begin{align*}
        \frac{\sum_{i=1}^n w_i \psi_m(K_i) \varepsilon_i}{\sqrt{\sum_{i=1}^n w_i^2 \psi_m^2(K_i) \sigma_i^2 }} 
        =
        \frac{\xi^D_{m,n}}{\sqrt{\sum_{i=1}^n w_i^2 \psi_m^2(K_i) \sigma_i^2 \Delta_n^2 }} \xrightarrow[]{d}  \mathcal{N}(0, 1).
    \end{align*}

    Finally, the discretization error $\zeta^D_{m,n}$ does not depend on the stochastic observation errors in option prices but introduces bias in the estimation of $\widehat{D}_m$:
    \begin{align*}
        \E\left[\widehat{D}_m - D_m \right] = \zeta^D_{m,n}.
    \end{align*}
    This bias, however, can be controlled by the choice of the numerical integration scheme and is of order $\mathcal{O}\left( \frac{m^{2+\iota}}{n^{\iota}} \right)$ for $\iota \geq 1$. The latter follows since $\psi_m^{(\iota)}(x) = \mathcal{O}(m^{2+\iota})$ and numerical integration error bounds. Therefore, for the fixed $m$ and $n \to \infty$, we obtain the limiting distribution for the option-implied cosine coefficients in the form:
    \begin{align*}
        \frac{\widehat{D}_m - D_m}{\sigma_{D}(m)}  \xrightarrow[]{d}  \mathcal{N}(0, 1)
    \end{align*}
    with $\sigma_D^2(m) := \sum_{i=1}^n w_i^2 \psi_m^2(K_i) \sigma_i^2 \Delta_n^2 $.

    \hfill $\square$

    \textit{Proof of Proposition \ref{prop:iCall_bar}:}

    We start by decomposing the difference between the option-implied call price estimator $\widehat{\overline{C}}(x)$ and its error-free counterpart $\overline{C}_0(x)$ for a fixed strike price $x \in[\alpha, \beta]$:
    \begin{align*}
        \widehat{\overline{C}}(x) - \overline{C}_0(x) &= \sideset{}{'}\sum_{m=0}^{N-1} \widehat{D}_m  H_m(x) - \sideset{}{'}\sum_{m=0}^{\infty} D_m  H_m(x)  \\
        &= 
        \sideset{}{'}\sum_{m=0}^{N-1} \sum_{i=1}^n w_i \psi_m(K_i) O(K_i) \Delta_n H_m(x) \\
        &
        -  \sideset{}{'}\sum_{m=0}^{N-1} \int_\alpha^\beta \psi_m(K)O_0(K) \diff K H_m(x)
        - \sum_{m=N}^{\infty} D_m  H_m(x) \\
        &= \sum_{i=1}^n w_i \underbrace{\sum_{m=1}^{N-1}  \psi_m(K_i) H_m(x) }_{=:\psi(x, K_i)}  O(K_i) \Delta_n \\ 
        &- \int_\alpha^\beta \underbrace{ \sum_{m=1}^{N-1} \psi_m(K) H_m(x) }_{=:\psi(x, K)} O_0(K) \diff K 
        \underbrace{ - \sum_{m=N}^{\infty} D_m  H_m(x) }_{=:\overline{\eta}(x)}\\
        &= \sum_{i=1}^n w_i \psi(x, K_i) O(K_i) \Delta_n - \int_\alpha^\beta \psi(x, K) O_0(K) \diff K + \overline{\eta}(x),
    \end{align*}
    where the difference between the first two terms involves the discretization and observation errors, and the third term is due to the truncation of the cosine series.
    Similar to the proof of Proposition \ref{prop:Dm}, we have the following decomposition for the difference between the first two terms:
    \begin{align*}
        &\sum_{i=1}^n w_i \psi(x, K_i) O(K_i) \Delta_n - \int_\alpha^\beta \psi(x, K) O_0(K) \diff K \\
        =
        &\underbrace{\sum_{i=1}^n w_i \psi(x, K_i) \varepsilon_i \Delta_n }_{=:\xi(x)} 
        + \underbrace{\sum_{i=1}^n w_i \psi(x, K_i)O_0(K_i) \Delta_n - \int_\alpha^\beta \psi(x, K) O_0(K) \diff K}_{=:\zeta(x) },
    \end{align*}
    where $\xi(x)$ represents the observation and $\zeta(x)$ is the discretization error in the call price estimator $\widehat{\overline{C}}(x)$. 
    
    To get the orders of these errors that depend both on the number of options $n$ and the number of expansion terms $N$, we first notice that $\psi_m(x) = \mathcal{O}(m^2)$ and $H_m(x) = \mathcal{O}(m^{-2})$. Hence, $\psi(x, K_i) = \sum_{m=1}^{N-1} \psi_m(K) H_m(x) = \mathcal{O}(N)$. Therefore, similar to $\xi^D_{m,n} = \mathcal{O}_p(n^{-1/2})$, the order of the observation error $\xi(x) = \mathcal{O}_p(N n^{-1/2})$.

    As for the discretization error, we can further express it as
    %
    %
    \begin{align*}
        \zeta(x) 
        &= \sum_{i=1}^n w_i \psi(x, K_i)O(K_i) \Delta_n - \int_\alpha^\beta \psi(x, K) O_0(K) \diff K\\ 
        &= \sum_{i=1}^n w_i \sum_{m=1}^{N-1}  \psi_m( K_i) H_m(x) O_0(K_i) \Delta_n - \int_\alpha^\beta \sum_{m=1}^{N-1}  \psi_m(K) H_m(x) O_0(K) \diff K\\ 
        &=  \sum_{m=1}^{N-1} \left( \sum_{i=1}^n w_i \psi_m(K_i)O_0(K_i) \Delta_n - \int_\alpha^\beta \psi_m(K) O_0(K) \diff K \right)  H_m(x) \\
        &= \sum_{m=1}^{N-1} \zeta^D_{m,n} H_m(x).
    \end{align*}
    That is, the discretization error in the call price estimator arises from the discretization errors in the option-implied cosine coefficients $\widehat{D}_m$ with  $m=1,\dots,N-1$. Therefore, the order of $\zeta(x)=  \mathcal{O}\left(\frac{N^{1+\iota}}{n^{\iota}} \right)$ is also determined by the chosen numerical integration scheme with $\iota \geq 1$.


    Finally, the order of the truncation error $\overline{\eta}(x)$ depends on the smoothness of the density function.
    From \citeA{fang2008novel}, Lemma 4.2 and Lemma 4.3, we have that $\overline{\eta}(x) = \mathcal{O}\left(e^{-N\tilde{\nu}}\right)$ with $\tilde{\nu} >0$ in the case of infinitely differentiable density functions and $\overline{\eta}(x) = \mathcal{O}\left(N^{1-p}\right) $ under Assumption \ref{assumption:smoothness}. 
    
    Given the orders of these errors, we see that the truncation error decreases with $N$, while the observation and discretization errors can also increase with $N$. However, if $N$ increases slower than $\sqrt{n}$, i.e., if $Nn^{-1/2} \to 0 $ as $N \to \infty$ and $n \to \infty$, the observation and discretization errors (asymptotically) converge to zero. In other words, under this condition, the option-implied call price estimator $\widehat{\overline{C}}(x)$ is asymptotically unbiased.

    Finally, similar arguments as in the proof of Proposition \ref{prop:Dm} allows us to apply the Lyapunov CLT to the observation errors $\xi(x)$ and obtain the following asymptotic distribution for $\widehat{\overline{C}}(x)$:
    \begin{align*}
        \frac{ \widehat{\overline{C}}(x) - \overline{C}_0(x)}{\overline{\sigma}_c(x)}  \xrightarrow[]{d}  \mathcal{N}\left(0, 1 \right)
    \end{align*}
    as $n \to \infty$ and $N \to \infty$ with $Nn^{-1/2} \to 0 $, where
    \begin{align*}
        \overline{\sigma}_c^2(x) := \sum_{i=1}^n w_i^2 \psi^2(x, K_i) \sigma_i^2 \Delta_n^2 
        = \sum_{i=1}^n w_i^2 \left( \sum_{m=1}^{N-1}  \psi_m(K_i) H_m(x) \right)^2 \sigma_i^2 \Delta_n^2.
    \end{align*}


    \hfill $\square$

    \begin{lemma}\label{lemma:ols}
        The OLS estimator $\widehat{\boldsymbol{\theta}}$ of the linear regression of $ C(K_i) - \widehat{\overline{C}}(K_i) - C(\beta)$ on $Z_{c}^N(K_i)$ and $Z_{p}^N(K_i)$ with an intercept is consistent, asymptotically unbiased, and normally distributed as $n\to \infty$ and $N \to \infty$ with $Nn^{-1/2} \to 0$. 
    \end{lemma}

    \textit{Proof:}
    For ease of notation, let us define a vector of dependent variables as
    \begin{align*}
        \boldsymbol{c}:= \left(C(K_1) - \widehat{\overline{C}}(K_1) - C(\beta),  \dots, C(K_n) - \widehat{\overline{C}}(K_n) - C(\beta)\right)',
    \end{align*}
    a $n\times 3$ matrix of regressors as
    \begin{align*}
        \boldsymbol{Z}:= \left(\boldsymbol{z}_1, \dots, \boldsymbol{z}_n\right)^\prime  
        \mbox{ with } \boldsymbol{z}_i := \left(1, Z_c^{N}(K_i), Z_p^{N}(K_i) \right)^\prime, \quad i=1,\dots,n,
    \end{align*}
    a parameter vector as $\boldsymbol{\theta}:= (\bar\theta, \theta_c, \theta_p)'$,
    and vectors of observation, discretization and series truncation errors:
    \begin{align*}
        \boldsymbol{\epsilon}:&= (\varepsilon_1, \dots, \varepsilon_n)', \\
        \boldsymbol{\xi}:&= (\xi_1, \dots, \xi_n)', \quad \xi_i:= \xi(K_i)= \sum_{j=1}^n w_j \psi_{ij} \varepsilon_j \Delta_n,\ \psi_{ij} := \psi(K_i, K_j), \\
        \boldsymbol{\zeta}:&= (\zeta_1, \dots, \zeta_n)', \quad \zeta_i:=\zeta(K_i)= \sum_{j=1}^n w_j \psi(K_i, K_j)O_0(K_j) \Delta_n - \int_\alpha^\beta \psi(K_i, K) O_0(K) \diff K, \\
        \boldsymbol{\eta}:&= (\eta_1, \dots, \eta_n)',\quad \eta_i:= \eta(K_i) 
        = -\overline{\eta}(K_i) + \left( \sum_{m=N}^{\infty} (-1)^m  H_m(K_i) \right) \theta_c - \left( \sum_{m=N}^{\infty} H_m(K_i) \right) \theta_p.
    \end{align*}

    Note that the series truncation error 
    \begin{align*}
        \eta(x)
        = -\overline{\eta}(x) + \left( \sum_{m=N}^{\infty} (-1)^m  H_m(x) \right) \theta_c - \left( \sum_{m=N}^{\infty} H_m(x) \right) \theta_p
        = e^{-rT} \sum_{m=N}^{\infty} A_m H_m(x).
    \end{align*}

    Let the intercept $\bar\theta$ capture the average of finite-sample biases due to the discretization and truncation errors, i.e., 
    \begin{align*}
        \bar \theta = \frac{1}{n} \sum_{i=1}^n \eta_i - \frac{1}{n} \sum_{i=1}^n \zeta_i.
    \end{align*}
    %
    We note that $\bar \theta $ converges to zero asymptotically, but we include it to reduce the finite-sample bias.
    Then the OLS regression of the form 
    \begin{align}\label{ols}
        \boldsymbol{c} = \boldsymbol{Z} \boldsymbol{\theta} + \boldsymbol{e}, \mbox{ with } \boldsymbol{e}:= \boldsymbol{\epsilon} - \varepsilon_n - \boldsymbol{\xi} - \boldsymbol{\zeta} + \boldsymbol{\eta} -  \bar \theta,
    \end{align}
    gives consistent and asymptotically unbiased estimator: 
    \begin{align*}
        \widehat{\boldsymbol{\theta}} = \left(\boldsymbol{Z}' \boldsymbol{Z}\right)^{-1}\boldsymbol{Z}'\boldsymbol{c} 
        = \boldsymbol{\theta} + \left(\boldsymbol{Z}' \boldsymbol{Z}\right)^{-1}\boldsymbol{Z}'(\boldsymbol{\epsilon} - \varepsilon_n  - \boldsymbol{\xi} - \boldsymbol{\zeta} + \boldsymbol{\eta} -  \bar \theta ).
    \end{align*}
    Indeed, the consistency follows since each of the error term (asymptotically) converges to zero under the joint asymptotic scheme, i.e. as $n\to \infty$ and $N \to \infty$ with $Nn^{-1/2} \to 0$. In fact, for the observation errors $\boldsymbol{\xi}$ we have that $ \frac{1}{n} \boldsymbol{Z}' \boldsymbol{\xi} = \mathcal{O}_p(Nn^{-1/2})$ due to the same order of each component~$\xi_i$. Furthermore, we have that
    \begin{align*}
        \frac{1}{n} \boldsymbol{Z}' \boldsymbol{\xi} 
        &= \frac{1}{n} \sum_{i=1}^n \boldsymbol{z}_i \xi_i
        = \frac{1}{n} \sum_{i=1}^n \boldsymbol{z}_i \left(\sum_{j=1}^n w_j \psi_{ij} \varepsilon_j \Delta_n \right) \\
        &=  \frac{1}{n} \sum_{j=1}^n \sum_{i=1}^n \boldsymbol{z}_i w_j \psi_{ij} \varepsilon_j \Delta_n 
        = \frac{1}{n} \sum_{j=1}^nw_j \left( \sum_{i=1}^n \boldsymbol{z}_i  \psi_{ij} \Delta_n \right) \varepsilon_j  \xrightarrow[]{a.s.} 0.
    \end{align*}
    Also, since $\zeta_i = \mathcal{O}\left( \frac{N^{1+\iota}}{n^{\iota}} \right)$ with $\iota \geq 1$, which is controlled by the choice of the numerical integration (see Table \ref{tab:integrations}), we have
    \begin{align*}
        \frac{1}{n} \boldsymbol{Z}' \boldsymbol{\zeta} - \frac{1}{n} \sum_{i=1}^n  \zeta_i 
        = \frac{1}{n} \sum_{i=1}^n (\boldsymbol{z}_i - 1) \zeta_i \xrightarrow[]{} 0
    \end{align*}
    as $n\to \infty$ and $N \to \infty$ with $Nn^{-1/2} \to 0$. Finally, as the number of expansion terms increases, for the truncation errors we have
    \begin{align*}
        \frac{1}{n} \boldsymbol{Z}' \boldsymbol{\eta}  - \frac{1}{n} \sum_{i=1}^n \eta_i
        = \frac{1}{n} \sum_{i=1}^n (\boldsymbol{z}_i - 1) \eta_i 
        %
        \xrightarrow[]{} 0.
    \end{align*}

    Therefore, under Assumption \ref{assumption:observation_errors} on the observation error scheme, the OLS estimates $\widehat{\theta}$ are consistent and asymptotically unbiased.

    To derive the limit distribution, we first introduce a matrix $ \boldsymbol{\tilde\Psi}:= \{w_j\psi_{ij} \Delta_n\}_{i,j=1,\dots,n} $ with its $j$-th column denoted as $\boldsymbol{\tilde\Psi}_{\cdot j}$.
    Rearranging the terms, we have:
    \begin{align*}
        \boldsymbol{Z}' \boldsymbol{\xi} 
        = \sum_{i=1}^n \boldsymbol{z}_i \xi_i 
        = \sum_{j=1}^n \sum_{i=1}^n \boldsymbol{z}_i w_j \psi_{ij} \Delta_n \varepsilon_j 
        = \sum_{j=1}^n \boldsymbol{Z}' \boldsymbol{\tilde\Psi}_{\cdot j} \varepsilon_j 
        = \boldsymbol{Z}' \boldsymbol{\tilde\Psi}  \boldsymbol{\epsilon}.
    \end{align*}
    Note that each term $\boldsymbol{Z}' \boldsymbol{\tilde\Psi}_{\cdot j} = \sum_{i=1}^n \boldsymbol{z}_i w_j \psi_{ij} \Delta_n = \mathcal{O}(1)$ since it converges to a finite interval $w_j\int_{\alpha}^{\beta} \boldsymbol{z}(x) \psi(x,K_j) \diff x$. 
    Finally, to incorporate the observation error of the last call option price~$\varepsilon_n$, we modify the last column of the matrix $\boldsymbol{\tilde\Psi}$ by adding the unit vector $\boldsymbol{1}_n$ of length $n$, i.e., $$\boldsymbol{\Psi} := \left[ \boldsymbol{\tilde\Psi}_{\cdot 1:n{-}1} , \boldsymbol{\tilde\Psi}_{\cdot n} + \boldsymbol{1}_n \right].$$
    Therefore, the limit distribution of $\widehat{\boldsymbol{\theta}}$ is given by the CLT as
    \begin{align*}
        n^{-1/2} \boldsymbol{Z}' \left(  \boldsymbol{\epsilon} - \varepsilon_n - \boldsymbol{\xi} \right)
        = n^{-1/2} \boldsymbol{Z}' \left(  \boldsymbol{\epsilon} - \boldsymbol{\Psi} \boldsymbol{\epsilon} \right) \xrightarrow[]{d} \mathcal{N}(0, \boldsymbol{M}),
    \end{align*}
    with
    \begin{align*}
        \boldsymbol{M} = \plim n^{-1} \boldsymbol{Z}' ( \boldsymbol{\epsilon} - \boldsymbol{\Psi} \boldsymbol{\epsilon} ) ( \boldsymbol{\epsilon} - \boldsymbol{\Psi} \boldsymbol{\epsilon} )' \boldsymbol{Z} 
        =\lim n^{-1} \boldsymbol{Z}' \left( \boldsymbol{\Sigma_\epsilon} - 2 \boldsymbol{\Psi} \boldsymbol{\Sigma_\epsilon} + \boldsymbol{\Psi} \boldsymbol{\Sigma_\epsilon}\boldsymbol{\Psi}' \right) \boldsymbol{Z},
    \end{align*}
    where $\boldsymbol{\Sigma_\epsilon} := \mbox{diag}(\sigma_1^2, \dots, \sigma_n^2) $. Denoting $\boldsymbol{M_{zz}} := \lim n^{-1} \boldsymbol{Z}'\boldsymbol{Z} $, we get the limit distribution 
    \begin{align*}
        \sqrt{n}(\widehat{\boldsymbol{\theta}} - \boldsymbol{\theta}) \xrightarrow[]{d} \mathcal{N}\left(0,\boldsymbol{M_{zz}}^{-1} \boldsymbol{M} \boldsymbol{M_{zz}}^{-1}\right).
    \end{align*}

    Hence, the variance of $\widehat{\boldsymbol{\theta}}$ is given by 
    \begin{align*}
        \mbox{Var}(\widehat{\boldsymbol{\theta}}) 
        = \left(\boldsymbol{Z}' \boldsymbol{Z}\right)^{-1}\boldsymbol{Z}' \left( \boldsymbol{\Sigma_\epsilon} - 2 \boldsymbol{\Psi} \boldsymbol{\Sigma_\epsilon} + \boldsymbol{\Psi} \boldsymbol{\Sigma_\epsilon}\boldsymbol{\Psi}' \right) 
        \boldsymbol{Z} (\boldsymbol{Z}' \boldsymbol{Z})^{-1}.
    \end{align*}
    \hfill $\square$

    \begin{lemma}\label{lemma: feasible ols}
        Feasible estimates for the variance matrix of the observation errors in option prices can be obtained from the residuals of the OLS regression  $\widehat{\boldsymbol{e}} := \boldsymbol{c} - \widehat{\boldsymbol{c}}$ from Lemmas \ref{lemma:ols} as $ \widehat{\boldsymbol{\Sigma}}_\epsilon = \frac{n}{\nu} \mbox{diag}(\widehat{e}_1^2,\dots, \widehat{e}_n^2)$ with 
        $ \nu 
        = \mbox{tr}\left( \boldsymbol{Q} \right) - 2\mbox{tr}\left( \boldsymbol{Q}\boldsymbol{\Psi} \right) + \mbox{tr}\left( \boldsymbol{Q}\boldsymbol{\Psi}\boldsymbol{\Psi}' \right)$ with $\boldsymbol{Q} = \boldsymbol{I} - \boldsymbol{Z} (\boldsymbol{Z}' \boldsymbol{Z})^{-1} \boldsymbol{Z}'$. 
    \end{lemma}

    \textit{Proof:}
    In the standard linear regression setting, the feasible version of the covariance matrix can be obtained using the OLS residuals defined as $\widehat{\boldsymbol{e}}:= \boldsymbol{c} - \boldsymbol{Z}\widehat{\boldsymbol{\theta}}$.
    However, some finite sample correction might be needed to obtain unbiased estimates of the error variances. Using the projection matrix argument with $\boldsymbol{Q} = \boldsymbol{I} - \boldsymbol{Z} (\boldsymbol{Z}' \boldsymbol{Z})^{-1} \boldsymbol{Z}'$, we can express the OLS residuals as:
    \begin{align*}
        \widehat{\boldsymbol{e}} = \boldsymbol{c} - \widehat{\boldsymbol{c}} = \boldsymbol{Q}\boldsymbol{e} = \boldsymbol{Q}(\boldsymbol{I} - \boldsymbol{\Psi})\boldsymbol{\epsilon},
    \end{align*} 
    where we ignore asymptotically vanishing terms. If option errors were homoskedastic with an error variance term $\sigma^2_\varepsilon$, the sum of squared residuals would estimate
    \begin{align*}
        \E[\widehat{\boldsymbol{e}}' \widehat{\boldsymbol{e}}] 
        = \E\left[ \mbox{tr}\left( \boldsymbol{\epsilon} \boldsymbol{\epsilon}'  (\boldsymbol{I} - \boldsymbol{\Psi})^\prime \boldsymbol{Q}(\boldsymbol{I} - \boldsymbol{\Psi}) \right) \right] = \nu \sigma^2_\varepsilon 
    \end{align*}
    with
    \begin{align*}
        \nu:= \mbox{tr}\left( (\boldsymbol{I} - \boldsymbol{\Psi})^\prime \boldsymbol{Q}(\boldsymbol{I} - \boldsymbol{\Psi}) \right)
        = \mbox{tr}\left( \boldsymbol{Q} \right) - 2\mbox{tr}\left( \boldsymbol{Q}\boldsymbol{\Psi} \right) + \mbox{tr}\left( \boldsymbol{Q}\boldsymbol{\Psi}\boldsymbol{\Psi}' \right),
        %
        %
    \end{align*}
    where $\mbox{tr}\left( \boldsymbol{Q} \right) = n-3$, and the traces of the last two terms can be easily computed in practice. 
    
    Hence, using the degrees of freedom correction term $n / \nu$ yields unbiased estimates for the variance of observation errors under the homoskedastic error assumption. 
    Although we do not impose homoskedasticity, in practice, we adjust the squared residuals for degrees of freedom $\nu$ to obtain estimates of the error variances in finite samples, that is, we use
    $\widehat{\boldsymbol{\Sigma}}_\epsilon := \frac{n}{\nu} \mbox{diag}(\widehat{e}_1^2,\dots, \widehat{e}_n^2)$. This is similar to the one conventionally used to obtain heteroskedasticity-consistent standard errors of parameter estimates in the context of linear regression \cite{mackinnon1985some}. Importantly, the matrix $\boldsymbol{\Psi}$ and, hence, its trace are functions of the number of expansion terms~$N$. Therefore, we correct for degrees of freedom with every cosine expansion term.

    \hfill $\square$

\textit{Proof of Proposition \ref{prop:iCall}:}

    Using results and notations defined in Lemma \ref{lemma:ols}, we have 
    \begin{align*}
        \widehat{C}(x) - C_0(x) & = \left(\widehat{\overline{C}}(x) + C(\beta) + \boldsymbol{z}(x)^\prime\widehat{\boldsymbol{\theta}} \right) - \left(\overline{C}_0(x) + C_0(\beta) + Z_c(x)\theta_c + Z_p(x)\theta_p \right) \\
        & = \widehat{\overline{C}}(x) - \overline{C}_0(x)  + \boldsymbol{z}(x)^\prime\widehat{\boldsymbol{\theta}} + \bar{\theta} - \bar{\theta} +  \varepsilon_n - Z_c(x)\theta_c - Z_p(x)\theta_p\\
        &= \xi(x) +  \varepsilon_n + \zeta(x) - \eta(x) + \bar{\theta} + \boldsymbol{z}(x)^\prime(\widehat{\boldsymbol{\theta}} - \boldsymbol{\theta}).
    \end{align*}

    The last equality follows since 
    \begin{align*}
        - \eta(x) &= \overline{\eta}(x) - \left( \sum_{m=N}^{\infty} (-1)^m  H_m(x) \right) \theta_c + \left( \sum_{m=N}^{\infty} H_m(x) \right) \theta_p\\
        &=- \sum_{m=N}^{\infty} D_m H_m(x) - \sum_{m=N}^{\infty} b_m  H_m(x) = - e^{-rT} \sum_{m=N}^{\infty} A_m  H_m(x).     
    \end{align*}


    Given Proposition \ref{prop:iCall_bar} and Lemma \ref{lemma:ols}, we obtain a consistent and asymptotically unbiased call price estimator~$\widehat{C}(x)$. A finite-sample bias in this estimator arises from the deviation of the discretization and truncation errors for an option with a strike price $x$ from the average bias across all contracts~$\bar{\theta} = \frac{1}{n} \sum_{i=1}^n \eta_i - \frac{1}{n} \sum_{i=1}^n \zeta_i$. Therefore, the magnitude of the bias term is sufficiently smaller than that in the $\widehat{\overline{C}}(x)$ estimator. The order, however, is the same since $\overline{\eta}(x)$ dominates the other term in $\eta(x)$ as the density is smoother than the payoff function of the call option. Therefore, we have that $\eta(x)  - \frac{1}{n} \sum_{i=1}^n \eta_i = \mathcal{O}\left( N^{1-p} \right)$ and $\zeta(x)  - \frac{1}{n} \sum_{i=1}^n \zeta_i =  \mathcal{O}\left(\frac{N^{1+\iota}}{n^{\iota}} \right)$ with $\iota \geq 1$. That is, the call price estimator is in fact asymptotically unbiased.

    The limiting distribution of the call price estimator is determined by the first and the last terms since $\sqrt{n}\xi(x) = \mathcal{O}_p(1)$ and so is the OLS estimator. Hence, we need to combine the limiting distributions resulting from $\widehat{\overline{C}}(x)$ and $\widehat{\theta}$.
    For that, let us first denote 
    \[
        \boldsymbol{\psi}_w(x) := \left( w_1\psi(x,K_1) \Delta_n , \dots, w_n\psi(x,K_n) \Delta_n +1 \right).
    \] 
    That is, if $x=K_i$ for some $i=1,\dots,n$, then $\boldsymbol{\psi}_w(K_i) = \boldsymbol{\Psi}_{i\cdot}$ is the $i$-th row of the matrix $\boldsymbol{\Psi}$.
    Then, we can express the first and the last terms as 
    \begin{align*}
        \sqrt{n}\left( \xi(x) + \varepsilon_n + \boldsymbol{z}(x)^\prime(\widehat{\boldsymbol{\theta}} - \boldsymbol{\theta}) \right)
        &= \sqrt{n}\left( \boldsymbol{\psi}_w(x) \boldsymbol{\epsilon} + \boldsymbol{z}(x)^\prime \left( \boldsymbol{Z}' \boldsymbol{Z}\right)^{-1}\boldsymbol{Z}' (\boldsymbol{I} - \boldsymbol{\Psi}) \boldsymbol{\epsilon} \right)\\
        &= \sqrt{n}\left( \boldsymbol{\psi}_w(x)  + \boldsymbol{z}(x)^\prime \left( \boldsymbol{Z}' \boldsymbol{Z}\right)^{-1}\boldsymbol{Z}' (\boldsymbol{I} - \boldsymbol{\Psi})  \right) \boldsymbol{\epsilon},
    \end{align*}
    which converges by the CLT to the normal distribution with mean zero and the asymptotic variance matrix 
    \begin{align*}
        \boldsymbol{M_c} 
        &= \plim n 
        \left[ 
            \boldsymbol{\psi}_w(x)  + \boldsymbol{z}(x)^\prime \left( \boldsymbol{Z}' \boldsymbol{Z}\right)^{-1}\boldsymbol{Z}' (\boldsymbol{I} - \boldsymbol{\Psi}) 
        \right] 
        \boldsymbol{\epsilon} \boldsymbol{\epsilon}' 
        \left[ 
            \boldsymbol{\psi}_w(x)^\prime  + (\boldsymbol{I} - \boldsymbol{\Psi}') \boldsymbol{Z} \left( \boldsymbol{Z}' \boldsymbol{Z}\right)^{-1} \boldsymbol{z}(x)
        \right] \\
        &= \lim n \boldsymbol{\psi}_w(x) \boldsymbol{\Sigma_\epsilon} \boldsymbol{\psi}_w(x)^\prime + \boldsymbol{z}(x)' \boldsymbol{M_{zz}}^{-1} \boldsymbol{M} \boldsymbol{M_{zz}}^{-1} \boldsymbol{z}(x)\\
        &\ + 2\lim n \boldsymbol{z}(x)^\prime \left( \boldsymbol{Z}' \boldsymbol{Z}\right)^{-1}\boldsymbol{Z}' (\boldsymbol{I} - \boldsymbol{\Psi}) \boldsymbol{\Sigma_\epsilon} \boldsymbol{\psi}_w(x)^\prime.
    \end{align*}
    
    Therefore, the variance matrix of the call price estimator is given by
    \begin{align}\label{var call}
        \sigma_c^2(x):
        =&\ \boldsymbol{\psi}_w(x) \boldsymbol{\Sigma_\epsilon} \boldsymbol{\psi}_w(x)^\prime  
        + \boldsymbol{z}(x)' \mbox{Var}(\widehat{\boldsymbol{\theta}}) \boldsymbol{z}(x) 
        + 2 \boldsymbol{z}(x)^\prime \left( \boldsymbol{Z}' \boldsymbol{Z}\right)^{-1}\boldsymbol{Z}' (\boldsymbol{I} - \boldsymbol{\Psi}) \boldsymbol{\Sigma_\epsilon} \boldsymbol{\psi}_w(x)^\prime \notag\\
        =&\ \overline{\sigma}_c^2(x)
        + \boldsymbol{z}(x)' \mbox{Var}(\widehat{\boldsymbol{\theta}}) \boldsymbol{z}(x) 
        + 2 \boldsymbol{z}(x)^\prime \left( \boldsymbol{Z}' \boldsymbol{Z}\right)^{-1}\boldsymbol{Z}' (\boldsymbol{I} - \boldsymbol{\Psi}) \boldsymbol{\Sigma_\epsilon} \boldsymbol{\psi}_w(x)^\prime .
    \end{align}
    The feasible version of this variance can be obtain by using $ \widehat{\boldsymbol{\Sigma}}_\epsilon $ from Lemma~\ref{lemma: feasible ols}.

    \hfill $\square$

\textit{Proof of Proposition \ref{prop:iRND}:}

    The proof closely follows the proofs of Propositions \ref{prop:iCall_bar} and \ref{prop:iCall}. Therefore, for ease of notation, we first define analogous objects:
    \begin{align*}
        H_m^f(y) &:= \cos\left(u_m y - u_m \log \alpha \right),\  
        \psi^f(y, K):= \sum_{m=1}^{N-1} \psi_m(K) H_m^f(y), \\
        \boldsymbol{\psi}^f_w(y) &:= \left( w_1\psi^f(y,K_1) \Delta_n , \dots, w_n\psi^f(y,K_n) \Delta_n \right),\\
        \eta^f(y)&:= e^{-rT} \sum_{m=N}^{\infty } A_m H_m^f(y),\ 
        \xi^f(y):= \sum_{i=1}^n w_i \psi^f(x, K_i) \varepsilon_i \Delta_n = \boldsymbol{\psi}^f_w(y) \boldsymbol{\epsilon}, \\
        \zeta^f(y)&:= \sum_{i=1}^n w_i \psi^f(x, K_i)O_0(K_i) \Delta_n - \int_\alpha^\beta \psi^f(x, K) O_0(K) \diff K = \sum_{m=1}^{N-1}\zeta^D_{m,n} H_m^f(y),\\
        Z_c^f(y) &:= \sideset{}{'}\sum_{m=0}^{N-1} (-1)^m H_m^f(y),\ 
        Z_p^f(y) := \sideset{}{'}\sum_{m=0}^{N-1} H_m^f(y),\ \boldsymbol{z}^f(y):= \left(0, Z_c^f(y), Z_p^f(y)\right)'.
        %
    \end{align*}

    Repeating the steps from the proof of Proposition \ref{prop:iCall_bar}, we can decompose the (scaled) difference between the non-parametric RND estimator, defined in equation \eqref{icos-rnd-estimator} and the true RND given by the relation \eqref{icos-rnd} for a given point $y$:
    \begin{align*}
        \frac{\widehat{f}(y) - f(y) }{\nu_f }
        &=  \sideset{}{'}\sum_{m=0}^{N-1} \left(\widehat{D}_m + (-1)^m \widehat{\theta}_c - \widehat{\theta}_p \right) \cos\left(u_m y - u_m\log \alpha \right)\\
        & \quad - e^{-rT} \sideset{}{'}\sum_{m=0}^{\infty } A_m \cos\left(u_m y - u_m\log \alpha \right) \\
        &= \sideset{}{'}\sum_{m=0}^{N-1} \widehat{D}_m H_m^f(y) - \sideset{}{'}\sum_{m=0}^{N-1} D_m H_m^f(y) + \eta^f(y) + Z_c^f(y)(\widehat{\theta}_c - \theta_c)  + Z_p^f(y)(\widehat{\theta}_p - \theta_p)\\
        &= \xi^f(y) + \zeta^f(y) - \eta^f(y) + \boldsymbol{z}^f(y)^{\prime}(\widehat{\boldsymbol{\theta}} - \boldsymbol{\theta}).
    \end{align*}
    
    We notice, however, differently from Proposition \ref{prop:iCall_bar}, the order of the cosine coefficient $H_m^f(y) = \mathcal{O}(1)$ and, hence, $\psi^f(y, K) = \mathcal{O}(N^3)$. This further results in different orders for the observation and discretization errors, namely $\xi^f(y) = \mathcal{O}_p\left(N^3 n^{-1/2}\right)$ and $\zeta^f(y) = \mathcal{O}\left(N^{3+\iota}n^{-\iota}\right)$, respectively. Therefore, to guarantee the asymptotic unbiasedness of the RND estimator, we require $N$ to grow slower than $n^{-1/6}$, i.e. $Nn^{-1/6} \to 0$.

    Finally, using similar arguments as in the proof of Proposition \ref{prop:iCall}, we can establish the limiting distribution of the non-parametric RND estimator:
    \[
        \frac{ \widehat{f}(y) - f(y) }{ \nu_f \sigma_f(y)}  \xrightarrow[]{d} \mathcal{N} \left( 0 , 1 \right),
    \]
    as $n \to \infty$ and $N \to \infty$ with $Nn^{-1/6} \to 0$ and the covariance matrix
    \begin{align}\label{var rnd}
        \sigma_f^2(y) := 
        \boldsymbol{\psi}^f_w(y) \boldsymbol{\Sigma_\epsilon} \boldsymbol{\psi}^f_w(y)^\prime  
        + \boldsymbol{z}^f(y)' \mbox{Var}(\widehat{\boldsymbol{\theta}}) \boldsymbol{z}^f(y) 
        + 2 \boldsymbol{z}^f(y)^\prime \left( \boldsymbol{Z}' \boldsymbol{Z}\right)^{-1}\boldsymbol{Z}' (\boldsymbol{I} - \boldsymbol{\Psi}) \boldsymbol{\Sigma_\epsilon} \boldsymbol{\psi}_w^f(y)^\prime.
        \end{align}
    \hfill $\square$

    \textit{Proof of Proposition \ref{prop:iDelta}:}

    Similarly, we can decompose the scaled difference between the non-parametric delta estimator and the true value given by the spanning result \eqref{iDelta} under Assumption \ref{assumption:delta}:
     \begin{align*}
        S_0 \left(  \widehat{\delta}(x) - \delta(x) \right)
         &= -\sum_{m=1}^{N-1} u_m \widehat{B}_m  H_m(x) + \left(C(\beta) - \beta \widehat{\theta}_c \right) 
         + \sum_{m=1}^{\infty} u_m B_m  H_m(x) - \left(C_0(\beta) - \beta \theta_c \right) \\ 
         &= -\sum_{m=1}^{N-1} u_m \sum_{i=1}^n w_i \widetilde{\psi}_m(K_i) O(K_i) \Delta_n H_m(x) 
         +  \sum_{m=1}^{N-1} u_m \int_\alpha^\beta \widetilde{\psi}_m(K)O_0(K) \diff K H_m(x) \\
         &\quad + \sum_{m=N}^{\infty} u_m B_m  H_m(x) + C(\beta) - C_0(\beta) - \beta( \widehat{\theta}_c - \theta_c ) \\
         &\quad + \sum_{m=1}^{N-1} \frac{u_m^2}{\beta}(-1)^m H_m(x) \left( C(\beta) - C_0(\beta) \right) - \sum_{m=1}^{N-1} \frac{u_m^2}{\alpha} H_m(x) \left( P(\alpha) - P_0(\alpha) \right),\\
         %
     \end{align*}
     which can further be written as 
     \begin{align*}
        S_0 \left(  \widehat{\delta}(x) - \delta(x) \right)
         %
         &= -\sum_{i=1}^n w_i \underbrace{\sum_{m=1}^{N-1} u_m \widetilde{\psi}_m(K_i) H_m(x) }_{=:\widetilde{\psi}(x, K_i)}  O(K_i) \Delta_n  
         + \int_\alpha^\beta \underbrace{\sum_{m=1}^{N-1} u_m \widetilde{\psi}_m(K) H_m(x) }_{=:\widetilde{\psi}(x, K)} O_0(K) \diff K \\
         &\quad + \underbrace{ \sum_{m=N}^{\infty} u_m B_m  H_m(x) }_{=:\widetilde{\eta}(x)}
         + \underbrace{ \left( 1 + \sum_{m=1}^{N-1} \frac{u_m^2}{\beta}(-1)^m H_m(x) \right) }_{=:\widetilde{Z}^N_c(x)}  \varepsilon_n \underbrace{ - \sum_{m=1}^{N-1} \frac{u_m^2}{\alpha} H_m(x) }_{ =:\widetilde{Z}^N_p(x) } \varepsilon_1\\
         &\quad - \beta( \widehat{\theta}_c - \theta_c ) \\
         &= -\sum_{i=1}^n w_i \widetilde{\psi}(x, K_i)  O(K_i) \Delta_n  
            + \int_\alpha^\beta \widetilde{\psi}(x, K) O_0(K) \diff K + \widetilde{\eta}(x) \\
         &\quad + \widetilde{Z}^N_c(x) \varepsilon_n + \widetilde{Z}^N_p(x) \varepsilon_1 - \beta( \widehat{\theta}_c - \theta_c ).
     \end{align*}

    The difference between the first two terms can be similarly decomposed into the observation and discretization errors. Therefore, we have
    \begin{align*}
        S_0 \left(  \widehat{\delta}(x) - \delta(x) \right)
        = \widetilde{\xi}(x) + \widetilde{\zeta}(x) + \widetilde{\eta}(x) + \widetilde{Z}^N_c(x) \varepsilon_n + \widetilde{Z}^N_p(x) \varepsilon_1 - \beta( \widehat{\theta}_c - \theta_c ).
    \end{align*}

    Similar to the proofs of Propositions \ref{prop:iCall} and \ref{prop:iRND}, we can establish the limiting distribution of the non-parametric delta estimator:
    \begin{align*}
        \frac{ \widehat{\delta}(x) - \delta(x) }{ \tfrac{1}{S_0} \sigma_\delta(y)}  \xrightarrow[]{d} \mathcal{N} \left( 0 , 1 \right)
    \end{align*}
    as $n \to \infty$ and $N \to \infty$ with $Nn^{-1/2} \to 0$ since the order of $\widetilde{\psi}_m(x)$ is the same as of $\psi_m(x)$ and $\widetilde{\eta}(x)$ decreases with an increase of $N$.
    Here, the variance matrix estimator of $\widehat{\delta}(x)$ is given by 
    \begin{align*}
       \sigma_\delta^2(x) := 
       \boldsymbol{\widetilde{\psi}}_w(x) \boldsymbol{\Sigma_\epsilon} \boldsymbol{\widetilde{\psi}}_w(x)^\prime  
       + \beta^2 \mbox{Var}(\widehat{\theta}_c) 
       + 2 \boldsymbol{z}^\delta(x)^\prime \left( \boldsymbol{Z}' \boldsymbol{Z}\right)^{-1}\boldsymbol{Z}' (\boldsymbol{I} - \boldsymbol{\Psi}) \boldsymbol{\Sigma_\epsilon} \boldsymbol{\widetilde{\psi}}_w(x)^\prime 
    \end{align*}
    with $\boldsymbol{z}^\delta(x)^\prime = (0,-\beta, 0)$ and
    \begin{align*}
        \boldsymbol{\widetilde{\psi}}_w(x) &:= \left( w_1 \widetilde{\psi}(x,K_1) \Delta_n + \widetilde{Z}^N_p(x), w_2 \widetilde{\psi}(x,K_2) \Delta_n, \dots, w_{n-1} \widetilde{\psi}(x,K_{n-1}) \Delta_n, w_n \widetilde{\psi}(x,K_n) \Delta_n + \widetilde{Z}^N_c(x) \right).
    \end{align*}
    \hfill $\square$

        \section{Additional results}\label{appendix:additional}

    \subsection{Put options}

    In this section we provide the results for put options.
    In particular, for the plain vanilla put option with a strike price $x>\alpha$, the value on the interval $(0,\alpha)$ is given by 
    \begin{align*}
        P_0^{(0,\alpha)}(x) &= e^{-rT}\E^{\Q}[\max(x-S_T, 0)\mathbf{1}_{\{S_T < \alpha\}}] = e^{-r\tau} \int_0^\alpha \max(x - S_T, 0) f_S(S_T) \diff S_T\\
        &= e^{-rT}(x -\alpha)\int_0^\alpha f_S(S_T) \diff S_T + e^{-rT}\int_0^\infty \max(\alpha - S_T,0) f_S(S_T) \diff S_T\\
        &= (x -\alpha)P_K'(\alpha) + P_0(\alpha),
    \end{align*}
    where $P_0(\alpha)$ is the put price with the strike $\alpha$ and $P_K'(\alpha)$ is its derivative with respect to the strike price evaluated at $\alpha$. $P_0^{(0,\alpha)}(x)$ represents the price of a so-called \textit{gap put option} with a strike price $x$ and a trigger price $\alpha$.

    Therefore, for the put option price with strike price $x$ such that $\alpha \leq x \leq \beta$, we have:
    \begin{align}
        P_0(x) = P_0^{[\alpha, \beta]} (x) +(x -\alpha)P_K'(\alpha) + P_0(\alpha).
    \end{align}

    The value of the put contract with a payoff restricted to the interval $[\alpha, \beta]$ is given by the COS formula, similar to the call counterpart, as
    \begin{align}
        P_0^{[\alpha, \beta]} (x) 
        = \sideset{}{'}\sum_{m=0}^{\infty} D_m  H^p_m(x) + \left(\sideset{}{'}\sum_{m=0}^{\infty} (-1)^m  H^p_m(x) \right) \theta_c 
        - \left(\sideset{}{'}\sum_{m=0}^{\infty}  H^p_m(x) \right) \theta_p,
    \end{align}
    where $H^p_m(x)$ is the cosine series coefficient for the put payoff function with the strike price $x$ and $\theta_c = C_K'(\beta) $ and $ \theta_p = P_K'(\alpha) $ are the first-order derivatives of the call and put options evaluated at the boundaries of the interval. Therefore, the put option can be represented as the following portfolio
    \begin{align}
        P_0(x) &= P_0^{[\alpha, \beta]} (x) + (x -\alpha)\theta_p  + P_0(\alpha)\notag \\
        &= \underbrace{\sideset{}{'}\sum_{m=0}^{\infty} D_m  H^p_m(x)}_{\strut =:\overline{P}_0(x)} + P_0(\alpha) 
        + \underbrace{\left( \sideset{}{'}\sum_{m=0}^{\infty} (-1)^m  H^p_m(x) \right)}_{\strut =:Z^p_c(x) }  \theta_c + 
        \underbrace{  \left( x - \alpha -  \sideset{}{'}\sum_{m=0}^{\infty}  H^p_m(x) \right)}_{\strut =:Z^p_p(x)} \theta_p \notag \\
        \label{iPut}
        &= \overline{P}_0(x) + P_0(\alpha) + Z^p_c(x) \theta_c + Z^p_p(x) \theta_p.
    \end{align}
    This spanning result for the put option is analogous to the call price representation given by equation \eqref{iCall}. This result can be used to further construct the computationally feasible non-parametric put price estimator similar to the call options.

\subsection{Cosine coefficients}
    The cosine coefficients $H_m$ that correspond to the cosine transformation of the call payoff with the strike price $x$ and transformed variable $y = \log \frac{S_T}{x}$ are given by
    \begin{align*}
        H_m(x) &= \frac{2}{b-a} \int_a^b x\max\{e^y -1, 0\} \cos(u_m y - u_m a) \diff y\\
        &= \frac{2}{b-a} \int_0^b x (e^y -1 ) \cos(u_m y - u_m a) \diff y.
    \end{align*}
    Using the result of \citeA[eq.\ 24]{fang2008novel} and the interval bounds $a=\log \frac{\alpha}{x}$ and $b=\log \frac{\beta}{x}$ from Section \ref{sec:iCOS}, we have for $m>0$
    \begin{align*}
        H_m(x) &= \frac{2x}{b-a}\left[\frac{1}{1 + u_m^2}\left( \cos(m\pi) e^{b} - \cos(-u_m a) - u_m \sin(-u_m a)\right) + \frac{1}{u_m} \sin(-u_m a) \right]\\
        &= \frac{2x}{\log \frac{\beta}{\alpha}}\left[\frac{1}{1 + u_m^2}\left( \cos(m\pi) \frac{\beta}{x} - \cos\left(u_m \log \frac{\alpha}{x}\right) + u_m \sin\left(u_m \log \frac{\alpha}{x}\right)\right) - \frac{1}{u_m} \sin\left(u_m \log \frac{\alpha}{x}\right) \right]\\
        &= \frac{2x}{u_m(1 + u_m^2) \log \frac{\beta}{\alpha}} 
        \left( (-1)^m \frac{u_m \beta }{x} - u_m \cos\left(u_m \log \frac{\alpha}{x}\right) - \sin\left(u_m \log \frac{\alpha}{x}\right)\right) .
    \end{align*}

    A similar result can be obtained for the put options. These cosine coefficients are deterministic functions of the strike price and the fixed interval bounds $\alpha$ and $\beta$.

\subsection{Optimal number of expansion terms}

    In this section we provide a rule-of-thumb algorithm for the optimal number of expansion terms~$N$. 
    The algorithm operationalizes the result that if $A_N^2 - \mbox{Var}(\widehat{A}_N) > 0$, then $\mbox{AMISE}_{N+1} < \mbox{AMISE}_{N}$ following the discussion in Section \ref{sec:optimal N}.

    The estimated cosine coefficients $\widehat{A}_m$ are given by 
    \begin{align}\label{Am_hat}
        \widehat{A}_m = e^{rT} \left( \widehat{D}_m + (-1)^m \widehat{\theta}_c - \widehat{\theta}_p \right),
    \end{align}
    where $\widehat{D}_m $ is given by equation \eqref{iDm_hat}, and $\widehat{\theta}_c$ and $\widehat{\theta}_p$ are the OLS estimates from regression \eqref{ols}. Following the proof steps and notations in Lemma \ref{lemma:ols} and Proposition \ref{prop:iCall}, the variance of the cosine coefficients is given by
    \begin{align}\label{var Am}
        \sigma_A^2(m) := \mbox{Var}\left(\widehat{A}_m\right) 
        &= \boldsymbol{\psi}_m \boldsymbol{\Sigma_\epsilon} \boldsymbol{\psi}_m^\prime  
        + \boldsymbol{z}_A' \mbox{Var}(\widehat{\boldsymbol{\theta}}) \boldsymbol{z}_A 
        + 2 \boldsymbol{z}_A^\prime \left( \boldsymbol{Z}' \boldsymbol{Z}\right)^{-1}\boldsymbol{Z}' (\boldsymbol{I} - \boldsymbol{\Psi}) \boldsymbol{\Sigma_\epsilon} \boldsymbol{\psi}_m^\prime \notag\\
        &= \sigma_D^2(m)
        + \boldsymbol{z}_A^\prime \mbox{Var}(\widehat{\boldsymbol{\theta}}) \boldsymbol{z}_A 
        + 2 \boldsymbol{z}_A^\prime \left( \boldsymbol{Z}' \boldsymbol{Z}\right)^{-1}\boldsymbol{Z}' (\boldsymbol{I} - \boldsymbol{\Psi}) \boldsymbol{\Sigma_\epsilon} \boldsymbol{\psi}_m^\prime,
    \end{align}
    where $\boldsymbol{z}_A := \left(0, (-1)^m, -1\right)^\prime$, $\boldsymbol{\psi}_m := \left( w_1\psi_m(K_1) \Delta_n , \dots, w_n\psi(K_n) \Delta_n \right)$, and $\sigma_D^2(m)$ is the variance of the estimator $\widehat{D}_m$ given in Proposition \ref{prop:Dm}. The feasible variances are obtained by using the feasible covariance matrix $\widehat{\boldsymbol{\Sigma}}_\epsilon$ as discussed in the proof of Lemma \ref{lemma:ols}.

    For a given number of expansion terms $N$, we collect the estimated cosine coefficients and their feasible standard deviations into the vectors:
    \begin{align*}
        \boldsymbol{\widehat{A}} := \left( \widehat{A}_1, \dots, \widehat{A}_N \right)^\prime \mbox{ and }
        \boldsymbol{\widehat{\sigma}}_A := \left( \widehat{\sigma}_A(1), \dots, \widehat{\sigma}_A(N) \right)^\prime.
    \end{align*}
    In the algorithm below, we denote the calculation of these vectors $\boldsymbol{\widehat{A}}$ and $\boldsymbol{\widehat{\sigma}}_A$ by the function $iCOSA(\boldsymbol{K}, \boldsymbol{C}, N)$, which takes as inputs the data vectors $\boldsymbol{K}$ and $\boldsymbol{C}$ of strike prices and call prices, and the number of expansion terms $N$. 

    Since the true value of the cosine coefficient $A_N$ is unknown, we use the estimated value $\widehat{A}_N$ instead, averaged over the adjacent coefficients. 
    In particular, we notice that the true cosine coefficients $A_m$ exponentially decay to zero as $m$ increases with a rate depending on the smoothness of the density. Therefore, the coefficients are averaged out after taking the logarithm of they absolute values. This average is then compared to the logarithm of the standard deviation of the coefficients. This rule-of-thumb is sketched in Algorithm \ref{alg:optimal N}.

    \begin{algorithm}[ht]
        \caption{Optimal number of expansion terms}
        \label{alg:optimal N}
        \begin{algorithmic}
            \Require $\boldsymbol{K}, \boldsymbol{C}$
            \Ensure $N^*$
            \State $N = 5,\ N_{max} = 50$
            \State $\bar{a} = 1,\ s_a = 0$
            \While{$\bar{a} > s_a$ \& $N < N_{max}$} 
                \State $N = N + 1$
                \State $\boldsymbol{\widehat{A}},\ \boldsymbol{\widehat{\sigma}_A} = iCOSA(\boldsymbol{K}, \boldsymbol{C}, N)$
                \State $\bar{a} = \frac{1}{3} (\log | \boldsymbol{\widehat{A}}_{[N-2]} | + \log | \boldsymbol{\widehat{A}}_{[N-1]} | + \log | \boldsymbol{\widehat{A}}_{[N]} |)$
                \State $s_a = \log \boldsymbol{\widehat{\sigma}_A}_{[N-1]}$
            \EndWhile
            \State $N^* = N - 1$
        \end{algorithmic}
    \end{algorithm}

    \end{appendices}
    
    \clearpage
    \bibliography{lit}
    \bibliographystyle{apacite}

\end{document}